\documentclass[a4paper,11pt]{article}
\pdfoutput=1

\usepackage{jheppub}

\title{\boldmath Real-space RG, error correction and Petz map}
\author[a]{Keiichiro Furuya,}
\author[a,b,1]{Nima Lashkari\note{Corresponding author.}}
\author[a]{and Shoy Ouseph}

\affiliation[a]{Department of Physics and Astronomy, Purdue University, West Lafayette, IN
47907, USA}
\affiliation[b]{School of Natural Sciences, Institute for Advanced Study, Princeton, New Jersey 08540, USA}

\emailAdd{kfuruya@purdue.edu}
\emailAdd{nima@purdue.edu}
\emailAdd{souseph@purdue.edu}

\abstract{There are two parts to this work: 

First, we study the error correction properties of the real-space renormalization group (RG). The long-distance operators are the (approximately) correctable operators encoded in the physical algebra of short-distance operators.  This is closely related to modeling the holographic map as a quantum error correction code. As opposed to holography, the real-space RG of a many-body quantum system does not have the complementary recovery property. We discuss the role of large $N$ and a large gap in the spectrum of operators in the emergence of complementary recovery.

Second, we study the operator algebra exact quantum error correction for any von Neumann algebra. We show that similar to the finite dimensional case, for any error map in between von Neumann algebras the Petz dual of the error map is a recovery map if the inclusion of the correctable subalgebra of operators has finite index.}

\usepackage{natbib}
\usepackage{braket}
\newtheorem{theorem}{Theorem}
\usepackage[utf8]{inputenc}
\usepackage{todonotes}
\usepackage{amssymb,amsmath,amsbsy}
\usepackage{braket}
\usepackage{graphicx,color}
\usepackage{hyperref}
\usepackage{multirow}
\usepackage{mathtools}

\DeclarePairedDelimiter\floor{\lfloor}{\rfloor}
\newcommand{\mA}{\mathcal{A}}

\newcommand{\mB}{\mathcal{B}}
\newcommand{\mC}{\mathcal{C}}
\newcommand{\mS}{\mathcal{S}}

\newcommand{\mT}{\mathcal{T}}

\newcommand{\mK}{\mathcal{K}}
\newcommand{\mE}{\mathcal{E}}
\newcommand{\p}{\partial}

\newcommand{\mI}{\mathbb{I}}
\newcommand{\mH}{\mathcal{H}}
\newcommand{\nn}{\nonumber}
\newcommand{\lb}{\left(}
\newcommand{\rb}{\right)}
\newcommand{\ee}{\end{equation}}
\newcommand{\bea}{\begin{eqnarray}}
\newcommand{\eea}{\end{eqnarray}}
\newcommand{\mO}{\mathcal{O}}
\newcommand{\mR}{\mathcal{R}}
\newcommand{\mcD}{\mathcal{D}}

\newcommand{\tr}{\text{tr}}

\makeatletter
\begin{document} 

\maketitle

\flushbottom

\section{Introduction}\label{sec:intro}

In quantum computing, we use the Hilbert space of a quantum system to encode and process information. The interactions with the environment lead to errors and an important challenge is to protect our information from the errors. One of the main goals of the theory of quantum error correction (QEC) is to identify the subalgebra of correctable operators associated to an error model, and construct the recovery map that undoes the errors.\footnote{For completeness, we have included a review of the theory of operator algebra error correction in appendix \ref{app:errorcorrection}. See also \cite{kribs2005unified,beny2007quantum,beny2007generalization}.} 

\begin{figure}[t]
    \centering
    \includegraphics[width=0.5\linewidth]{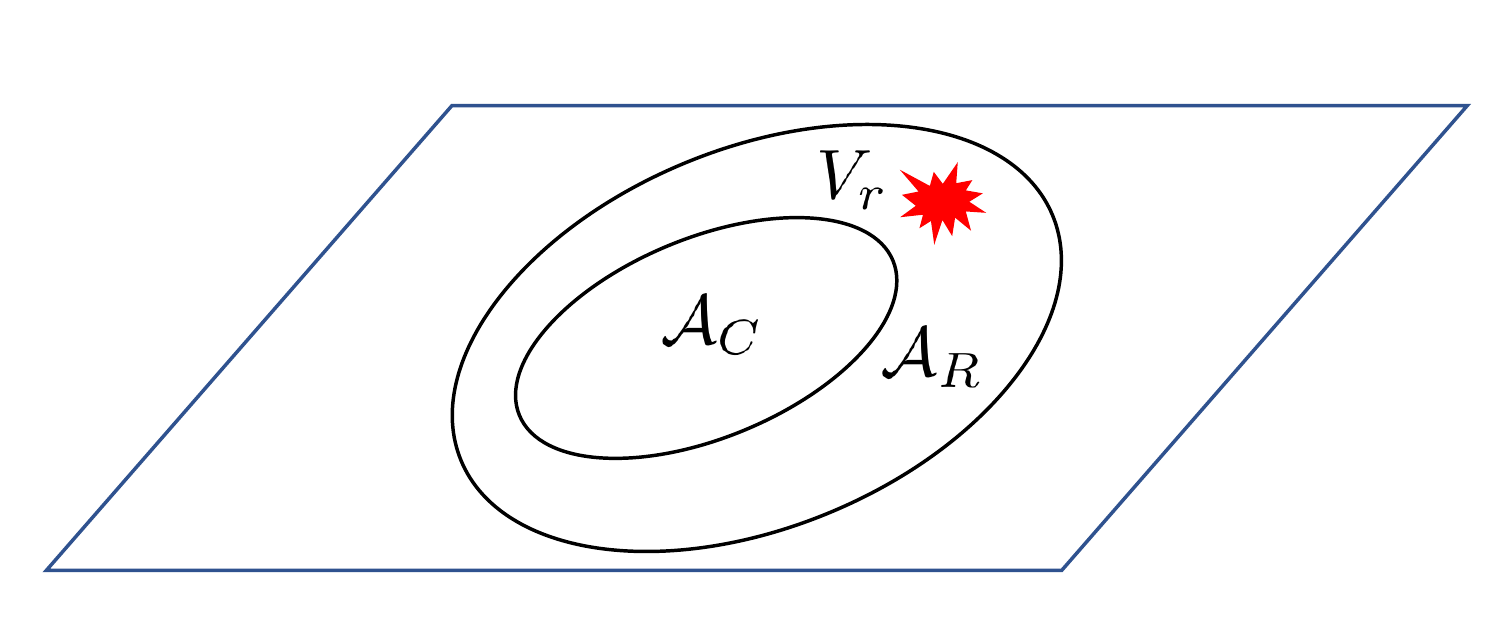}
    \caption{The local algebra of region $C$ is a subalgebra of the algebra of a larger region. Any error $V_r$ that acts on the relative commutant $\mA_R$ does not disturb the encoded information in $\mA_C$.}
    \label{fig0u}
\end{figure}

In local many-body quantum systems, to every subregion of space $A$ we associate an algebra of observables $\mathcal{A}_A$ that includes the identity operator.\footnote{In interacting relativistic theories, we associate an algebra to the causal development of every ball.} A manifestation of the principle of locality is that if the region $C$ is inside $A$ we have the inclusion of algebras $\mA_C\subseteq \mA_A$. If we have a lattice the algebra $\mA_A$ factors as $\mA_A=\mA_C\otimes \mA_R$ for some $\mA_R$ that is called the relative commutant of $\mA_C$ in $\mA_A$.\footnote{The relative commutant algebra $\mA_C$ in $\mA_A$ is the set of all operators in $\mA_A$ that commute with every operator in  $\mA_C$. The commutant of an algebra $\mA$ that we denote by $\mA'$ is the set of all operators in the Hilbert space that commute with all operators in $\mA$.} The relative commutant is the algebra associated to the region $A\cap C'$. Any such inclusion is trivially an exact quantum error correction code in the following sense: the physical operators are $\mA_A$ and the logical operators are encoded in the subalgebra $\mA_C$. The errors act on the relative commutant $\mA_R$, and by locality, the errors do not disturb the encoded information because $[a,V_r]=0$ for all $a\in \mA_C$ and any error $V_r\in \mA_R$; see figure \ref{fig0u}. 

\begin{figure}[t]
    \centering
    \includegraphics[width=0.8\linewidth]{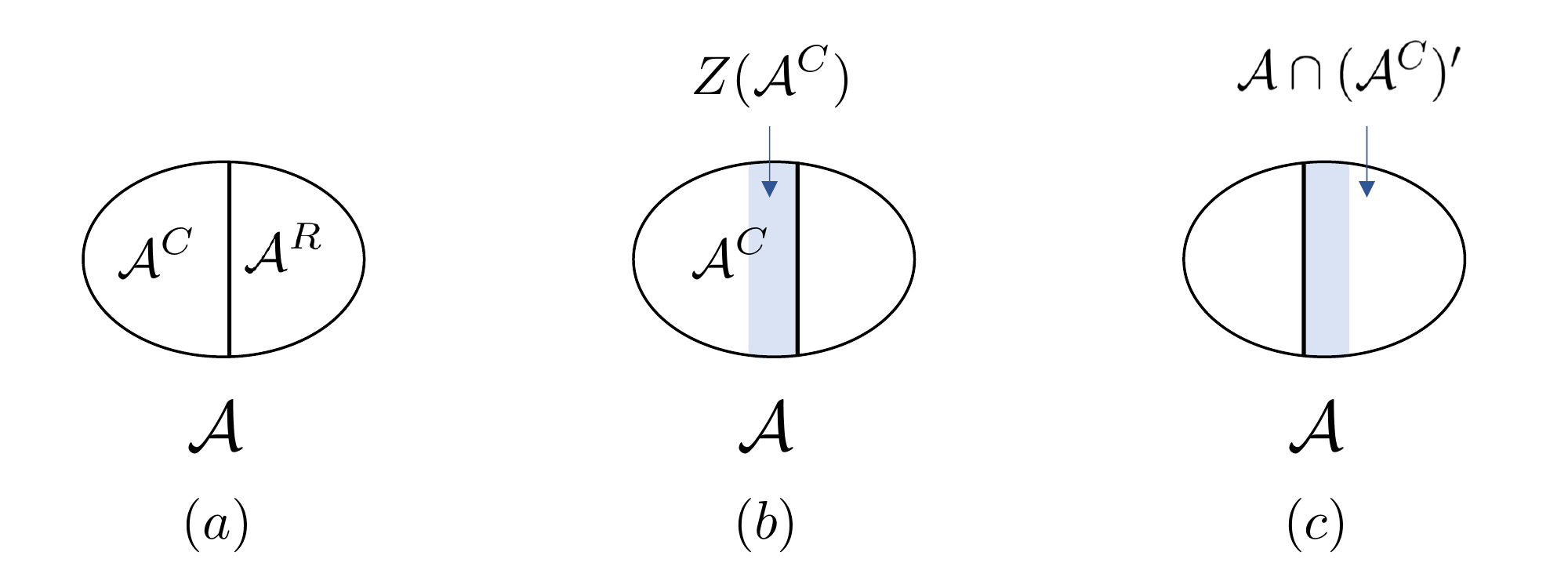}
    \caption{(a) If $\mA^C$ with trivial center is a subalgebra of a finite dimension algebra $\mA$, then we have the tensor product factorization $\mA=\mA^C\otimes \mA^R$. (b) If $\mA^C$ has a non-trivial center $Z(\mA^C)$, we modify the diagram to represent the center as a blue stripe. (c) The center is part of both $\mA^C$ and the relative commutant.}
    \label{fig3u}
\end{figure}

Let us apply a unitary rotation in $\mA_A$. We obtain a new algebra inclusion $U\mA_C U^\dagger\subset U\mA_A U^\dagger$ and a new error correction code; however, the unitary can obscure locality. In fact, every algebra inclusion is an exact quantum error correction code and, if finite dimensional, can be trivialized by a choice of unitary on $A$. Intuitively, this means that there is a hidden notion of locality in the inclusion of any subalgebra $\mA^C\subset \mA$.\footnote{With an abuse of notation, we have denoted a general subalgebra that includes the identity operator as $\mA^C$ because, in this work, the upper index $C$ in $\mA^C$ will stand for ``correctable subalgebra''.} Consider a finite dimensional matrix algebra with a trivial center (the observable algebra of a qudit). If the subalgebra $\mA^C$ also has a trivial center there exists a unitary $U$ in $\mA$ such that $U\mA U^\dagger =U\mA^CU^\dagger\otimes \mA^R$ where $\mA^R$ is the relative commutant of $U\mA^CU^\dagger$ in $U\mA U^\dagger$. If $\mA^C$ is a subalgebra with a non-trivial center $Z(\mA^C)$ then up to the choice of a unitary the algebra $\mA$ factors as the direct sum $\oplus_q \mA_C^{(q)}\otimes \mA_R^{(q)}$ and $\mA^C=\oplus_q \mA_C^{(q)}\otimes \mI_R^{(q)}$. To visualize this structure we use the diagrams in figure \ref{fig3u}. 
In this work, we argue that the inclusion of algebras that share the identity operator appear naturally in renormalization group (RG) and holography, however, in these cases the inclusions are not due to any obvious locality principle. 

There are two parts to this work.
In the first part, in section \ref{sec:realRG}, we argue that the real-space RG can be modeled as an approximate error correction code that encodes the long-distance operators in the algebra of the short-distance operators. In this picture, the short distance local perturbations are the errors and the long-distance operators (or a subset of them) are the correctable operators. This is closely related to modeling the holographic map as a quantum error correction code \cite{almheiriharlowdong2015bulk,Harlow2017,faulkner2020holographic}.

The connection between the RG and error correction can be seen even in classical systems \cite{Furuya:2021lgx}. The intuition is that exciting a long-range degree of freedom requires acting on a macroscopically large number of short-distance degrees of freedom. The disturbance caused by a local short-distance error cannot alter long-distance modes. Under the RG, local ultra-violet (UV) operators become exponentially weak in the infra-red (IR). Deep in the IR, the UV errors are negligible, and in fact, there is no need to actively correct for them. Low energy states of a gapped system, do not have excitations at distances much larger than the correlation length. To make our connection concrete, we focus on real-space RG in systems near critical points where the long range modes of arbitrary wave-length are excited.

As a concrete model of real-space RG that applies to the quantum system near a critical point, in section \ref{sec:realRG}, we consider the multi-scale renormalization ansatz (MERA) tensor network for lattice models.
MERA has found many applications in the study of quantum field theory (QFT) and gravitational theories in AdS/CFT correspondence \cite{evenbly2011tensor,swingle2012entanglement}. To our knowledge, the connection between MERA and error correction codes was first discussed in \cite{kim2017entanglement}. This connection was extended to continuous MERA (cMERA) in \cite{Furuya:2021lgx}.
The error correction property of MERA is similar to the holographic map modeled as an error correction code with the difference that in a general RG flow we do not have complementary recovery property.\footnote{See figure \ref{fig5u} for complementary recovery in holography. Note that, even in holography, the complementary recovery is an approximate notion. It is known to fail in situations where the code subspace is large \cite{hayden2019learning,akers2019large}.} 
Holography suggests that complementary recovery has to emerge in a special class of theories with a large number of local degrees of freedom (large $N$) and are strongly interacting (large gap). We discuss the role of large $N$ and large gap in complementary recovery.

\begin{figure}[t]
    \centering
    \includegraphics[width=0.25\linewidth]{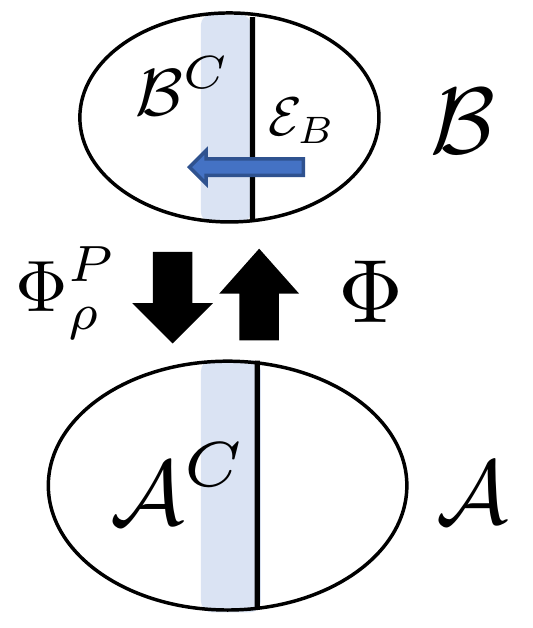}
    \caption{We encode the algebra $\mB$ in the physical algebra $\mA$. If the correctable subalgebra $\mB^C\subset \mB$ is strictly smaller than $B$ we use a conditional expectation $\mE_B$ to project $\mB$ down to $\mB^C$. Absorbing $\mE_B$ in the error map $\Phi$ we are back to the case where the whole algebra is correctable.}
    \label{fig4u}
\end{figure}
\begin{figure}[t]
    \centering
    \includegraphics[width=0.4\linewidth]{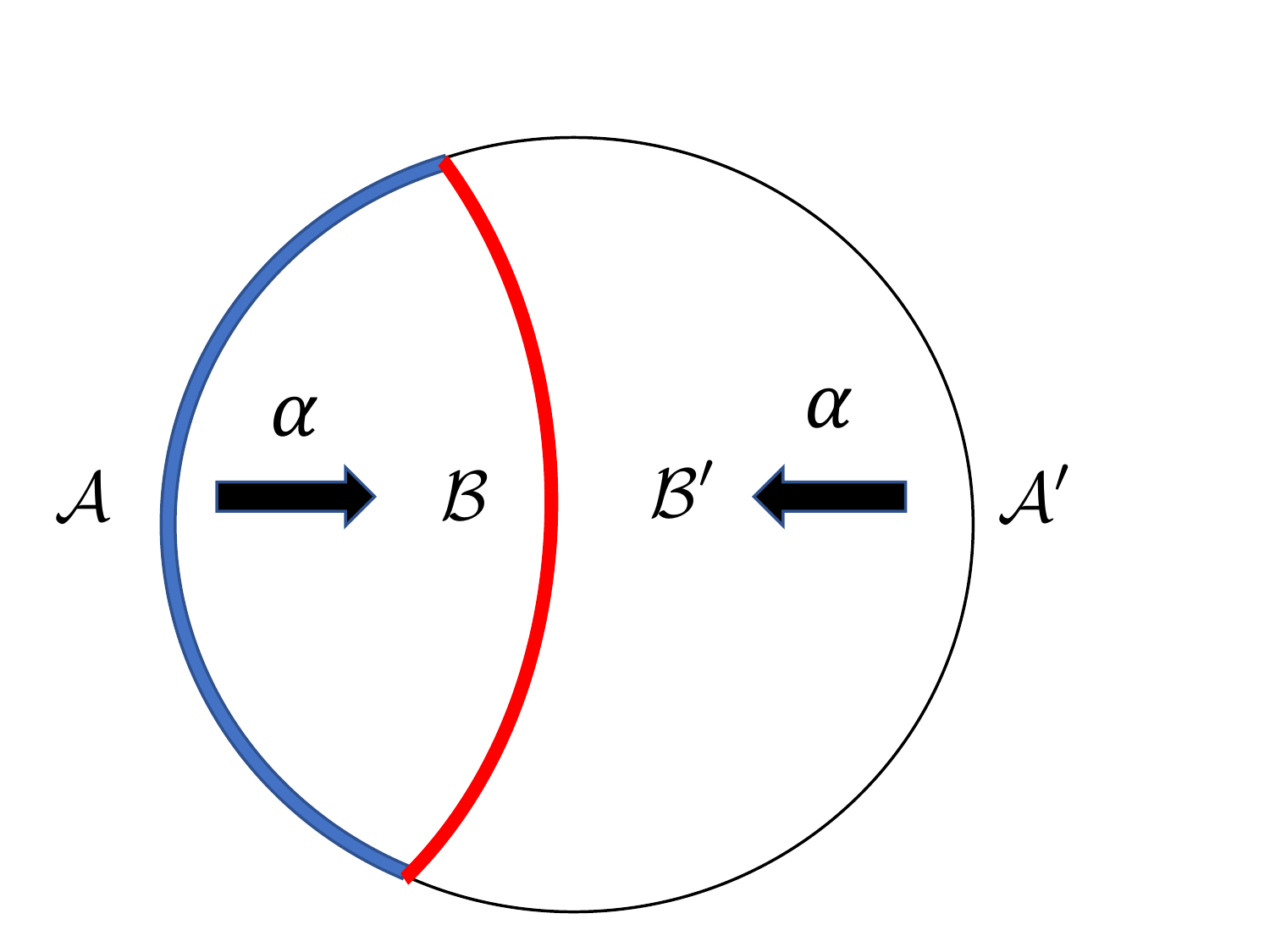}
    \caption{The subsystem error correction code in holography satisfies complementary recovery.}
    \label{fig5u}
\end{figure}

Motivated by the connection between the RG and error correction, in the second part of this work in section \ref{sec:QFT}, we study the operator algebra error correction for an arbitrary von Neumann algebra as a mathematical framework for error correction in continuum quantum field theory (QFT). The error map is modeled by a normal unital completely positive (CP) map $\Phi:\mA\to \mB$; see figure \ref{fig4u}. When the whole algebra $\mB$ is correctable and the error map has no kernel the recovery map is unique and given by the Petz dual of the error map. It isometrically embeds $\mB$ in $\mA$. 
More generally, we consider the setup where only a subalgebra $\mB^C$ of the logical operators $\mB$ is correctable.\footnote{For instance, in holography, this situation arises when the reconstructable wedge is smaller than the entanglement wedge.} Then, the recovery map restricted to the correctable operators is still the Petz dual of the error map. Any unital CP map that projects $\mB$ down to $\mB^C$ (i.e. any conditional expectation $\mE_B:\mB\to\mB^C$) can be used to redefine the error such that its full image is correctable. Such conditional expectations exist if the inclusion $\mB^C\subset \mB$ has finite index \cite{longo1995nets}.
 
For completeness, in the appendices, we have included a self-contained review of the mathematical and information-theoretic background needed for the second part of this work. In appendix \ref{sec:CPmaps&duals}, we review some information theory concepts such as the completely positive (CP) maps and their duals. Appendix \ref{sec:GNS} discusses the GNS Hilbert space which has the following two advantages: 1) linear maps on the algebra (superoperators) correspond to linear operators in the GNS Hilbert space. This simplifies the study of error correction. 2) The GNS Hilbert space can be constructed for all quantum systems (von Neumann algebra), including the local algebra of quantum field theory (QFT) that we are ultimately interested in. 
We show that insisting on the dual of a CP map to remain CP leads to two natural notions of dual maps: 1) the dual map of Accardi and Cecchini that we call the $\rho$-dual map and 2) Petz dual map.
Both of these maps play an important role in error correction. 
The Petz dual map can understood as the dual with respect to an alternate inner product that has already found several applications in QFT in the discussion of Rindler positivity \cite{casini2011wedge,hartman2017averaged}.  While our discussion applies to any quantum system, to help the readers less familiar with von Neumann algebras we mostly use the more familiar notation of finite quantum systems.


In appendix \ref{app:errorcorrection}, we review the Heisenberg picture of quantum error correction. 
We say a subalgebra $\mB^C$ is correctable if there exists a recovery map $\mathcal{R}:\mB^C\to \mA$ such that $\Phi(\mathcal{R}(c))=c$ for all $c\in \mB^C$. We call the constraint $\Phi\circ \mathcal{R}=\text{id}$ the error correction equation. The recovery map is non-unique because any $\mathcal{R}+\mathcal{X}$ satisfies the error correction equation as long as $\Phi(\mathcal{X}(c))=0$. In other words, the recovery is non-unique when the kernel of the error map is non-trivial. Another source of non-uniqueness comes from the fact that the error correction equation defines the recovery map from $\mB^C$ to $\mA$. Any extension of the domain of $\mathcal{R}$ from $\mB^C$ to $\mB$ can be also called a recovery map. We denote the range of the recovery map by $\mA^C\equiv \mathcal{R}(\mB^C)$. It is a subalgebra of the physical operators. The recovery map is an isometric embedding of the correctable algebra in $\mA$.

Conditional expectations are unital CP maps that project an algebra to a subalgebra that includes the identity. In finite dimension, there is a one-to-one correspondence between conditional expectations $\mathcal{E}_{\sigma}$ and unnormalized states $\sigma=\oplus_q \mI^q_1\otimes \sigma_2^q$ on the relative commutant of $\mA^C$ in $\mA$.\footnote{For examples and a more detailed discussion of conditional expectations see appendix \ref{sec:examplesoofCPmaps}.} All the density matrices that are preserved under a conditional expectation $\mE_\sigma$ take the separable form $\rho=\oplus_q p_q \rho_1^q\otimes \sigma_2^q$. In exact error correction, $\mR\circ \Phi$ is a conditional expectation and its invariant states are the correctable states.
The von Neumann entropy of a correctable state splits into two terms
\begin{eqnarray}\label{entropysplit}
    S(\oplus_q p_q \rho_1^q\otimes \sigma^2_q)=H(p)+\sum_q p_q (S(\rho^q_1)+S(\sigma_2^q))=S(\rho_1)+\sum_q p_q S(\sigma_2^q)\ .
\end{eqnarray}
Note that the second term is a property of the correctable subalgebra and not the correctable state.

In holography, the boundary algebra is our physical algebra, and the bulk is the code algebra. An isometry $W$ encodes the bulk Hilbert space on the boundary. In the Heisenberg picture, the map $\alpha(a)=W^\dagger a W$ maps the boundary operators to the bulk respecting the complementary recovery property: the boundary operators supported on region $A$ go to those in the bulk localized in $B$ and the operator supported on the complementary region $A'$ go to those in $B'$; see figure \ref{fig5u}.
The bulk operators localized in region $B$ of the bulk are protected against the erasure of $A'$.
The error map is $\Phi=\alpha\circ \tr_{A'}$ and its Petz dual is the recovery map $\mR:\mB\to \mA$. The complementary recovery implies that the composite map $\mR\circ \Phi$ is a conditional expectation.\footnote{A similar observation was made in \cite{faulkner2020holographic}.} In holography, the second term on the right-hand-side of (\ref{entropysplit}) is argued to be similar to the contribution of the area operator to the holographic entanglement \cite{Harlow2017}.

\section{Real-space RG as an error correction code}\label{sec:realRG}

\subsection{Conventional theory of QEC}

We start this section with a quick review of the conventional approach to quantum error correction.\footnote{For completeness, in section \ref{app:errorcorrection}, we have included a formal derivation of these results using the operator algebra quantum error correction that we generalize to arbitrary von Neumann algebras in section \ref{sec:QFT}} 
In the Schr\"{o}dinger picture of error correction, consider an encoding isometry $W:\mK_B\to \mH_A$ from the code Hilbert space $\mK_B$ to the physical Hilbert space $\mH_A$ and a decoding co-isometry $W^\dagger$. The projection operator $P_C=WW^\dagger$ projects to a subspace of $\mH_A$ called the code subspace because it is isomorphic to $\mK_B$. Throughout this work, we use the following notation: we denote an irreducible representations of an algebra $\mB$ by $\mK_B$, and a reducible representation (such as the GNS representation) of $\mB$ with $\mH_B$. In finite dimensional matrix algebras, we have $\mH_B=\mK_B\otimes \mK_{B'}$.

\begin{figure}
    \centering
    \includegraphics[width=\linewidth]{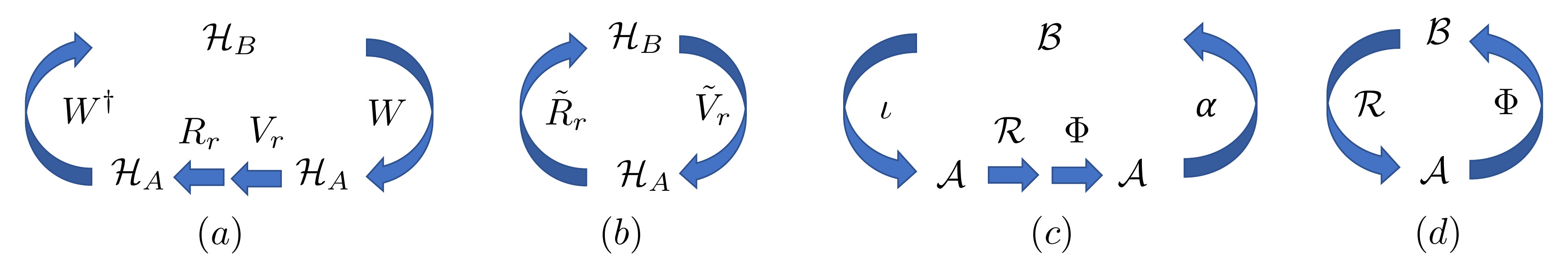}
    \caption{(a) Error correction in the Schr\"{o}dinger picture. The isometry $W$ is the encoding and $W^\dagger$ is the decoding. The errors are $V_r$ and the correction operators are $R_r$. (b) We can absorb the $W$ and $W^\dagger$ in the definition of the errors and the correction operators. (c) Error correction in the Heisenberg picture. The order of operations is reversed. Both the error map $\Phi$ and the recovery map $\mathcal{R}$ are unital completely positive maps. (d) The encoding $\iota$ and decoding $\alpha$ can be absorbed in the definition of the error and the recovery maps.}
    \label{fig6u}
\end{figure}

A collection of error operators $V_r$ corrupt the physical states and a collection of recovery operators $R_r$ correct the errors; see figure \ref{fig6u}. In the simple case where the errors $V_r$ are unitary operators we can undo the error using the correction operators $R_r=V_r^\dagger$. Even when the error is not unitary the correction operator is still made out of the conjugate of the error; see appendix \ref{app:errorcorrection}.
For general errors $V_r$, the necessary and sufficient condition for the recovery to be possible is the {\it Knill-Laflamme condition} \footnote{The physical intuition behind the Knill-Laflamme  condition can be seen by defining a set of basis states $\{ | C_i \rangle \}$ in the code subspace $P_C \mH_A$. Then,
	\begin{equation}
		P_C V_r^{\dagger} V_s P_C = \sum_{ij} | C_i \rangle \langle C_i |   V_r^{\dagger} V_s |C_j \rangle \langle C_j | =  \sum_{ij} \langle C_i |   V_r^{\dagger} V_s | C_j \rangle  | C_i \rangle \langle C_j | .
	\end{equation}
We satisfy Knill-Laflamme condition if $\langle C_i |   V_r^{\dagger} V_s | C_j \rangle = \lambda_{rs} \delta_{ij}$. This   condition implies that the two orthogonal code vectors $\ket{C_i}$ and $\ket{C_j}$ remain orthogonal after the action of the error operators. This ensures that the distinguishable states remain distinguishable despite the errors. } \cite{knill-laflamme1997qec}
\begin{eqnarray}\label{schrodingererror}
P_CV_r^\dagger V_sP_C\propto P_C\ .
\end{eqnarray}
When this condition is satisfied the recovery map is $R_r\propto P_CV_r^\dagger$.

For example, consider the 3-qutrit code where the code Hilbert space $\mK_B$ is a single qutrit spanned by $\ket{i}$ with $i=0,1,2$ that is mapped by an isometry $W$ to the subspace $\ket{\bar{i}}=W\ket{i}$:
\begin{eqnarray}
&&\ket{\bar{0}}=\frac{1}{\sqrt{3}}(\ket{000}+\ket{111}+\ket{222})\nn\\
&&\ket{\bar{1}}=\frac{1}{\sqrt{3}}(\ket{012}+\ket{120}+\ket{201})\nn\\
&&\ket{\bar{2}}=\frac{1}{\sqrt{3}}(\ket{021}+\ket{102}+\ket{210})\ .
\end{eqnarray}
An error that occurs on the third qutrit $V_3$ can be corrected using the $R_3\propto P_C V_3^\dagger$ because
\begin{eqnarray}
W^\dagger R_3 V_3W\ket{i}\propto \ket{i}
\end{eqnarray}
where we have used (\ref{schrodingererror}).
It is convenient to absorb the encoding isometry $W$ in the definition of the errors and the decoding co-isometry $W^\dagger$ in the definition of the recovery operators
\begin{eqnarray}
&&\tilde{V}_r=V_rW,\qquad \tilde{R}_r= W^\dagger R_r\ .
\end{eqnarray}
See figure \ref{fig6u} (a) and (b). There exists a unitary $U$ and a factorization of the Hilbert space $\mH_A=\mK_A\otimes \mK_{A'}$ such that 
\begin{eqnarray}\label{unitcode}
    U\ket{\bar{i}}=\ket{i}_A\ket{\chi}_{A'}
\end{eqnarray}
for some state $\ket{\chi}_{A'}$. The unitary trivializes the encoding such that the information is encoded in $A$ and the errors act on $A'$. The error correction is guaranteed by the locality property $[a,V_r]=0$ for all $a$ acting on $A$ and error $V_r$ acting on $A'$.


In the Heisenberg picture of error correction, we have the algebra of code operators $\mB$ and that of the physical operators $\mA$. An error correction code is a collection of four CP maps $(\iota,\mathcal{R},\Phi,\alpha)$, where $\iota:\mB\to \mA$ is an isometric embedding of $\mB$ in $\mA$ and $\alpha:\mA\to \mB$ undoes it. The recovery map is $\mathcal{R}:\mA\to \mA$ and the error map $\Phi:\mA\to \mA$ is unital. These maps have the Kraus representation 
\begin{eqnarray}\label{Krausforms}
&&\alpha(a)=W^\dagger a W,\qquad \iota(b)=WbW^\dagger\nn\\
&&\Phi(a)=\sum_r V_r^\dagger a V_r,\qquad \mathcal{R}(a)=\sum_r R_r^\dagger a R_r\ .
\end{eqnarray}
We have an error correction if for all the code operators $b\in \mB$ we have
\begin{eqnarray}
\alpha\circ\Phi\circ\mathcal{R}\circ\iota(b)=b\ .
\end{eqnarray}
See figure \ref{fig6u} (c).
The error correction condition above implies the Knill-Laflamme condition in (\ref{schrodingererror}) as a special case, but it is more general.
To simplify the notation, it is often convenient to absorb $\iota$ in the definition of the recovery map and $\alpha$ in the definition of the error map. In this way, an error correction code is a doublet $(\mathcal{R},\Phi)$ where $\Phi:\mA\to \mB$ is the error and $\mathcal{R}:\mB\to \mA$ is the recovery map; see figure \ref{fig6u} (d):
\begin{eqnarray}
&&\Phi(a)=\sum_r \tilde{V}_r^\dagger a \tilde{V}_r,\nn\\
&&\mathcal{R}(b)=\sum_r \tilde{R}_r^\dagger b \tilde{R}_r .
\end{eqnarray}
The map $\mathcal{R}\circ \Phi:\mA\to \mA^C$ projects the physical operators to the subalgebra of correctable operators $\mA^C$. 
These operators are invariant under the action of $\Phi\circ\mathcal{R}$. 


A special error channel relevant to the RG flow and holography is erasure.
In finite dimensional matrix algebras, the erasure is the error map that acts as\footnote{In the Schr\"{o}dinger picture, the erasure channel acts on the density matrices according to $\mE_{\sigma'}^*(\rho_{AA'})=\rho_A\otimes \sigma'$.}\footnote{This is the simplest example of a conditional expectation that preserves the states $\rho\otimes \sigma'$.}
\begin{eqnarray}
    \mathcal{E}_{\sigma'}(a\otimes a')=(a\otimes \mI')\tr(\sigma' a')\ .
\end{eqnarray}
Any operator $a'\in \mA'$ is an error and the necessary and sufficient condition for recovery similar to (\ref{schrodingererror}) is
\begin{eqnarray}\label{erasureerror}
\forall a'\in \mA':\qquad P_C a'P_C\propto P_C\ .
\end{eqnarray}
This is equivalent to the statement that for any operator $b$ there exists an operator $\mathcal{R}(b)$ acting in subsystem $A$ such that \footnote{We prove this for an error correction code in a general von Neumann algebra in section \ref{sec:QFT}. For a proof in finite dimensional matrix algebras see, for instance, theorem 3.1 of \cite{Harlow2017}}
\begin{eqnarray}\label{mathcalRi}
\mathcal{R}(b)W\ket{i}=W b \ket{i},\qquad \mathcal{R}(b^\dagger)W\ket{i}=W b^\dagger \ket{i}\ .
\end{eqnarray}
Since $P_C[\mathcal{R}(b),a']P_C=0$ any error $V'_r$ supported on $A'$ satisfies
\begin{eqnarray}
\mathcal{R}(b)V'_r W=V'_r W b\ .
\end{eqnarray}
Defining the errors $\tilde{V}'_r=V'_rW$ we have
\begin{eqnarray}
\Phi(\mR(b))=\sum_r (\tilde{V}_r')^\dagger \mathcal{R}(b) \tilde{V}'_r=b
\end{eqnarray}
which is the error correction condition in the Heisenberg picture. 


\begin{figure}
    \centering
    \includegraphics[width=0.8\linewidth]{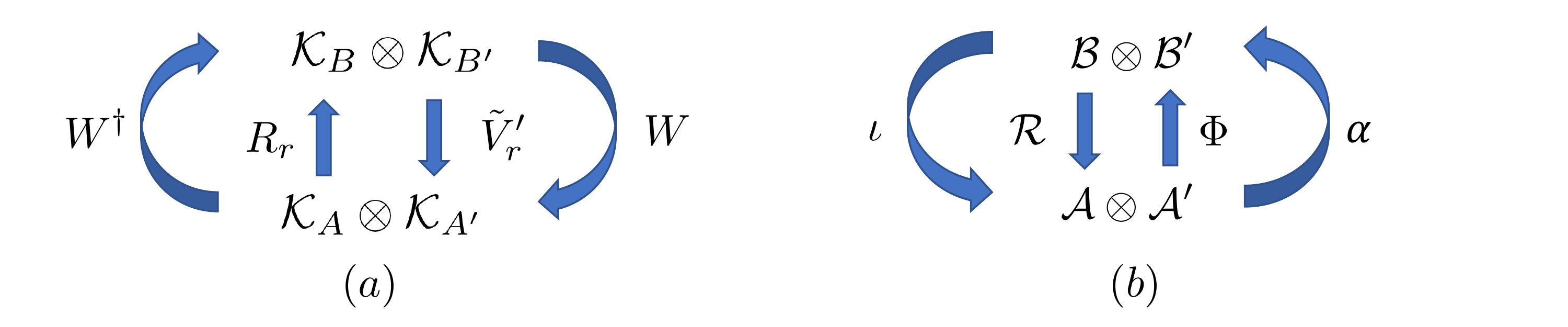}
    \caption{The subsystem error correction in (a) the Schr\"{o}dinger picture (b) the Heisenberg picture.}
    \label{fig7u}
\end{figure}

To see how the Heisenberg picture error correction goes beyond the equation (\ref{schrodingererror}) we consider the subsystem error correction. This is the setup where both the physical Hilbert space and the code Hilbert space admit tensor product forms, respectively $\mH_A=\mK_A\otimes \mK_{A'}$ and $\mH_B=\mK_B\otimes \mK_{B'}$. The goal is to encode the operators $b$ supported on $B$ in the physical Hilbert space such that they are protected against the erasure of $A'$. In this case, the necessary and sufficient condition generalizes the Knill-Laflamme conditions in (\ref{schrodingererror}) to
\begin{eqnarray}\label{subsystemcond}
W^\dagger a'W\in \mB'\ .
\end{eqnarray}
This is to be compared with the condition in (\ref{erasureerror}) that can be written as
\begin{eqnarray}
W^\dagger a' W=\lambda \mI\ .
\end{eqnarray}
 It is a standard result in quantum error correction that (\ref{subsystemcond}) is equivalent to the existence of a map $\mathcal{R}:\mB\to \mA$ such that
\begin{eqnarray}
\mathcal{R}(b)W=W b\ .
\end{eqnarray}
We provide a proof of this for any von Neumann algebra in section \ref{sec:QFT}.
Since $P_C[\mathcal{R}(b),a']P_C=0$ for any error $\tilde{V}'_r=V'_r W$ we have
\begin{eqnarray}\label{erroreq}
\mathcal{R}(b)\tilde{V}'_r=\tilde{V}'_r b
\end{eqnarray}
or equivalently $\Phi(\mathcal{R}(b))=b$; see figure \ref{fig7u}. 

\subsection{Entanglement renormalization}

\begin{figure}[t]
    \centering
    \includegraphics[width=0.8\linewidth]{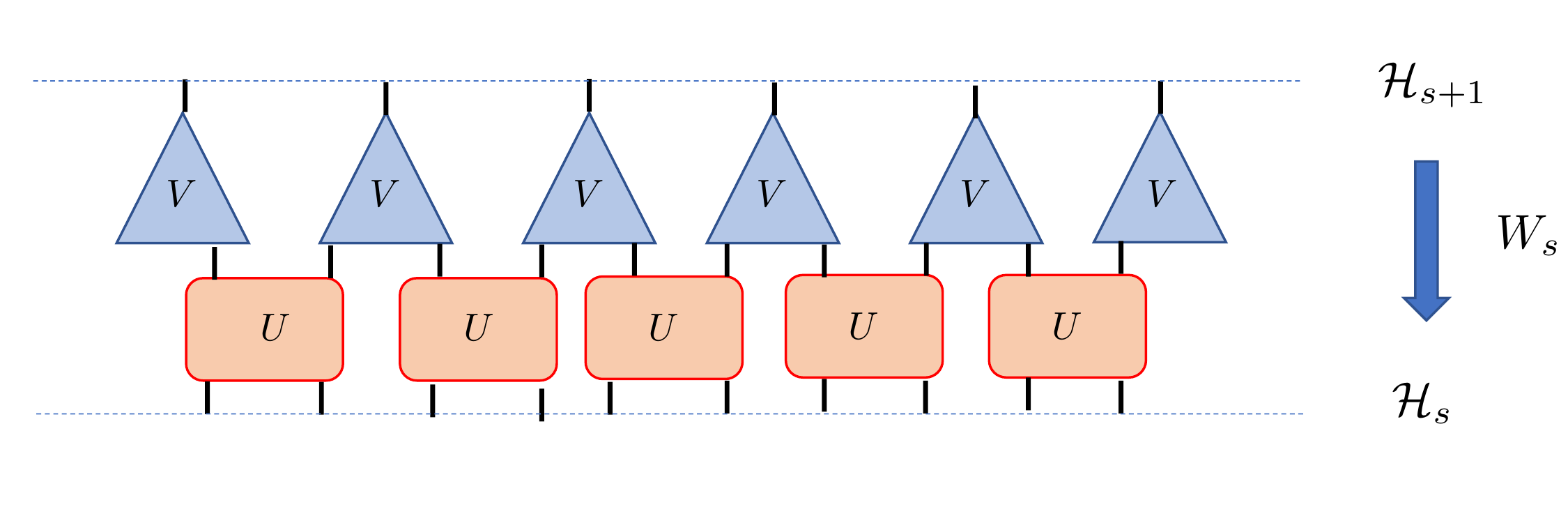}
    \caption{A layer of MERA is an isometry $W_s:\mH_{s+1}\to \mH_s$ that is comprised of two layers: the coarse-graining isometries $V$ and the local disentangling unitaries $U$.}
    \label{fig8u}
\end{figure}

As an explicit example of the connection between the real-space renormalization and the quantum error correction codes we consider a MERA tensor network. A MERA is a sequence of increasingly coarse-grained lattices $\{\mathcal{L}_0,\mathcal{L}_1,\cdots, \mathcal{L}_n\}$ and their corresponding Hilbert spaces $\{\mH_0,\mH_1,\cdots , \mH_n\}$. The Hilbert space $\mH_s$ describes the states of the theory at length scale $l_s$ and $l_0<l_1<\cdots <l_n$. The states of $\mH_0$ are deep in the UV, and the states of $\mH_n$ are in the IR. At each site of every lattice $\mathcal{L}_s$ we have a local Hilbert space that we take to be a qudit for simplicity. A sequence of isometries $W_s:\mH_{s+1}\to \mH_{s}$ embed $\mH_{s+1}$ into the Hilbert space of less coarse-grained states $\mH_{s}$. In the standard MERA, each such isometry is comprised of a layer of local coarse-graining isometries $V$ followed by a layer of disentangling unitaries $U$; see figure \ref{fig8u}. 
The hierarchical structure of correlations in MERA allows for states with long-range correlations. The isometries $W_s$ can be understood as maps that prepare the states $W_1W_2\cdots W_n\ket{\Psi_n}$ with long-range correlations. Below, we summarize the argument presented in \cite{kim2017entanglement} for the error correction properties of MERA.

\begin{figure}[t]
    \centering
    \includegraphics[width=0.8\linewidth]{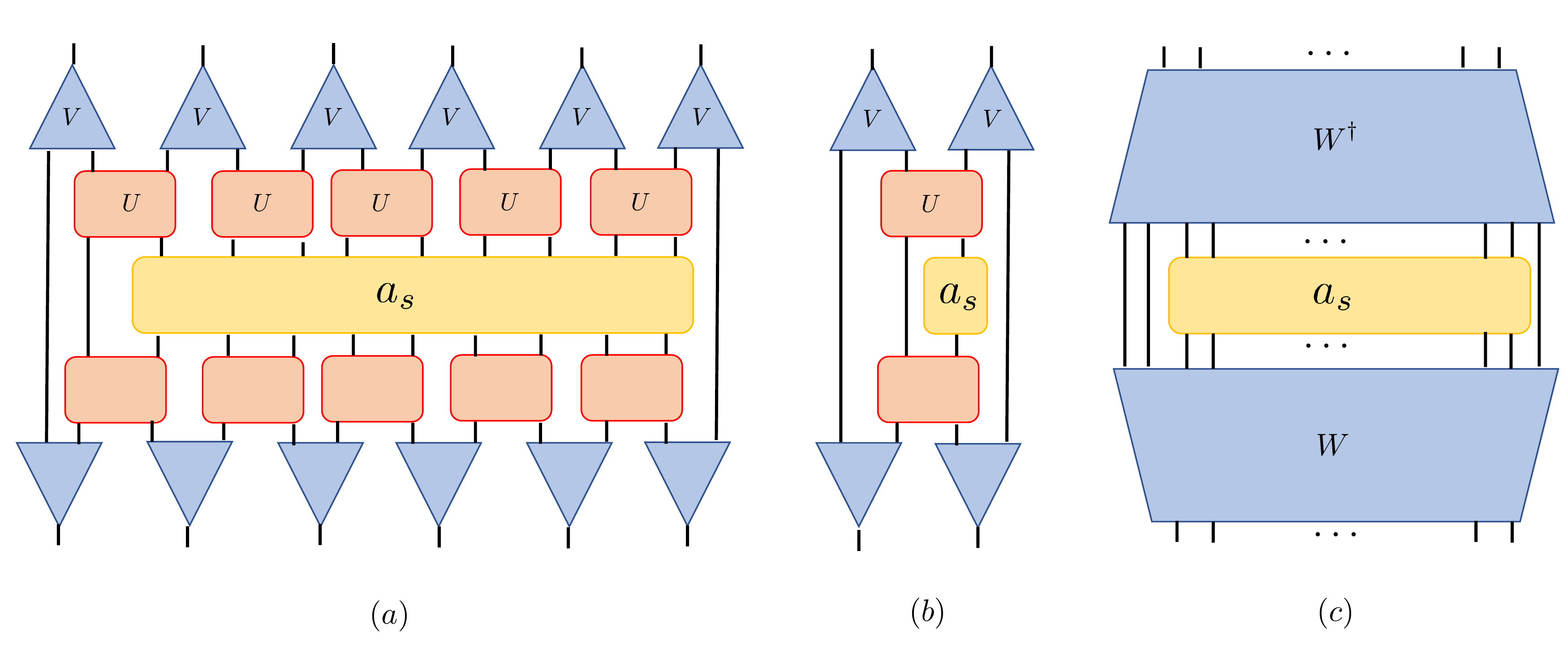}
    \caption{(a) One step of RG for a $9$-local operator $a_s$ turns it into a $6$-local operator acting on $\mH_{s+1}$. (b) The support of operators supported on a few sites fluctuates but remains almost constant. (c)  The support of $k$-local operators with $k\gg 1$ shrinks under the RG. For instance, the support of a $k$-local operator shrinks to at most $\floor{k/2}+2$. In general, the expectation is that the support of operators shrinks by the coarse-graining factor except for some boundary effects that become important when the operator has support on $O(1)$ number of sites.}
    \label{fig9u}
\end{figure}

In the Heisenberg picture, MERA is a renormalization map for the operators: $\mA_s\to \mA_{s+1}$ where $\mA_s$ is the algebra of observables of the Hilbert space $\mH_s$;
see figure \ref{fig9u}:
\begin{eqnarray}
\alpha(a_s)=W_s^\dagger a_s W_s\ .
\end{eqnarray}
The most important property of MERA for us is that it shrinks the support of local operators in the following sense: if $a_s$ is supported on $k$ adjacent sites with $k\gg 1$ on $\mathcal{L}_s$  and the isometries cut down the number of sites by a factor $\gamma>1$ then the operator $\alpha(a_s)$ is supported on approximately $k/\gamma$ sites of $\mathcal{L}_{s+1}$ \cite{swingle2012entanglement}; see figure \ref{fig9u}. 
This is not exactly true because of the boundary effects. For instance, for the MERA in figure \ref{fig9u}, for any $k$-local operator $a$ the support of $\alpha(a)$ is at most $\floor{k/2}+2$.
For $k=O(1)$ the support of the operator almost remains the same.\footnote{It can fluctuate up and down but it can never grow much.}
In higher dimensions, the number of sites in a region scales like the volume of the region and the number of the sites at the boundary scales like the area of the region therefore it is natural to expect that the volume term in the support $a$ shrinks by $\gamma$ up to potential area corrections.

A UV operator $a_0$ supported on region $A_0$ under the RG flow is mapped to the operator $a_s$ whose support we define to be $A_s$. After $s$ layers of RG the linear size of $A_s$ is order $\gamma^{-s}|A_0|$. When $s$ becomes comparable to $\log|A_0|$ the support of the operator reaches a few sites. At this scale, the second stage of the RG flow starts. As we flow further into the IR, the operator remains local on a few sites, however its norm falls exponentially fast. This is because, in the Heseinberg picture, the RG flow map is a quantum channel and hence a  contraction: its eigenvalues have norm smaller than one; see appendix \ref{sec:fixed}. The operators that are invariant under the RG flow survive deep in the IR forming a subalgebra of exactly correctable operators. These are the eigenoperators with eigenvalue one. All the other operators decay exponentially fast with the exponent set by $h_{min}=-\log|\lambda|$ where $\lambda$ is the largest eigenvalue of the RG channel with norm less than one \cite{kim2017entanglement}.\footnote{In principle, there can be eigenoperators whose eigenvalues are a phase $e^{i\theta}$. If such operators exist, under the RG flow they will show recurrences. We expect a generic RG flow to not have such recurrences.}

\begin{figure}
    \centering
    \includegraphics[width=0.45\textwidth]{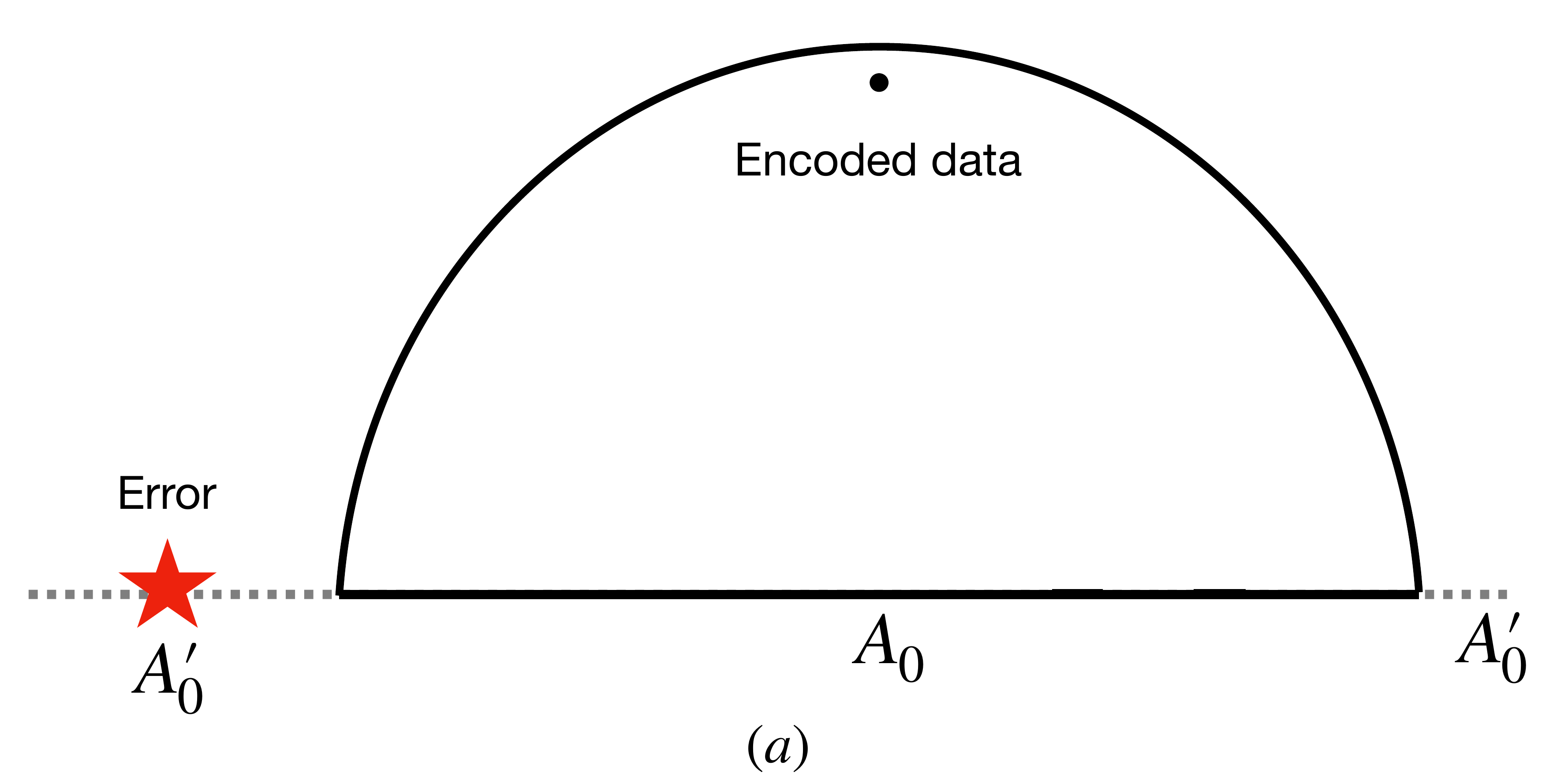}
    \includegraphics[width=0.45\textwidth]{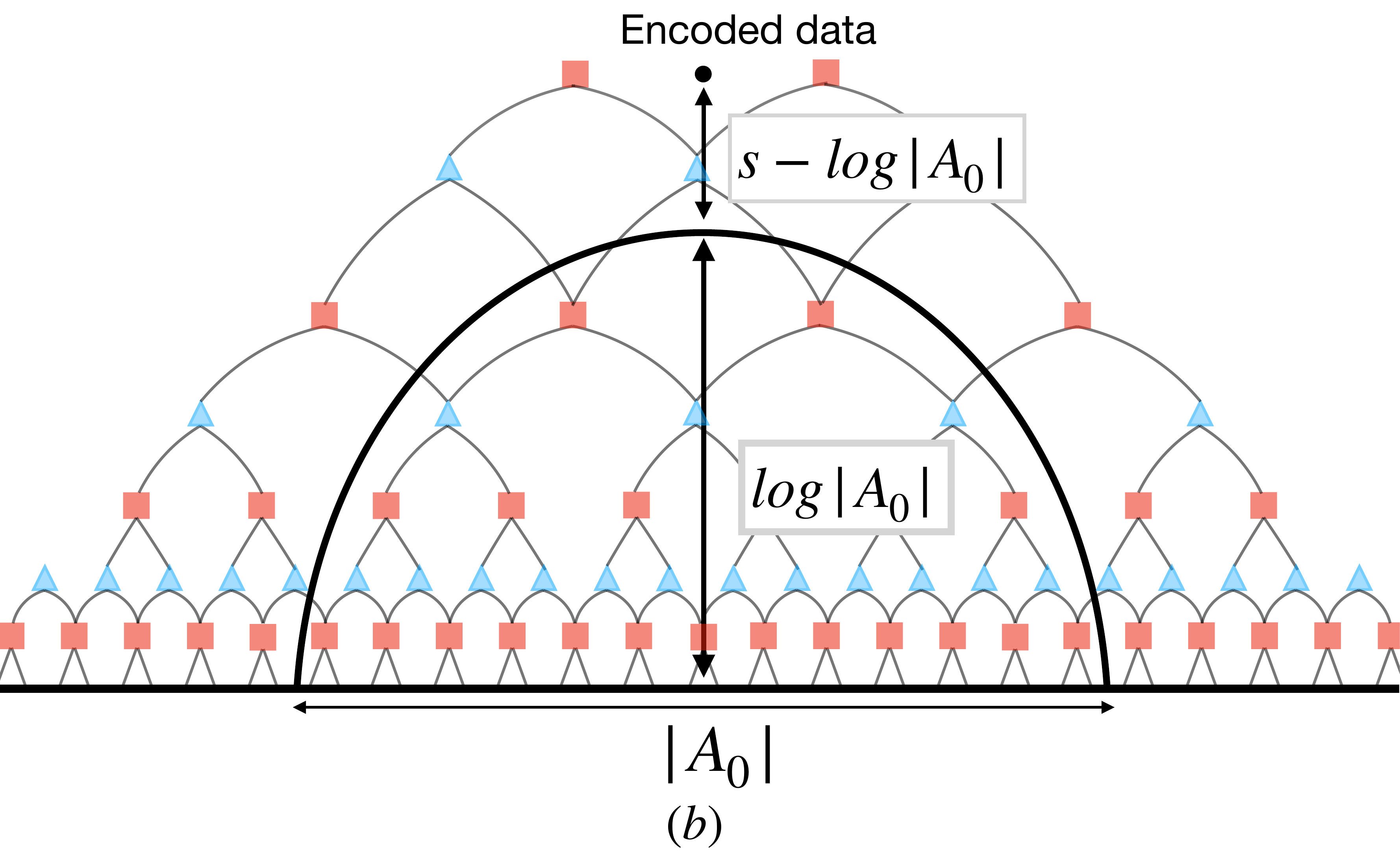}
    \caption{(a) Any UV errors $a'_0$ (red star) supported on $A'_0$ do not disturb the IR operators that are originally supported on $A_0$ before the RG flow. The black dot denoted as the encoded data represents $a_s$.   (b) In the figure, there are $s$ layers between UV and IR where the encoded data is sitting. The size of the support of UV operators shrinks as $\log|A_0|$, though the size drawn in the figure is schematic. }
    \label{fig:figMERAu}
\end{figure}

We split the ultra-violet lattice $\mathcal{L}_0$ into a simply connected region $A_0$ and the complement $A'_0$. 
The RG flow respects locality in the sense that operators supported on $A_0$ are mapped to operators supported on $A_s$. Therefore, the UV errors $a'_0$ localized on $A'_0$ does not disturb the IR operators $a_s$ in $A_s$: $[\alpha^s(a'_0),a_s]=0$. This is a trivial subsystem error correction code. As we flow further into the IR the support $A_s$ shrinks until it reaches a few sites. At this point, the support of the operator no longer shrinks, instead under the RG flow the norm of the operator drops exponentially fast.  If there are $s$ layers of coarse-graining between the IR and the UV states a UV operator supported on a region of size $A_0$ becomes a local operator with a norm that is suppressed by $e^{-(s-\log |A_0|)}$; for a precise statement see lemma 3 in \cite{kim2017entanglement}. Deep in the IR ($s-\log |A_0|\gg 1$) the UV perturbations are vanishingly small. They do not disturb the IR physics; see figure \ref{fig:figMERAu}.

\subsection{Real-space RG in QFT}

In this section, we generalize the connection between MERA and error correction to the RG flow of continuous Poincare invariant QFT. It was shown in \cite{Furuya:2021lgx}, that in continuous MERA (cMERA) \cite{haegeman2013entanglement}, the RG flow of massive free fields is an approximate quantum error correction code. We comment on the emergence of the complementary recovery in holographic code.

The canonical quantization of a QFT that is a perturbation of massive free fields uses the constant time field operator $\varphi(x)$ and its momentum conjugate $\pi(x)$. For simplicity, we set the mass scale to one. As instructed by cMERA \cite{zou2019magic}, to model the RG flow, we deform the Hamiltonian by adding the irrelevant operator $e^{2s}\p_i\pi(x)\p_i \pi(x)$ where the index $i$ runs over spatial directions only and the summation over $i$ is implicit. This term acts as an effective cut-off at the length scale $e^s$. For $f^\pm(x)$ real test function on the space, we define the annihilation operators $a(f)=\int d^{d-1}x \: (f^-(x)+i f^+(x)) a(x)$. Under the RG flow this operator renormalizes to $a_s(f_s)$ where $a_s$ is the annihilation operator at scale $e^s$ and the test function $f_s$ is \cite{Furuya:2021lgx}
\begin{eqnarray}\label{RGfpm}
    f^\pm_s(x)=(1-e^{2s}\nabla^2)^{\pm 1/4} f^\pm(x)\ .
    \end{eqnarray}
Deep in the UV ($s\to -\infty$) the functions $f^\pm$ are supported on region $A$. For smooth enough $f^\pm$ as long as $s\ll \log|A|$ the term $e^{2s}\nabla^2f^\pm$ in (\ref{RGfpm}) is smaller than $f^\pm$ and the renormalization of the field is negligible. This is analogous to the stage one of the RG flow of the operators in MERA. Here, the support does not change but the cut-off length is growing exponentially fast. The cut-off length is analogous to a single site in MERA (the lattice spacing), therefore the support of $f$ in units of the cut-off length is shrinking exponentially fast.

The support of the operator, in units of the cut-off, shrinks until $e^s\sim |A|$ at which point the operator is supported on a region of cut-off length, and the second stage starts. In the second stage, the second term on the right-hand-side of (\ref{RGfpm}) is no longer negligible. It was shown in \cite{Furuya:2021lgx} that for large $s$ the projection of the UV coherent operators to the code subspace becomes approximately proportional to the projection to the code subspace:
\begin{eqnarray}
    P_C e^{a_s^\dagger(f_s)-a_s(f^*_s)}P_C\simeq P_C
\end{eqnarray}
which is the Knill-Laflamme condition for approximate error correction.
 More generally, we can directly analyze the spectrum of the RG quantum channel.
 Deep in the IR, the eigen-operators of the RG quantum channel with the largest eigenvalues are the conformal primaries of the IR fixed point \cite{vidal2007entanglement,vidal2008class}
\begin{eqnarray}
    e^{-s\mathcal{D}}(a_h)=e^{-s h } a_h
\end{eqnarray}
where we have defined the superoperator $\mathcal{D}$ that generates the RG flow from the unit length scale to $e^s$. Here, $h\geq 0$ is the scaling dimension of the eigen-operator. The norm of a non-identity operator decays fast with scale. This implies that any local perturbation in the UV becomes exponentially weak in the IR. The only UV operators that survive the RG flow to the low energies are supported on macroscopically large number of degrees of freedom.\footnote{In principle, it is plausible that the RG map has invariant local eigen-operators. Such operators would have vanishing conformal dimensions.} The parameter $h_{min}(s-\log |A|)$ where $h_{min}$ is the dimension of the lightest primary controls how well this error correction code works.

Quantum error correction makes a surprising appearance in quantum gravity and the AdS/CFT duality \cite{almheiriharlowdong2015bulk}. 
The discovery of the Ryu-Takayanagi (RT) formula in holography led to an understanding of the duality at the level of subregion density matrices \cite{nishioka2009holographic,czech2012gravity}. It revealed that the map that encodes the bulk operators in the Hilbert space of the boundary theory defines an error correction code. These error correction properties have been used to develop toy models of holography using finite dimensional quantum systems \cite{Pastawski-preskill2015toymodel}. It was recently shown that the Petz map gives a reconstruction of the bulk operators in terms of the boundary observables \cite{chen2020entanglement}. See \cite{penington2019replica} for a recent discussion of the Petz map in the reconstruction of operators behind the horizon of a black hole.

At first look, it appears that the approximate error correction in RG is not related to the exact error correction realized in holography because making the error correction above exact requires the conformal dimension of the lightest primary to go to infinity. The holographic QEC code has the complementary recovery property which means that the operators supported on $A_0$ are mapped to those in $A_s$ and the operators on the complementary region $A'_s$ are encoded in those in the complementary region $A'_0$.\footnote{We will use the Latin letters $A$ and $A'$ to refer a region and its complement and $\mA_A$ and $\mA_{A'}$ to refer to their corresponding algebra of operators. Note that in the presence of conserved charges $\mA_{A'}\neq \mA'_A$. This happens because the local algebras have non-trivial centers. We assume periodic boundary conditions so that both $A'$ and its complement $A$ can be chosen to be simply connected.} 
In general, the approximate QEC in RG does not have complementary recovery. This property has to emerge in holographic theories. 

The connection with holography becomes clearer when we consider an RG with two groups of primaries: light primaries with conformal dimensions $h_L\ll \Delta$ and heavy primaries with $h_H\geq \Delta$ for some large parameter $\Delta$. If we choose our code subspace to be the theory at length scale $e^s l$ with $s=\log|A|+\epsilon$ and $l$ some fixed length scale then any noise $\mO_H(A)$ caused by an integrals of heavy operators supported on $A$ can be corrected as long as $\epsilon \Delta\gg 1$. As the gap $\Delta$ goes to infinity, the error correction becomes exact and we obtain complementary recovery. Note that there is no need for a recovery map as the errors simply do not perturb the code subspace. The commutator between the heavy UV operators on $A$ and any local IR operators $a_{IR}(x)$ vanishes simply because their correlation function vanishes $\braket{\mO_H(A,l)a_{IR}(e^sl)}\simeq e^{-\Delta(s-\log |A|)}$.  

In holography, we can correct for the erasure of region $A$. The error operators include the light operators supported on $A$ in addition to the heavy operators.
As opposed to the heavy operators, the light operators on $A$ have non-vanishing correlations with the IR operators. To argue that their effect is correctable in the IR we need a new mechanism in specific to holographic theories. Such a mechanism is provided in theories with $N\times N$ matrix degrees of freedom at large $N$. The light primaries are $k$ trace operators of the form $\tr(X_1)\cdots \tr(X_k)$ with dimension $O(N^0)$. The heavy operators have large dimension $O(N^2)$ that is the size of the gap $\Delta$ in holography. It follows from the large $N$ factorization that the commutator of light operators are $1/N$ suppressed.\footnote{We thank Venkatesa Chandrasekaran for insightful conversations about the role of large $N$ in error correction.} A small commutator is sufficient for the effect of light operators in $A$ to be correctable in the IR.

\section{Error correction in arbitrary von Neumann algebra}\label{sec:QFT}

The local algebra of quantum field theory is different from the matrix algebras in two important ways: 1) It has no irreducible representations.
2) It does not admit a trace. We need to generalize our discussion of error correction to the GNS Hilbert space to include the local algebra of QFT.\footnote{See appendix \ref{sec:GNS} for a review of the GNS Hilbert space.}
In part two of this work, we generalize the formalism of operator algebra error correction to arbitrary von Neumann algebras. To help the reader, we have included a self-contained review of the mathematical background needed for this section in the appendices \ref{sec:CPmaps&duals} and \ref{sec:GNS} that we refer to frequently in the text. Appendix \ref{app:errorcorrection} reviews the theory of operator algebra error correction.

To define the code and the physical GNS Hilbert spaces we need a state $\rho_B$ of $\mB$ \footnote{A state is a normal positive functional of the algebra. When the algebra has a trace it is a density matrix. See appendix \ref{sec:GNS} for more information.}. After the action of the error map this state becomes $\rho_A=\Phi^*(\rho_B)$.\footnote{In the Schr\"{o}dinger picture, the error map corresponds to a quantum channel $\Phi^*$ that sends the states of $\mB$ to those of $\mA$.} We will choose $\rho_B$ to be full rank (a faithful state). 
If the error map has a kernel the state $\rho_A$ is no longer faithful. This means that the errors have erased some information permanently and there will not exist any state that is fully correctable. One way to deal with this is to define a projection to the kernel of the error map and use it explicitly in the recovery map. The recovery map will no longer be unital. Another approach is to enlarge the algebra $\mB$ by including the degrees of freedom until the extended error map has trivial kernel.
Physically, an error occurs because of the interaction with some environment degrees of freedom. If there is a kernel for the error map $\Phi:\mA\to \mB$ it is because the information has left $\mB$ and entered the environment.  If we add to $\mB$ the degrees of freedom of the environment that contain the information that has left $\mB$ the extended error map will have a trivial kernel.\footnote{In the extreme case where we include the whole environment in $\mB$ the error map is a simple unitary rotation, and completely correctable.}
In the real-space RG in QFT, and in holography, the kernel of the error map is empty. This is because the state $\rho_A$ (the vacuum state of short-distance theory in QFT or the boundary state in holography restricted to a region $A$) is faithful. In this section, in generalizing our discussion of error correction to an arbitrary von Neumann algebra, we will focus on the case where the kernel of the error map is empty.

To get oriented, let us start with matrix algebras. In finite dimensional systems, the GNS Hilbert space of a full rank density matrix $\rho_A$ is a double copy Hilbert spaces $\mH_{\rho_A}\equiv \mK_A\otimes \mK_{A'}$ with a distinguished vector $\ket{\rho_A^{1/2}}\in \mH_{\rho_A}$ whose density matrix on both $A$ and $A'$ is equal to $\rho_A$; see appendix \ref{sec:GNS}. Such a vector is called cyclic and separating.
Given a state $\rho_B$ an arbitrary error map $\Phi:\mA\to \mB$ is represented in the GNS Hilbert space as a contraction $F:\mH_{\rho_A}\to \mH_{\rho_B}$\footnote{A contraction is an operator with $\|F\|_\infty\leq 1$.}. 
We assume that the state $\rho_A$ is also full rank therefore the purification of $\rho_A$ is cyclic and separating. There is a one-to-one correspondence between the linear operators in the GNS Hilbert space and the linear superoperators on the algerbra; see appendix \ref{sec:supervsoperator}.
The operator $F^\dagger$ corresponds to the super-operator $\Phi'_\rho:\mB'\to \mA'$ that we call the $\rho$-dual map and the operator $J_A F^\dagger J_B$ corresponds to the Petz dual map $\Phi^P_\rho:\mB\to \mA$ (see section \ref{sec:petz}). 
Here, $J_A$ and $J_B$ are the modular conjugation operators corresponding to $\ket{\rho_A^{1/2}}$ and $\ket{\rho_B^{1/2}}$, respectively. 

In the special case $F$ is a co-isometry we call the problem of solving for the recovery map a {\it reconstruction problem}. Both real-space RG and holography are reconstruction problems. In theorem \ref{theoremPetz}, we show that any error correction problem where the whole image of the error map is correctable is a reconstruction problem. In reconstruction, the operator $F$ is a co-isometry.
In von Neumann algebras, the analog of the Knill-Laflamme condition for exact error correction is the condition $F^\dagger J_B=J_AF^\dagger$ that we refer to as the {\it Takesaki condition}.\footnote{In the remainder of this work, we often denote isometries like $F^\dagger$ with letter $W$.}
Appendix \ref{app:QECintuition} explains intuitively why the Takesaki condition is necessary and sufficient for exact quantum error correction.

\section{Recovery map in von Neumann algebras}

Consider a unital normal CP error map $\Phi:\mA\to \mB$ between two von Neumann algebras. 
The Kraus representation $\Phi(a)=\sum_rV_r^\dagger a V_r$ of a CP map generalizes to infinite dimensions\footnote{In matrix algebras, the Kraus operators were maps from $\mK_A\to \mK_B$ where $\mK_A$ and $\mK_B$ were the irreducible representations of the algebras $\mA$ and $\mB$. A general von Neumann algebra does not admit an irreducible representation. As we discuss in appendix \ref{app:Krausinf} the generalization of the Kraus representation to an arbitrary von Neumann algebra is in terms of the Kraus operators $V_r:\mH_{\rho_B}\to \mH_{\rho_A}$.} .
A recovery map is the isometric embedding of the correctable von Neumann subalgebra\footnote{A recovery map satisfies $\mathcal{R}(c)V_r=V_r c, \forall c\in \mB^C$.
 Therefore, $\mathcal{R}(c_1)\mathcal{R}(c_2)V_r\ket{\rho_A^{1/2}}=\mathcal{R}(c_1c_2)\ket{\rho_A^{1/2}}$. Since we assumed that the kernel of $\Phi$ is empty so the union of the range of all $V_r$ cover the whole Hilbert space and we find that a recovery map is multiplicative: $\mathcal{R}(c_1c_2)=\mathcal{R}(c_1)\mathcal{R}(c_2)$. Since it is CP it becomes an isometric embedding.}.
The CP map $\Phi$ corresponds to a contraction $F:\mH_A\to \mH_B$:\footnote{We simplify our notation from $\mH_{\rho_A}$ to $\mH_A$.}
\begin{eqnarray}
\Phi(a)\ket{\rho_B^{1/2}}=Fa\ket{\rho_A^{1/2}}
\end{eqnarray}
and if the whole algebra $\mB$ is correctable a recovery map corresponds to an isometry $W:\mH_B\to \mH_A$. Below, we collect all the theorems we need to generalize our discussion of error correction to arbitrary von Neumann algebra.

We start with the definition of the $\rho$-dual of $\Phi$ and its properties.
\begin{theorem}[$\rho$-dual map: proposition 3.1 \cite{accardi1982conditional}]\label{thm:rhodual}

Let $\Phi:\mA\to\mB$ be a positive map between von Neumann algebras. Let $\rho_B$ and $\rho_A=\rho_B\circ \Phi$ be faithful states of $\mB$ and $\mA$. Denote by $\ket{\rho_A^{1/2}}$ and $\ket{\rho_B^{1/2}}$ the cyclic and separating vectors that represent $\rho_A$ and $\rho_B$ in their corresponding Hilbert spaces $\mH_A$ and $\mH_B$. There exists a unique normal positive linear map between the commutants $\Phi'_\rho:\mB'\to \mA'$ defined by
 \begin{eqnarray}
 \braket{\Phi'_\rho(b')\rho_A^{1/2}|a \rho_A^{1/2}}=\braket{b'\rho_B^{1/2}|\Phi(a)\rho_B^{1/2}},\qquad \forall a\in \mA, b'\in \mB'\ .
 \end{eqnarray}
If $\Phi$ is CP so is $\Phi'_\rho$, and if $\Phi$ is unital $\Phi'_\rho$ is unital and faithful.
\end{theorem}

\begin{figure}[t]
    \centering
    \includegraphics[width=0.3\linewidth]{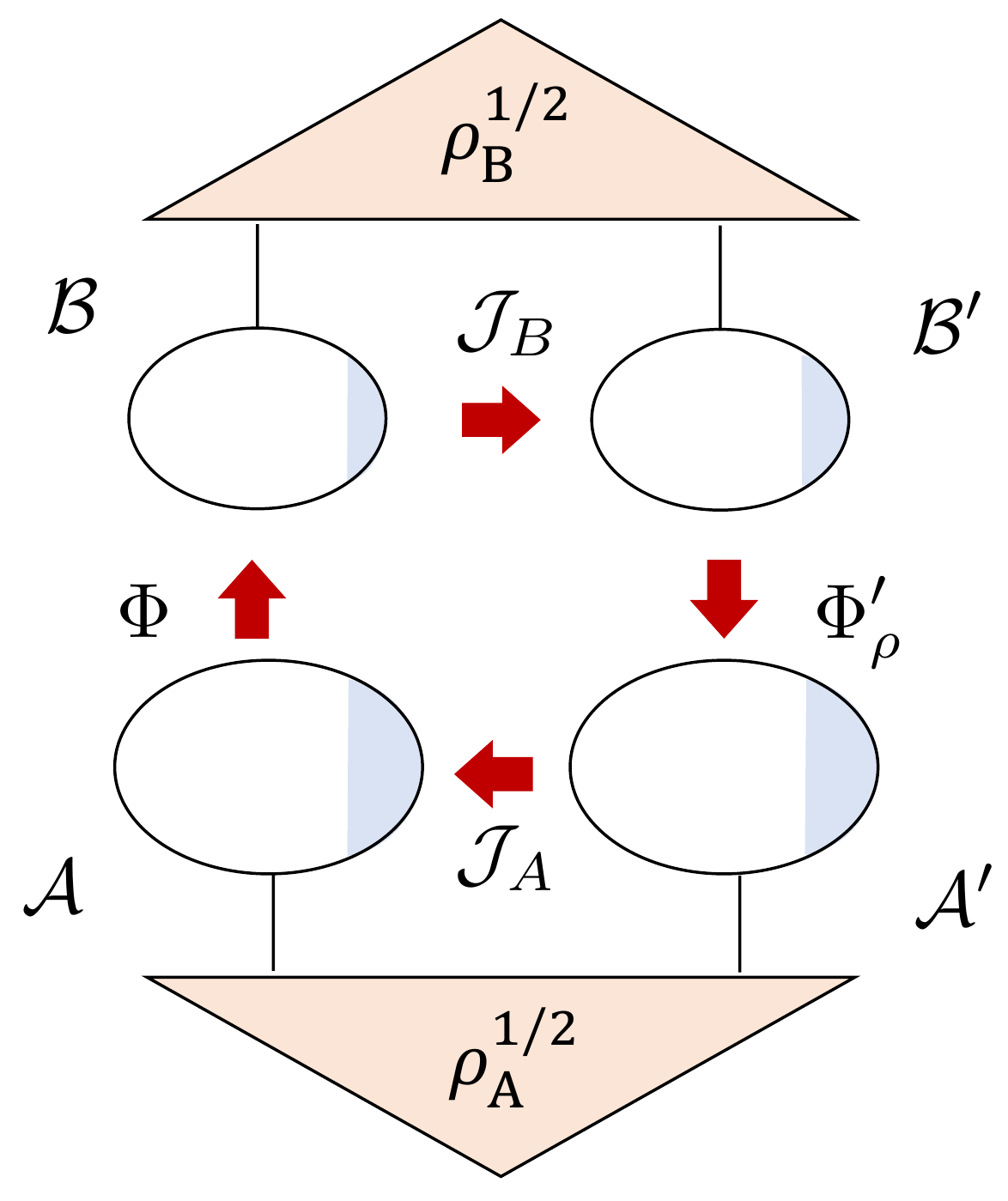}
    \caption{The figure shows $\rho$-dual map of $\Phi$ determined by the cyclic and separating vectors $\ket{\rho_A^{1/2}}$ and $\ket{\rho_B^{1/2}}$ as in theorem \ref{thm:rhodual}. The sequences of $\mathcal{J}_B$, $\mathcal{J}_A$, and $\Phi'_{\rho}$ appears to be a Petz dual map constructed in theorem \ref{theoremPetz}.}
    \label{fig16u}
\end{figure}

First, consider the case where the whole algebra $\mB$ is correctable. This means that there exists a recovery map $\mathcal{R}:\mB\to \mA$ that isometrically embeds $\mB$ in $\mA$
\begin{eqnarray}
\mathcal{R}(b)\ket{\rho_A^{1/2}}=W b \ket{\rho_B^{1/2}}
\end{eqnarray}
with $W:\mH_B\to \mH_A$ an isometry. The map $\Phi\circ\mathcal{R}=\text{id}$ and $\mathcal{R}\circ\Phi: \mA\to \mathcal{R}(\mB^C)\equiv \mA^C\subset \mA$ is a conditional expectation that preserves the faithful state $\rho_A$.

Theorem \ref{thmTakesaki2} tells us that the necessary and sufficient condition for the existence of such a conditional expectation is $J_AW=WJ_B$ that we call the Takesaki condition. We use this property in the next theorem to  establishes that the recovery map is the Petz dual of the error map, see figure \ref{fig16u}:
\begin{theorem}[Petz dual]\label{theoremPetz}

 Let $\Phi:\mA\to\mB$ be a unital completely positive map between von Neumann algebras. Let $\rho_B$  and $\rho_A=\rho_B\circ \Phi$ be faithful states. Denote by $\ket{\rho_A^{1/2}}$ and $\ket{\rho_B^{1/2}}$ the cyclic and separating vectors that represent $\rho_A$ and $\rho_B$ in their corresponding Hilbert spaces $\mH_A$ and $\mH_B$.
 If there exists a normal faithful representation $\mathcal{R}:\mathcal{B}\to \mathcal{A}$ that satisfies $\Phi\circ \mathcal{R}=\text{id}$, it is the {\it Petz dual} of the error map
 \begin{eqnarray}
 \mathcal{R}(b)=\Phi^P_\rho(b)\equiv \mathcal{J}_A\circ \Phi'_\rho\circ \mathcal{J}_B\ .
 \end{eqnarray}
 where $\mathcal{J}_A:\mA'\to \mA$ and $\mathcal{J}_B:\mB\to \mB'$ are the modular conjugation maps corresponding to $\ket{\rho_A^{1/2}}$ and $\ket{\rho_B^{1/2}}$, respectively.
\end{theorem}
{\bf Proof:}
The superoperator $\Phi$ is unital and CP, therefore it corresponds to a contraction $F:\mH_A\to \mH_B$. First, we prove that if the whole algebra $\mB$ is correctable $F$ is a co-isometry. 
The image of the recovery map $\mA^C\equiv \mathcal{R}(\mB)$ is a subalgebra of $\mA$. The composite map $\mathcal{E}=\mathcal{R}\circ \Phi:\mA\to \mA^C$ is unital, CP and preserves every operator in $\mA^C$, hence it is a conditional expectation. The operator corresponding to this conditional expectation is a projection to the range of $W$: $WW^\dagger$. Therefore,
\begin{eqnarray}
\mR\circ \Phi(a)\ket{\rho_A^{1/2}}=WFa\ket{\rho_A^{1/2}}=WW^\dagger a\ket{\rho_A^{1/2}}\ .
\end{eqnarray}
Since $\ket{\rho_A^{1/2}}$ is cyclic and separating we have $WF=WW^\dagger$ or equivalently $F=W^\dagger$ is a co-isometry. 
Since this conditional expectation preserves $\rho_A$ we have the Takesaki condition $J_AW=WJ_B$. 

Now, consider the Petz dual map $\Phi^P_\rho(b)$. We check that it satisfies the recovery equation
\begin{eqnarray}
\Phi\circ \Phi^P_\rho(b)\ket{\rho_B^{1/2}}=W^\dagger J_A W J_B b \ket{\rho_B^{1/2}}=b \ket{\rho_B^{1/2}}
\end{eqnarray}
where we have used the Takesaki condition for $\rho_A$. Since $\ket{\rho_B^{1/2}}$ is cyclic and separating this implies that $\Phi\circ \Phi^P_\rho(b)=b$ for all $b\in \mB$. In the absence of a kernel for the error map this is the unique recovery map from $\mB\to \mA^C$. $\Box$

Next, consider the reconstruction problem where only a proper subalgebra $\mB^C\subset \mB$ is correctable. The Hilbert space $\mH_B$ is a representation of $\mB^C$ but the vector $\ket{\rho^{1/2}_B}$ is no longer a cyclic and separating vector for $\mB^C$. We can use the theorem below to show that the recovery map is dual to $\Phi(a')=W^\dagger a' W\in (\mB^C)'$:

\begin{theorem}[Reconstruction maps: theorem 1 of \cite{faulkner2020holographic}]\label{theoremerror}

Let $W:\mH_B\to \mH_A$ be an isometry in between Hilbert spaces that represent von Neumann algebras $\mB$ and $\mA$, respectively. The following two statements are equivalent:
\begin{enumerate}
\item For all $a\in \mA$ we have $\alpha(a)=W^\dagger aW\in \mB$.
\item There exists a normal isometric embedding (injective $*$-homomorphism) $\alpha':\mB'\to \mA'$ such that $\alpha'(b')W=Wb'$ for all $b'\in\mB'$.
\end{enumerate}
When there exists a vector $W\ket{\rho_B^{1/2}}$ that is cyclic and separating for $\mA$,  $\alpha$ is faithful and the map $\alpha'$ is the unique $\rho$-dual and is unital.
\end{theorem}

The recovery map satisfies the statement (2) therefore it is dual to the map $W^\dagger a'W\in(\mB^C)'$ that we call $\Phi$ with an abuse of notation.  
The map $\Phi$ acts as $\Phi:\mA\to \mB$ and $\Phi:\mA'\to (\mB^C)'$. Since $\mB^C$ is smaller than $\mB$ we do not have complementary recovery. 
We cannot combine $\Phi:\mA\to \mB$ and $\mathcal{R}:\mB^C\to \mA$ to get a conditional expectation.
A simple solution is to look for conditional expectations that project from $\mB$ to $\mB^C$. As we review in appendix \ref{app:errorcorrection}, in finite dimensions, there is a one-to-one correspondence between the conditional expectations from $\mB$ to $\mB^C$ and the states on the relative commutant of $\mB^C$ in $\mB$. With any conditional expectation $\mE_B:\mB\to \mB^C$ we can redefine the error map to $\Phi\to \mathcal{E}_B\circ \Phi$. We are back to the case where the whole image of the error map is correctable, and the recovery map is the Petz dual of the new error map.

If the inclusion of $\mB^C\subset \mB$ has finite index there always exists a conditional expectation from $\mB\to \mB^C$. 
Any von Neumann subalgebra $\mB^C$ is a direct integral of factors: $\mB^C=\int_q^{\oplus} \mC^q$.\footnote{A factor is a von Neumann algebra with trivial center.} Roughly speaking, the index of a subfactor $[\mC^q:\mB]$ is a measure of how many times the algebra $\mC^q$ fits inside $\mB$, and when there exists no conditional expectations from $\mB$ to $\mC^q$ this index is defined to be infinite. When the index is finite there are conditional expectations $\mE^q:\mB\to \mC^q$ \cite{kosaki1986extension}. If all the inclusion of all $\mC^q$ in $\mB$ have finite indices the direct integral of $\mE^q$ is a conditional expectation $\mE:\mB\to \mB^C$. 

\begin{figure}[t]
    \centering
    \includegraphics[width=0.7\linewidth]{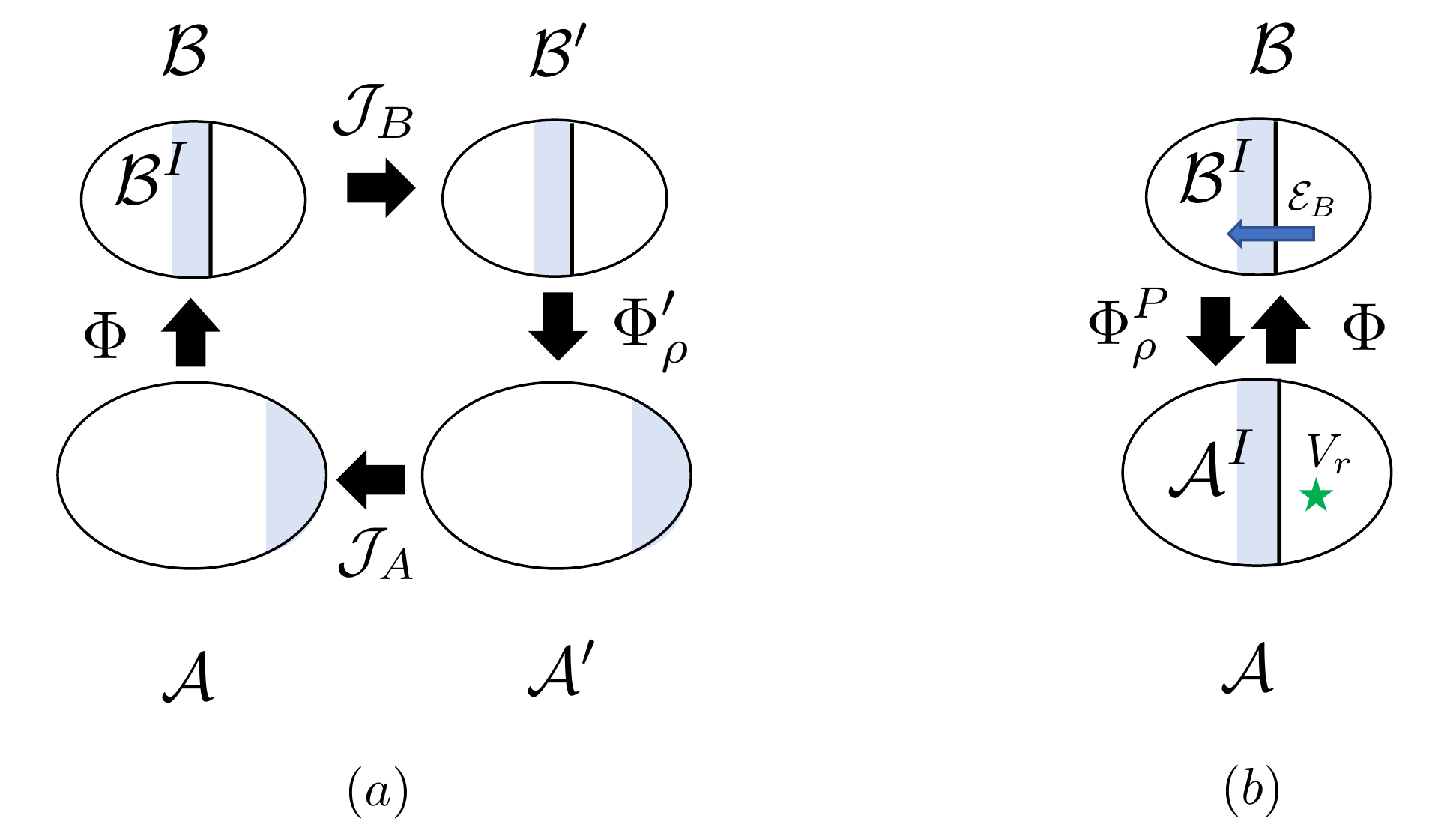}
    \caption{(a) Given a correctable state $\rho_A$ we can construct the conditional expectation that projects $\mB$ to the invariant subalgebra $\mB^I$ of $\Phi\circ \Phi^P_\rho$. (b) The Petz map $\Phi^P_\rho$ plays the role of the recovery map sending the operators in $\mB^I$ to the subalgebra $\mA^I$ that commutes with all errors. This is the von Neumann algebra generalization of the condition $[c,V_r^\dagger V_s]=0$ for the operators in the correctable subalgebra.}
    \label{fig17u}
\end{figure}

The correctable subalgebra is the subalgebra of operators that commute with $V_r^\dagger V_s$.\footnote{See appendix \ref{app:errorcorrection}.} We would like to generalize this to arbitrary von Neumann algebras.
If there exists no correctable states the correctable subalgebra is empty. Therefore, we consider 
 the case where we have an error map $\Phi:\mA\to \mB$ and a state $\rho_A$ that is correctable. We follow a strategy similar to the passive error correction in appendix \ref{passiveQEC}. The map $\Phi\circ\Phi^P_\rho:\mB\to \mB$ is unital and CP. We consider the conditional expectation that projects to its invariant subalgebra that we denote by $\mB^I$:
\begin{eqnarray}
\mathcal{E}_B=\lim_{N\to \infty}\frac{1}{N}\sum_{n=1}^N(\Phi\circ \Phi^P_\rho)^n\ .
\end{eqnarray}
This is an error correction code for the correctable algebra $\mB^I$ with the recovery map $\Phi^P_\rho$ because for all $c\in \mB^I$ we have $\Phi\circ \Phi_\rho^P(c)=c$. The range of the recovery map is a subalgebra in $\mA$ that we denote by  $\mA^I=\Phi^P_\rho(\mB^I)$. 
 The map $\mathcal{E}_A=\Phi^P_\rho\circ \mE_B\circ \Phi$ is a conditional expectation from $\mA$ down to $\mA^I$; see figure \ref{fig17u}. 
We can redefine the error map to $\mathcal{E}_B\circ \Phi:\mA\to \mB^I$. We are back to the standard case above, and the recovery map is once again the Petz dual of the error map.

\section{Discussion}\label{sec:discussion}

In summary, we argued that the renormalization group is an approximate error correction code. This is similar to modeling the holographic map as a subsystem error correction code, with the difference that we do not have complementary recovery. We discussed how the complementary recovery emerges in a theory with large $N$ and a large gap.

We studied the operator algebra quantum error correction for an arbitrary von Neumann algebra. If the error map has a kernel some information is irreversibly lost. In real-space RG, the vacuum vector of a QFT is cyclic and separating which implies that the kernel of the RG map is trivial. 
In von Neumann algebras, the analog of the Knill-Laflamme condition for exact error correction is the Takesaki condition. When recovery is possible, the recovery map is the Petz dual of the error map. 

If the kernel of the error map is not empty (we do not have a cyclic and separating vector) the composition of the recovery map and the error $\mathcal{R}\circ\Phi:\mA\to \mA^C$ is still a CP map that preserves every operator in $\mA^C$, but it is no longer unital. In the language of von Neumann algebras, such a map is an operator valued weight: an unbounded unnormalized positive map with dense domain in $\mA_+$ (the positive operators of $\mA$) that satisfies the bi-module property \footnote{See appendix \ref{sec:CPmaps&duals} for a discussion of the bi-module property.}. There exists a bijection in between the set of operator value weights from $\mA\to \mA^C$ and those from $(\mA^C)'$ to $\mA'$ \cite{connes1980spatial}. The study of operator valued weights could shed light on the problem of reconstruction in the absence of a faithful state. 

\begin{figure}[t]
     \centering
     \includegraphics[width=0.4\linewidth]{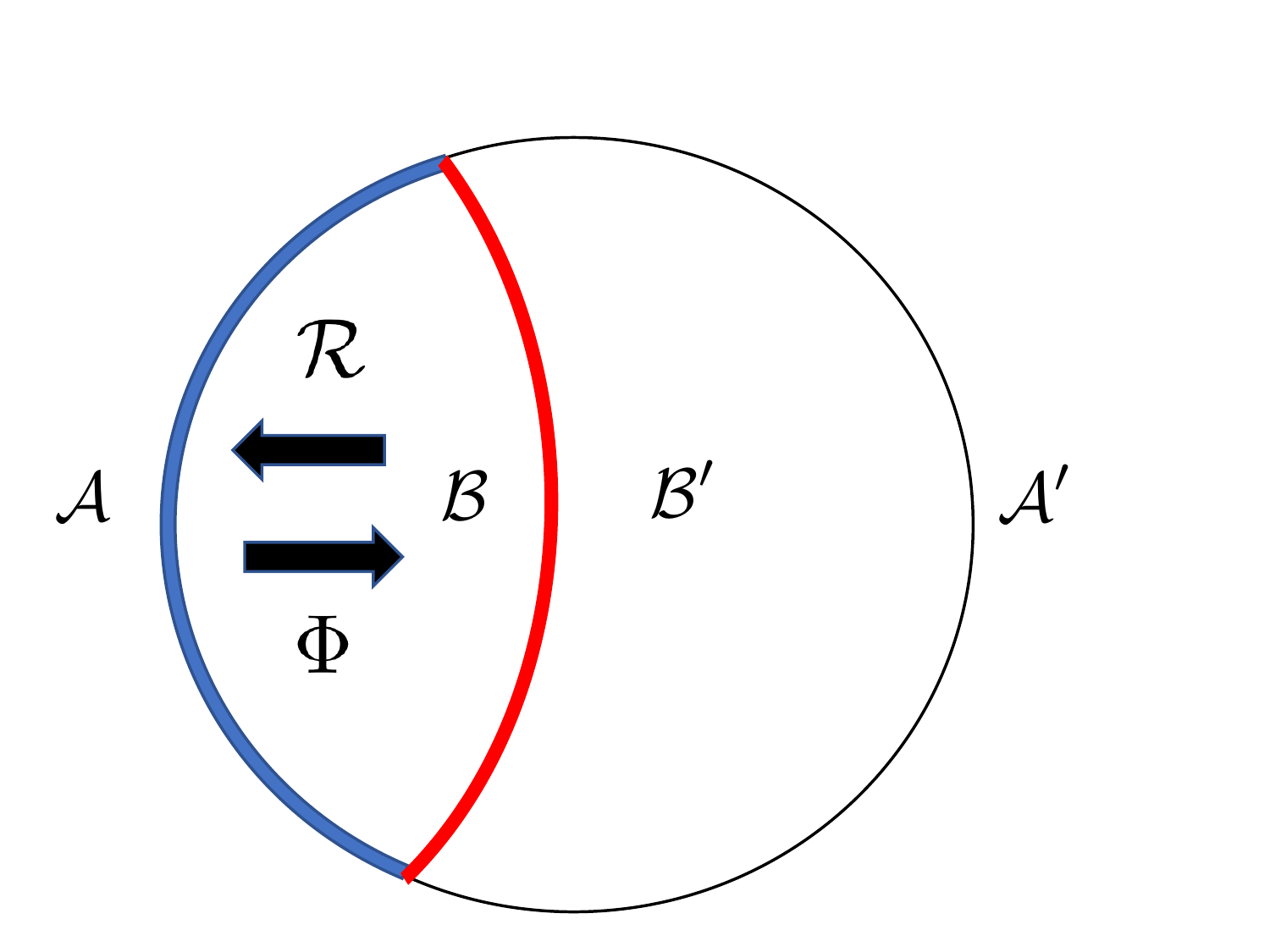}
     \caption{A time slice of anti-de Sitter space with $\mA$ the algebra of a region $A$ on the boundary and $\mB$ the algebra of the bulk region that is in between $A$ and the Ryu-Takayanagi surface of $A$. The CP map $\Phi$ maps the boundary local algebra to the bulk, whereas $\mathcal{R}$ reconstructs the bulk operators on the boundary.}
    \label{fig19u}
 \end{figure}

Consider the AdS$_{d+1}$/CFT$_d$ correspondence in $d>1$ and a simply connected region $A$.
In time-reversal symmetric geometries, the Rangamani-Takayanagi (RT) surface is the co-dimension two surface in the bulk that is anchored on the boundary of $A$, is homologous to $A$ and has minimal area; see figure \ref{fig19u}. Denote by $B$ the region in the bulk that is in between the RT surface and $A$. Consider the map $\mR$ that encodes the algebra $\mB$ of the bulk on the boundary (bulk reconstruction map). We choose the error map to be  $\Phi=\alpha\circ\tr_{A'}$ where $\alpha(\cdot)=W^\dagger (\cdot) W$ and $W:\mH_{bulk}\to \mH_{boundary}$ is the encoding isometry. All the bulk operators $b\in \mB$ satisfy the error correction condition $\Phi(\mathcal{R}(b))=b$ and the recovery map $\mR$ is an isometric embedding. The holographic map from the boundary algebra to the bulk algebra has no kernel because both of the bulk and boundary vectors are cyclic and separating with respect to their corresponding algebras. We have complementary recovery and the whole bulk algebra $\mB$ is reconstructable. The reconstruction map $\mR$ is the Petz dual of the holographic map $\Phi$.
A similar observation was discussed in a recent paper \cite{faulkner2020holographic}. 
Given a $\rho$-preserving conditional expectation we can define a measure of the information lost under the conditional expectation \cite{entanglemententropy_2020}. This leads to entropic uncertainty relations that play an important role in the derivation of the Ryu-Takayanagi formula in holography \cite{Harlow2017,faulkner2020holographic}. It has been argued that complementary recovery fails in some situations in holography \cite{akers2019large}. That brings the holography reconstruction problem closer to the real-space RG. 


Finally, we make the following observation: In AdS$_2$/CFT$_1$ the bulk reconstruction map cannot be a conditional expectation, because there exists no conditional expectations from a type I algebra (the boundary theory is $0+1$ dimensional) to a type III von Neumann algebra (the bulk theory is $1+1$ dimensional QFT). We believe that the resolution of this seeming paradox is that the bulk and boundary relative entropies match only up to $1/N$ corrections. The error correction properties of the holographic map are only approximate. A related observation is that we can define CP maps in between $*$-closed subspaces of observables (operator systems). This generalization can be helpful in moving away from the exact error correction in holography.

\section*{Acknowledgements}

We would like to thank Roy Araiza, Venkatesa Chandrasekaran, Shawn X. Cui, Mudassir Moosa, Kwing Lam Leung, Thomas Sinclair, and Edward Witten for valuable discussions. This work was supported, in part, by a grant-in-aid (PHY-1911298) from the National Science Foundation. NL would like to thank the Institute for Advanced Study for their hospitality during his visit where part of this work was completed.

\appendix

\section{Completely positive maps and their duals}\label{sec:CPmaps&duals}

Consider the algebra of $d\times d$ complex matrices acting irreducibly on the Hilbert space $\mK$ of a $d$-level system and denote it by $\mA=B(\mK)$.\footnote{In our notation, $B(\mK)$ is the algebra of all bounded operators on $\mK$.} Instead of this irreducible representation, we choose to use a reducible representation called the {\it standard representation} that realizes operators as vectors in a Hilbert space $\mH_e \equiv\mK \otimes \mK'$. The main  advantage of this representation is that linear maps from the algebra to itself ({\it superoperators}) correspond to linear operators in $B(\mH_e)$. Moreover, we are ultimately interested in the local algebra of quantum field theory which has standard representations but no irreducible representations.\footnote{To simplify the notation, we denote a Hilbert space by $\mK$ only if our algebra of interest $\mA$ acts irreducibly on it.} The Hilbert space $\mH_e$ is the simplest example of a GNS Hilbert space.

Given a choice of basis $\{\ket{i}\}$ in $\mK$ we construct the standard representation of operators $a\in \mA$ as a vector $\ket{a}\in \mH_e$ using the map 
\begin{eqnarray}
    a\to \ket{a}\equiv \sum_i (a\otimes \mI)\ket{i}\ket{i}\ .
\end{eqnarray}
The identity operator is represented by the unnormalized vector 
\begin{eqnarray}\label{evector}
\ket{e}=\sum_i \ket{i}\ket{i}\ .
\end{eqnarray}
The inner product of vectors $\ket{a}$ in $\mH_e$ is the 
Hilbert-Schmidt inner product for matrices
\begin{eqnarray}\label{inner}
    \braket{a_1|a_2}\equiv\text{tr}(a_1^\dagger a_2)=\braket{e|(a_1^\dagger a_2)\otimes \mI|e}\ .
\end{eqnarray}
We define $\mA'$, the {\it commutant} of $\mA$, to be the algebra of all operators in $B(\mK)$ that commute with $\mA$. Here, $\mA'=B(\mK')$. The map between operators and vectors is one-to-one because every operator is mapped to a unique vector $(a\otimes \mI)\ket{e}$ and every vector uniquely fixes an operator in the algebra.\footnote{If a vector corresponds to two distinct operators $a_1$ and $a_2$ in $\mA$ we have $(a_1-a_2)\otimes \mI\ket{e}=0$ which is not possible for our choice of $\ket{e}$.} The operators $a\in B(\mK)$ and $a^T\in B(\mK')$ create the same vector
\begin{eqnarray}
    (a\otimes \mI)\ket{e}=(\mI\otimes a^T)\ket{e}
\end{eqnarray}
where $a^T$ is transpose in the $\{\ket{i}\}$ basis. An operator $a'_m\in \mA'$ that creates the same vector as $a$ is called the {\it mirror operator} of $a$. 

The vector representative $(a\otimes 1)\ket{e}$ is a purification of the unnormalized density matrix $a a^\dagger$:
\begin{eqnarray}
    \braket{a|b|a}=\tr(a a^\dagger b)\ .
\end{eqnarray}
Consider the left polar decomposition of an operator $a=a_+U$ where $a_+$ is a non-negative positive operator and $U$ is a unitary. The unnormalized density matrix $aa^\dagger=a_+^2$ is independent of $U$. There is a one-to-one correspondence between positive operators $a_+\in \mA$, the vectors $\ket{a_+}\in \mH_e$ and unnormalized density matrices of $\mA$. The expectation value of an operator in a density matrix $\rho$ is the inner product
\begin{eqnarray}\label{rhoop}
    \tr(\rho a)=\braket{\rho|a}\ .
\end{eqnarray}
Alternatively, since $\rho>0$ we can use the cyclicity of trace to write it as
\begin{eqnarray}
 \tr(\rho a)=\braket{\rho^{1/2}|(a\otimes \mI)|\rho^{1/2}}\ .
\end{eqnarray}

In this work, we are primarily interested in linear maps from the algebra to itself $\mT:\mA\to \mA$ (superoperators). A superoperator $\mT$ is called {\it unital} if $\mT(\mI)=\mI$ and it is called {\it trace-preserving} if $\tr(\mT(a))=\tr(a)$ for all $a\in\mA$. In general, a map that satisfies $\tr(\rho \mT(a))=\tr(\rho a)$ for all $a\in\mA$ is called {\it $\rho$-preserving}.\footnote{Not to be confused with the map that satisfies $\mT(\rho)=\rho$.}
There is a one-to-one correspondence between superoperators $a\to \mT(a)\in\mA$ and linear operators $T$ acting on $\mH_e$.\footnote{If $(\mT_1(a)-\mT_2(a))\ket{e}=0$ for all $a$ we have $\mT_1=\mT_2$ and if $(T_1-T_2)\ket{a}=0$ for all $a$ we have $T_1=T_2$.} 
The operator $T$ that corresponds to the superoperator $\mT$ is called its {\it natural representation}.
For instance, the map $\mT(a)=x ay$ with arbitrary $x$ and $y$ matrices corresponds to the operator $T=x\otimes y^T$ acting on $\mH_e$, where $y^T$ is transpose in the $\{\ket{i}\}$ basis:
\begin{eqnarray}
&&(\mT(a)\otimes \mathbb{I})\ket{e}=T(a\otimes \mathbb{I})\ket{e}\ .
\end{eqnarray}
In this work, we frequently represent superoperators by their corresponding operators in the GNS Hilbert space. 
Table \ref{tab:supop-op} summarizes some of the important superoperators we use in this work and their corresponding operators. An entry in the table that plays an important role in this work is a conditional expectation that corresponds to a projection operator in $\mH_e$. We explain this in detail in section \ref{sec:GNS}. For now, we give a simple example of a conditional expectation and its corresponding projection in $\mH_e$.

Consider a linear map from $\mE:\mA\to \mA^C$, where $\mA^C\subset \mA$ is a subalgebra and we have $\mE(c)=c$ for all $c\in \mA^C$. The operator in $\mH_e$ that corresponds to this superoperator is a projection to the subspace $\mH_C$ spanned by vectors $c\ket{e}$ for all $c\in \mA^C$. 
If $\mE(a)=p a p$ for some projection $p\in \mA$ then $E=(p\otimes p^T)$. In this example, the subalgebra $\mA^C$ does not include the identity operator. A projections $E$ that satisfies $E\ket{e}=\ket{e}$ corresponds to a unital superoperators: $\mE(\mI)=\mI$. For instance, take the projection $E=\sum_i \ket{ii}\bra{ii}$. It preserves $\ket{e}$ and corresponds to the unital map $\sum_i \ket{i}\braket{i|a|i}\bra{i}$ that dephases in the basis of $\ket{e}$.\footnote{Clearly, no projection operator $E=p\otimes p'$ is going to leave $\ket{e}$ invariant.} This is the simplest example of a conditional expectation.

The range of a superoperator might be a different algebra: $\mT:\mA\to \mB$. In this case, we represent $\mA$ in Hilbert space $\mH_A$ and $\mB$ in $\mH_B$ defined using the vectors $\ket{e}_A$ and $\ket{e}_B$, respectively. We remind the reader that $\mH_A$ and $\mH_B$ are standard representations and reducible. The superoperator $\mT$ corresponds to an operator $T:\mH_A\to \mH_B$:
\begin{eqnarray}
   T(a\otimes \mathbb{I})\ket{e}_A=(\mT(a)\otimes \mathbb{I})\ket{e}_B\ .
\end{eqnarray}
If the map is unital $\mT(\mI_A)=\mI_B$, we have $T\ket{e}_A=\ket{e}_B$.

\subsection{Dual maps}

Complex conjugation in the Hilbert space $\mH_A$ defines for us a notion of a dual (transpose) map $\mT^*$:
\begin{eqnarray}\label{transpose}
    \braket{a_1|\mT(a_2)}=\braket{a_1|T a_2}=\braket{T^\dagger a_1|a_2}=\braket{\mT^*(a_1)|a_2}\ .
\end{eqnarray}
The dual of a unital map is trace-preserving and vice-versa.
For instance, the unitary evolution of a density matrix $\mT(\rho)=U\rho U^\dagger$ is dual to the unitary evolution of observables:
\begin{eqnarray}
    \braket{U\rho U^\dagger| a}=\braket{\rho| U^\dagger a U}
\end{eqnarray}
This is known to physicists as the equivalence of the Schr\"{o}dinger and the Heisenberg pictures. In this work, we frequently look at dual maps and it is helpful to have the Heisenberg-Schr\"{o}dinger duality in mind. An important property of the unitary maps is that they can be undone with no loss of information. The dual map $\mT^*$ reverses the unitary evolution making sure that $\mT^*(\mT(a))=a$ for all $a\in \mA$. Of course, if the linear map has a kernel then the information content of operators in its kernel is erased and cannot be recovered. 
The range of the dual map $\mT^*$ does not include the kernel of $\mT$. The dual map $T^*$ reverses the effect of $T$, for that reason they are often used in the construction of recovery maps in error correction.

In physics, the linear map $\mT$ models the evolution of observables. The evolution of a closed quantum system is a unitary map. For $\mT:\mA\to \mB$ the simplest example is an isometry. Consider a $d_A$-dimensional Hilbert space $\mK_A$ and a smaller Hilbert space $\mK_B$ with dimension $d_B$ spanned by an orthonormal basis $\{\ket{\alpha}\}$. 
Any isometry $V:\mK_B\to \mK_A$ ($V^\dagger V=\mI_B$ and $VV^\dagger=P$ where $P$ is a projection in $\mK_A$) can be written as
\begin{eqnarray}
   V=\sum_{\alpha=1}^{d_B}\ket{\psi_\alpha}\bra{\alpha}
\end{eqnarray}
where $\ket{\psi_\alpha}$ are orthonormal vectors in $\mK_A$.
The unital map $\mT(a)=V^\dagger a V$ is called a {\it compression}. 
The dual map $\mT^*(b)=V b V^\dagger$ is an {\it isometric embedding} of $B(\mK_B)$ in $B(\mK_A)$. It has the intertwining property
\begin{eqnarray}
   \mT^*(b) V=V b\ .
\end{eqnarray}
The dual map can no longer reverse the evolution: 
\begin{eqnarray}
   \mT^*\mT(a)=P a P\ .
\end{eqnarray}
If $aP =0$ then $\mT^*\mT(a)=0$ and some information is lost (erased).\footnote{The map $\mT^*\mT(a)$ does not preserve $\rho$ unless $\rho=P\rho P$.} 
The dual map recovers the information of operators that have both their domain and range in $P\mK_A$. 

Consider a general linear map $\mT$ that sends operators in $B(\mK_A)$ to operators in $B(\mK_B)$.
To ask about the information loss we need to compare the inner product before the evolution $\braket{a_1|a_2}_A$ and after the evolution $\braket{\mT(a_1)|\mT(a_2)}_B$. Alternatively, we can use the dual map to pull back $\mT(a)$ to $B(\mK_A)$ and compare $a$ with $\mT^*\mT(a)$:
\begin{eqnarray}
  \braket{\mT(a_1)| \mT(a_2)}_B=\braket{\mT^*(\mT(a_1))|a_2}_A\ .
\end{eqnarray}
We can recover the information of operator $a$ if $\mT^*\mT(a)=a$. 

\subsection{Completely positive maps}\label{sec:cp_maps}

An important class of linear maps for physics are completely positive (CP) maps. A positive map sends positive operators to positive operators. We introduce an auxiliary algebra of $n\times n$ complex matrices $M_n$.
A linear map $\Phi:\mA \to \mB$ is completely positive (CP) if $\Phi\otimes \text{id}_n:\mA\otimes M_n\to \mB\otimes M_n$ is positive for all $n$. In physics, $\mA$ and $\mB$ are the algebra of observables of our quantum system of interest before and after the evolution. We enlarge our algebra by modeling the environment degrees of freedom as an $n$-level quantum system  with the algebra $M_n$. In the Schr\"{o}dinger picture, the evolution is a trace-preserving CP map $\Phi^*:\mB\rightarrow\mA$ that acts on density matrices. We need the map to be trace-preserving so that the total probability is conserved $\tr(\Phi^*(\rho_B))=\tr(\rho_B)$. We will show below that the dual of a CP map is also CP. Therefore, in the Heisenberg picture, the algebra of observables evolves with a unital CP map. 
In this work, we mostly use the Heisenberg picture.

An important map that is positive but not CP is the Tomita superoperator, $\mS(\ket{i}\bra{j})=\ket{j}\bra{i}$. It is an anti-linear map that depends on the basis $\{\ket{i}\}$ with respect to which complex conjugation is defined. It is trivially positive. To see that it is not CP consider the positive operator $\ket{e}\bra{e}$. After applying the map we obtain 
$(\mS\otimes \mI)(\ket{e}\bra{e}) = \sum_{ij} \ket{ij}\bra{ji}$ which is the swap operator and non-positive.\footnote{The swap operator squares to identity and its eigenvalues are $\pm 1$.} 

Motivated by the example above, we consider the CP map $\Phi:B(\mK_A) \to B(\mK_B)$ for some Hilbert space $\mK_B$ with an orthonormal basis $\ket{\alpha}$ and define the {\it Choi} operator in the Hilbert space $\mK_B \otimes \mK_A$ to be
\begin{eqnarray}
    \sigma_\Phi=(\Phi\otimes \text{id})(\ket{e}\bra{e})=\sum_{ij} \Phi(\ket{i}\bra{j})\otimes \ket{i}\bra{j}\ .
\end{eqnarray}
The Choi operator carries all the information content of the CP map because
\begin{eqnarray}\label{Choielements}
    \Phi(\ket{i}\bra{j})=(\mI\otimes \bra{i})\sigma_\Phi(\mI\otimes \ket{j})\ .
\end{eqnarray}
The Choi operator $\sigma_\Phi$ is positive if $\Phi$ is CP. Below, we show the converse statement establishing a one-to-one correspondence between CP maps $\Phi:B(\mK_A)\to B(\mK_B)$ and positive operators in $B(\mK_B) \otimes B(\mK_A)$.\footnote{
From the definition of CP maps it appears that need to check the positivity $\Phi\otimes \text{id}_n$ for any $n$. However, this one-to-one correspondence implies that it is sufficient to check the positivity of the Choi operator. This one-to-one correspondence is sometimes called the Choi-Jamiolkowski isomorphism.}

\begin{figure}\label{Krausrep}
    \centering
    \includegraphics[width=0.8\linewidth]{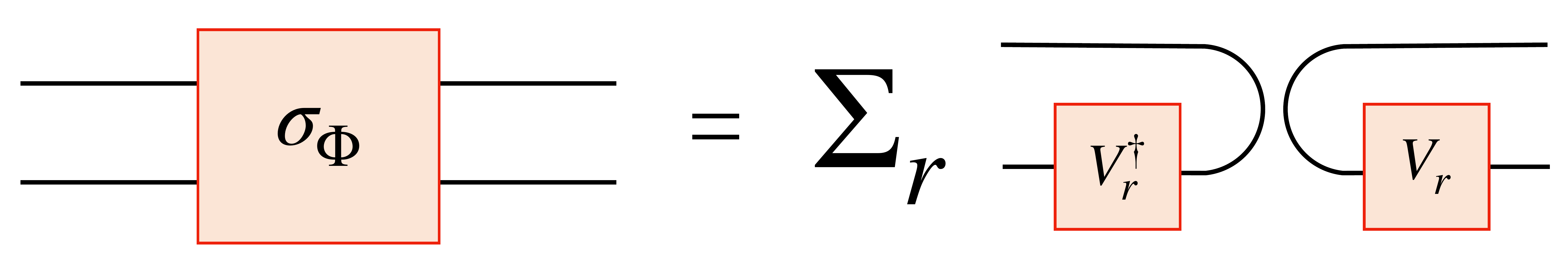}
    \caption{Choi-Jamiolkowski isomorphism: The tensor diagram for the Choi operator that is used to define the Kraus operators $V_r$.}
    \label{fig12u}
\end{figure}

If the Choi operator is positive it has a spectral decomposition in an orthonormal basis
\begin{eqnarray}
    &&\sigma_\Phi=\sum_{r=1}^{d_A d_B} \lambda_r \ket{\phi_r}\bra{\phi_r} \nn\\
    &&\ket{\phi_r}=\sum_{i\alpha}\varphi^{(r)}_{i\alpha}\ket{\alpha i}
\end{eqnarray}
with non-negative $\lambda_r$ and $\ket{\phi_r}\in\mK_B\otimes \mK_A$. 
Define the Kraus map $V_r:\mK_B\to \mK_A$ to be
\begin{eqnarray}
    V_r^\dagger=\sum_{i\alpha}\varphi^{(r)}_{i\alpha}\ket{\alpha}\bra{i}
\end{eqnarray}
so that $\ket{\phi_r}=(V_r^\dagger\otimes \mI)\ket{e}_A$. From the orthogonality of the basis it follows that
\begin{eqnarray}
    \braket{\phi_{r}|\phi_s}=\braket{e|V_{r} V_s^\dagger\otimes \mI)|e}=\tr(V_{r} V_s^\dagger)=\delta_{rs}\ .
\end{eqnarray}
The Choi operator becomes
\begin{eqnarray}
    \sigma_\Phi=\sum_r \lambda_r(V_r^\dagger\otimes \mI)\ket{e}\bra{e}(V_r\otimes \mI),
\end{eqnarray}
and from (\ref{Choielements}) it follows that
\begin{eqnarray}
    \Phi(\ket{i}\bra{j})=\sum_r \lambda_r V_r^\dagger\ket{i}\bra{j}V_r\ .
\end{eqnarray}
The map above is manifestly CP because for any $X\in B(\mK_A)\otimes M_n$ the operator
\begin{eqnarray}
    (\Phi\otimes \mI_n)(X^\dagger X)=\sum_r \lambda_r (V_r^\dagger\otimes  \mI_n)X^\dagger X (V_r\otimes \mI_n)
\end{eqnarray}
is manifestly positive.
See figure \ref{fig12u} for a tensor diagram of the Choi and Kraus operators of a CP map $\Phi$. In summary, a map $\Phi:B(\mK_A) \to B(\mK_B)$ is CP if and only if it has the Kraus decomposition
\begin{eqnarray}
    \Phi(a)=\sum_{r=1}^{d_A d_B}V_r^\dagger aV_r
\end{eqnarray}
where we have redefined $V_r\to \sqrt{\lambda_r}V_r$ to absorb the positive eigenvalues of the Choi operator in the Kraus operators. 

The dual of a CP map also has a Kraus representation, and therefore it is CP. As we said above, the physical relevance of this statement is that the dual to a CP map sends density matrices to density matrices, up to a normalization. Requiring the dual map $\Phi^*$ to preserve the normalization of density matrices (trace-preserving) restricts to the set of unital CP maps $\Phi$. Unital CP maps are sometimes called  coarse-grainings \cite{petz2007quantum}. The connection with the renormalization  justifies the name.\footnote{The dual of a coarse-graining is a trace-preserving CP map called a quantum channel.} As opposed to the unitary map, a general unital CP map leads to information loss. That is to say there exists no CP map $\tilde{\mT}$ that can perfectly reverse the evolution: $\tilde{\mT}\mT(a)=a$ for all operators in $\mA$. As we will see in section \ref{app:errorcorrection}, for an evolution described by a unital CP map $\Phi:\mA\to \mB$ we say an operator $b\in \mB$ is {\it correctable} if there exist an CP map $\mR$ ({\it recovery map}) that reverses the evolution: $\Phi(\mR(b))=b$. Finding recovery map for a given evolution $\Phi$ is one of the main goals of the theory of operator algebra error correction.

The Kraus representation of a CP map is non-unique. To understand this non-uniqueness we introduce an auxiliary Hilbert space $\mK_R$ of dimension $d_Ad_B$ with an orthonormal basis $\{\ket{r}\}$. We rewrite this CP map as
\begin{eqnarray}\label{nonunique}
 &&\Phi(a)=\sum_r V_r^\dagger aV_r=W^\dagger (a\otimes \mI_R)W\nn\\
 &&W=\sum_r V_r\otimes \ket{r}\ .
\end{eqnarray}
Sending $W\to (\mI\otimes U_R)W$ for unitary $U_R\in B(\mK_R)$ leaves the CP map invariant. 
Taking the inner product $(\mI\otimes \bra{r})W$, we see that any two  Kraus representations $\{V^{(1)}_r\}$ and $\{V^{(2)}_r\}$ of a CP map are related by the linear transformation
\begin{eqnarray}
V^{(1)}_r=\sum_s u_{rs} V^{(2)}_s\ .
\end{eqnarray}
where $u_{rs}$ are complex numbers and the matrix $(U_R)_{rs}=u_{rs}$ is unitary \cite{watrous_2018}.
However, as we saw above, there is a canonical choice for Kraus operators that comes from diagonalizing the Choi operator and satisfies 
\begin{eqnarray}\label{canonicalKraus}
\tr(V_r V_s^\dagger)=\delta_{rs}\ .
\end{eqnarray}

The Kraus representation makes it manifest that the composition of two CP maps is also CP. This brings up the question of whether there is a set of simple and physically relevant CP maps that generate all CP maps. The equation (\ref{nonunique}) suggests that it is always possible to write a CP map as a composition of a unital representation $a\otimes \mI_R$ followed by a compression.

The discussion above motivates the {\it Stinespring dilation} theorem that says every CP map $\Phi:\mA\to \mB$ admits the following decomposition
\begin{eqnarray}
\Phi(a)=W^\dagger\pi(a)W,
\end{eqnarray}
where $\pi:\mA\to B(\hat{\mH})$ is a unital representation\footnote{A representation of algebra $\mA$ in Hilbert space $\hat{\mH}$ is a map $\pi$ from the algebra to the bounded operators on $\hat{\mH}$ that is multiplicative $\pi(a_1a_2)=\pi(a_1)\pi(a_2)$. The observable algebra of a quantum system comes with a natural $*$-operation that $\pi$ represents as the Hermitian conjugation in the Hilbert space: $\pi(a^*)=\pi(a)^\dagger$. An injective multiplicative map is sometimes called an {\it embedding}. If it is invertible it is called an isometric embedding. 
A representation is a positive map because $\pi(a^* a)=\pi(a)^\dagger \pi(a)$. In fact, it is completely positive. 
It follows from the definition of a representation that the identity operator of $\mA$ is represented by a projection operator $\pi(\mI)$ in $\hat{\mH}$. If this projection is the identity of $\hat{\mH}$ the representation is unital. 
A faithful representation $\pi:\mA\to \mB$ is a $*$-isomorphism from $\mA$ to the image of $\pi$ in $\mB$, otherwise known as an isometric embedding of $\mA$ in $\mB$.} of $\mA$ in some large Hilbert space $\hat{\mH}$ and $W:\mK_B\to \hat{\mH}$. 
To prove the dilation theorem, we consider representations of $\mA\otimes\mB$.\footnote{For now we consider irreducible representation, however, the generalization to the standard representation is discussed in section \ref{sec:QFT}.} Choose two vectors $\ket{\phi}$ and $\ket{\psi}$ in $\mK_B$.
The standard inner product leads to the Hilbert space $\mH_A\otimes \mK_B$:
\begin{eqnarray}
    \braket{a_1,\phi| a_2,\psi}=\tr(a_1^\dagger a_2)\braket{\phi|\psi}=\braket{a_1|a_2}\braket{\phi|\psi}\ .
\end{eqnarray}
Given a CP map we can define a new inner product:
\begin{eqnarray}\label{newinnerphi}
    \braket{a_1,\phi|a_2,\psi}_\Phi\equiv\braket{\Phi(a_2^\dagger a_1)\phi|\psi}=\braket{\phi|\Phi(a_1^\dagger a_2)|\psi}\ .
\end{eqnarray}
The standard inner product is the special case when the CP map is $\Phi(a)=\tr(a)$. If there are $a\in \mA$ such that $\Phi(a^\dagger a)=0$ then the resulting vector  $\ket{a,\phi}$ has zero norm. We quotient by such zero norm vectors to obtain the Hilbert space $\hat{\mH}$.

When $\Phi$ is faithful $\hat{\mH}=\mH_A\otimes \mK_B$ and
the representation $\pi(a)=a\otimes \mI_{A'B}$. The isometry $W:\mK_B \to \mH_A\otimes \mK_B$ acts as 
\begin{eqnarray}
   && W\ket{\phi}=\ket{e,\phi}\nn\\
    &&\pi(a_1)\ket{a_2,\phi}=\ket{a_1a_2,\phi}\ .
\end{eqnarray}
From the inner product in (\ref{newinnerphi}) it follows that
$W^\dagger$ acts as
\bea
W^\dagger\ket{a,\phi}=\Phi(a)\ket{\phi}\ .
\eea
As a result, the CP map factors as
\begin{eqnarray}
    \Phi(a)=W^\dagger \pi(a)W\ .
\end{eqnarray}
Note that the projection $P=WW^\dagger$ satisfies
\begin{eqnarray}
 &&P\ket{a,\phi}=(\mI \otimes \Phi(a))\ket{1,\phi}\nn\\
 &&P(a\otimes \mI)P=\ket{e}\bra{e}\otimes \Phi(a)\ .
\end{eqnarray}
The CP map $Wb W^\dagger=\ket{e}\bra{e}\otimes b$ is an isometric embedding of $\mB$ in $B(\hat{\mH})$.

The take-home message from the Stinespring dilation theorem is that any unital CP map can be understood as a representation $\mA$ inside the bounded operators in $\hat{\mH}$ followed by an isometry $W$\footnote{One can always choose a representation for $\mA$ that is larger than the GNS Hilbert space $\mH_{A}$ by introducing new degrees of freedom. Then, the dilation theorem gives $(\pi,W,\hat{\mH})$ that are unnecessarily large. In particular, in this case the space of $\ket{a,\phi} = \pi(a)\ket{e,\phi}$ is not dense in $\hat{\mH}$. Given any such representation the restriction of it to the space of $\pi(a)\ket{e,\phi}$ is also a representation that is called the {\it minimal Stinespring representation}. The GNS Hilbert space $\mH_{A}$ gives a minimal Stinespring representation because $a\ket{\rho_A^{1/2}}$ is dense in $\mH_{A}$. 
Consider two Stinespring representations $(\pi_1,W_1,\hat{\mH}_1)$ and $(\pi_2,W_2,\hat{\mH}_2)$ for the same CP map $\Phi$. Then, the operator $v:\hat{\mH}_1\to \hat{\mH}_2$ is a partial isometry that intertwines the two representations:
\begin{eqnarray}
    v \pi_1(a)=\pi_2(a)v \ .
\end{eqnarray}
If both Stinespring representations are minimal then $v$ is a unitary.}. When $\mB=\mA$ the Stinespring representation is a familiar statement in physics. The representation $\pi(a)=a\otimes \mI_R$ introduces environment degrees of freedom modeled by a $d_A^2$ dimensional Hilbert space $\mH_R$. We let the system and environment interact via a unitary and finally we discard the environment degrees of freedom. The interaction and the restriction are described by the isometry $W:\mK_A\to \mH_A\otimes \mK_A$.

The dilation theorem tells us that for any $a\in \mA$ and unital CP map $\Phi$ we have
\begin{eqnarray}\label{schwarz}
   \Phi(a^\dagger)\Phi(a)=W^\dagger \pi(a^\dagger)P \pi(a)W\leq W^\dagger\pi(a^\dagger)\pi(a)W= \Phi(a^\dagger a)\ .
\end{eqnarray}
This is known as the {\it Schwarz inequality}.\footnote{This proof applies to the more general case where the CP map is not unital but satisfies $\Phi(\mI)=P$ for some projection $P$. Then, $W$ is a partial isometry and $\mI-WW^\dagger$ is another projection and the proof follows.}
The map $\Phi$ preserves a state $\rho$ if its corresponding Hilbert space operator $F$ satisfies $F^\dagger\ket{\rho^{1/2}}=\ket{\rho^{1/2}}$. 
When $\Phi$ preserves a state $\rho$ its corresponding operator $F$ in $\mH_A$ satisfies
\begin{eqnarray}\label{contraction}
   &&\|Fa\ket{\rho^{1/2}}\|^2=\braket{\rho^{1/2}|\Phi(a^\dagger)\Phi(a)|\rho^{1/2}}\leq \braket{\rho^{1/2}|\Phi(a^\dagger a)|\rho^{1/2}}\nn\\
   &&=\braket{\rho^{1/2}|a^\dagger a|\rho^{1/2}}=\|a\ket{\rho^{1/2}}\|^2 \ .
\end{eqnarray}
When $\Phi$ has a non-trivial kernel it cannot preserve any faithful states. If it preserves a state $\rho$ the vector representative of it $\ket{\rho^{1/2}}$ in the Hilbert space is cyclic but not separating. 
Therefore, in general, $F$ is a map from the Hilbert space $\mH_A$ to the subspace of $\mH_A$ spanned by $a\ket{\rho^{1/2}}$. Such an operator $F$ is called a {\it contraction}.
On the contrary when $\Phi$ has trivial kernel it can preserve faithful states (full rank density matrix in matrix algebras). Then $F:\mH\to \mH$ with $\|F\|\leq 1$. Note that in (\ref{contraction}) we have simplified our notation by replacing $a\otimes \mI$ with $a$. From here onward, we only write $a\otimes \mI$ when there is a chance of confusion.

The set of operators that are invariant under a unital CP map $\Phi$ that preserves an faithful (full rank) state $\rho$ form a subalgebra $\mA^I$ because if $\Phi(c)=c$ then
\begin{eqnarray}
\braket{\rho^{1/2}| c^\dagger c|\rho^{1/2}}=\braket{\rho^{1/2}|\Phi(c^\dagger c)|\rho^{1/2}}\geq \braket{\rho^{1/2}| \Phi(c^\dagger)\Phi(c)|\rho^{1/2}}=\braket{\rho^{1/2}| c^\dagger c|\rho^{1/2}}
\end{eqnarray}
which implies $\Phi(c^\dagger c)=c^\dagger c$. An operator $c$ is in the invariant subalgebra if
\begin{eqnarray}
    \forall r,\qquad [c,V_r]=[c,V_r^\dagger]=0
\end{eqnarray}
using the Kraus representation of $\Phi$ in (\ref{nonunique}). The converse also holds, and the invariant subalgebra is the commutant of all $V_r$ and $V_r^\dagger$.\footnote{The invariant subalgebra is a von Neumann algebra spanned by its projections. We only need to prove the converse for projection operators.
An invariant projection $p$ satisfies
\begin{eqnarray}
   p_\perp \Phi(p)p_\perp=0=\sum_r p_{\perp} V_r^\dagger p V_r p_{\perp}=0 \nn
\end{eqnarray}
where $p_\perp=\mI-p$. Since the expression above is the sum of positive operators that add up to zero each of them should individually be zero:
\begin{eqnarray}
   p_\perp V_r^\dagger p V_r p_\perp=0=(pV_r p_\perp)^\dagger(pV_r p_\perp)\nn\ .
\end{eqnarray}
As a result, we find $pV_rp_\perp=0$. For a unital map we have $\Phi(p_\perp)=\mathbb{I} - \Phi(p)$, too. Therefore, we also have $p_\perp V_r p=0$. Putting the two together we find that if $p$ is in the invariant subalgebra of $\Phi$ it commutes with all its Kraus operators \cite{kribs2003quantum}.}  In section \ref{app:errorcorrection} we will see that invariant subalgebra is the central object of interest in passive operator algebra error correction.

The set of operators $m\in \mA$ that saturate the Schwarz inequality form a subalgebra $\mA^{M}$ that is called the {\it multiplicative domain} of $\Phi$ \cite{choi1974schwarz}. The Schwarz inequality in (\ref{schwarz}) says that for operators $m\in\mA^M$ we have
\begin{eqnarray}
 W^\dagger \pi(m^\dagger)(\mI-P)\pi(m)W=0
\end{eqnarray}
which implies
\begin{eqnarray}
   (\mI-P)\pi(m)W=0\ .
\end{eqnarray}
It follows that the self-adjoint operators $m\in \mA^M$ satisfy:
\begin{eqnarray}
   \pi(m)P=P\pi(m)P=P\pi(m)
\end{eqnarray}
The converse is obviously true. An operator $m\in \mA^M$ if and only if $[\pi(m),P]=0$.
From the representation in (\ref{nonunique}) it follows that a self-adjoint operator $m$ is in the multiplicative domain of $\Phi$ if and only if
\begin{eqnarray}\label{multisubalgeb}
   \forall r,s\qquad [m,V_r V_s^\dagger]=0 \ .
\end{eqnarray}
As a result, the operators in $\mA^M$ form a subalgebra spanned by the commutant of $V_r V_s^\dagger$. This subalgebra plays an important role in active operator algebra error correction.
 In section \ref{app:errorcorrection}, we show that the multiplicative domain of a unital map $\Phi$ is equivalent to the correctable subalgebra of the dual map $\Phi^*$. The invariant subalgebra is a subalgebra of the multiplicative domain of $\Phi$, i.e. $\mA^I\subseteq\mA^M\subseteq \mA$. 

The multiplicative domain of $\Phi$ satisfies the {\it bi-module property}: for all $m\in \mA^M$ and all $a\in \mA$ we have:
\begin{eqnarray}\label{bimod}
   &&\Phi(m^\dagger  a)=\Phi(m^\dagger)\Phi(a)\nn\\
   &&\Phi(a^\dagger m)=\Phi(a^\dagger)\Phi(m)\ .
\end{eqnarray}
To prove this, we use the fact that $\Phi^{(2)}=\Phi\otimes \mathrm{id}_2$ is also a CP map that satisfies the Schwarz inequality. Consider the operator $X\in \mA\otimes M_2$ ($M_2$ is the algebra of complex $2\times 2$ matrices)
\begin{eqnarray}
    X=\begin{pmatrix}
0 & m^\dagger\\
m & a
\end{pmatrix}
\end{eqnarray}
for some $a\in\mA$ and $c\in\mA^M$. The Schwarz inequality gives
\begin{eqnarray}
\begin{pmatrix}
\Phi(m^\dagger m) & \Phi(m^\dagger a)\\
\Phi(a^\dagger m) & \Phi(m m^\dagger +a^\dagger a)
\end{pmatrix}
=\Phi^{(2)}(X^\dagger X)&&\geq \Phi^{(2)}(X^\dagger)\Phi^{(2)}(X)\\
&&=\begin{pmatrix}
\Phi(m^\dagger)\Phi(m) & \Phi(m^\dagger)\Phi(a)\\
\Phi(a^\dagger)\Phi(m) & \Phi(m)\Phi(m^\dagger)+\Phi(a^\dagger)\Phi(a)
\end{pmatrix}\nn
\end{eqnarray}
This implies that
\begin{eqnarray} \label{eq:2px}
   \begin{pmatrix}
0 & \Phi(m^\dagger a)-\Phi(m^\dagger)\Phi(a)\\
\Phi(a^\dagger m)-\Phi(a^\dagger)\Phi(m) & \Phi(a^\dagger a)-\Phi(a^\dagger)\Phi(a)
\end{pmatrix}\geq 0
\end{eqnarray}
which is possible if and only if its off-diagonal terms are exactly zero which proves (\ref{bimod}).
A  unital CP map $\mE$ from $\mA$ to its invariant subalgebra $\mA^I$ is called a {\it conditional expectation}. It satisfies the bi-module property that for all $c_1,c_2\in \mA^I$ and $a\in \mA$ we have
\begin{eqnarray}\label{bimodcond}
   \mE(c_1 ac_2)=c_1 \mE(a)c_2\ .
\end{eqnarray}

\subsection{Conditional expectations in matrix algebras}\label{sec:examplesoofCPmaps}

To make the discussion less abstract, in this subsection, we go over some important examples of conditional expectations in matrix algebras.
Our first example of a CP map is
$\iota_\sigma:\mA_1\to \mA_1\otimes \mA_2$ given by
\begin{eqnarray}\label{embedd}
    &&\iota_\sigma(a)=a\otimes \sigma,
\end{eqnarray}
where $\sigma$ is a positive operator with eigenvectors $\{\ket{k}\}$ and eigenvalues $\lambda_k^2$. The Stinespring dilation of this map factorizes as a representation on $\mK_1\otimes \mK_3$ and the isometry $W:\mK_1\otimes \mK_2\to \mK_1\otimes \mK_3$:
\begin{eqnarray}
    &&\iota_\sigma(a)=W^\dagger (a\otimes \mI_3)W\nn\\
   &&W=\sum_k \lambda_k (\mI_1\otimes \ket{k}_3\bra{k}_2)\nn\\
   &&\mI_3=\sum_k \ket{k}_3\bra{k}_3\ .
\end{eqnarray}
The Kraus operators are $V_k=\lambda_k (\mI_1\otimes\bra{k}_2)$. The dual map $\iota^*_\sigma:\mA_1\otimes \mA_2\to \mA_1$ is 
\begin{eqnarray}\label{iotastar}
    &&\iota^*_\sigma(a_1\otimes a_2)=\sum_k V_k(a_1\otimes a_2)V_k^\dagger=a_1\tr(\sigma a_2), \qquad 
\end{eqnarray}
with the Stinespring dilation
\begin{eqnarray}
    &&\iota^*_\sigma(a_1\otimes a_2)=W^\dagger (a_1\otimes a_2\otimes \mI_3)W\nn\\
    &&W=\sum_k \lambda_k (\mI_1\otimes  \ket{kk}_{23})
\end{eqnarray}
The map $\iota_\sigma$ is unital when $\sigma=\mathbb{I}_2$. 
In this case, it is an embedding of $\mA_1$ in $\mA_1 \otimes \mA_2$:
\begin{eqnarray}
    \iota_1(a_1a_2)=\iota_1(a_1)\iota_1(a_2)\ .
\end{eqnarray}
The dual $\iota^*_1$ is a quantum channel (trace-preserving CP map) $\mA_1\otimes \mA_2\to \mA_1$ that is partial trace over $\mA_2$:
\begin{eqnarray}
    &&\tr(\rho_{12}\: \iota_1(a))=\tr(\iota^*_1(\rho_{12}) a)\nn\\
    &&\iota^*_1(\rho_{12})=(\mI_1\otimes \bra{e}_{23})\rho_{12}(\mI_1\otimes \ket{e}_{23})=\rho_1\ .
\end{eqnarray}
The map $\iota_\sigma$ is a quantum channel when $\sigma$ is a density matrix: $\tr(\sigma)=1$. This channel prepares a density matrix $\sigma$ on $\mK_2$. 
The composition of two CP maps is also a CP map. 
For instance, the composite map $\iota_\sigma^*\circ\iota_\sigma (a_1)=a_1 \tr(\sigma)$ multiplies operators by a positive constant,  whereas $\iota_\sigma\circ\iota_\sigma^*(a_1\otimes a_2)=(a_1\otimes \sigma) \tr(\sigma a_2)$. 
An important composite map for us is 
\begin{eqnarray}\label{Condexpsigma}
&&\mE_\sigma\equiv\iota_1\circ \iota^*_\sigma:\mA_1\otimes \mA_2\to \mA_1\otimes \mI_2\nn\\
&&    \mE_\sigma(a_1\otimes a_2)=(a_1\otimes \mI_2)\tr(\sigma a_2)\ .
\end{eqnarray}
It has the property that when $\sigma$ is a density matrix it leaves the subalgebra $\mA_1\otimes \mI_2$ invariant 
\begin{eqnarray}
    \mE_\sigma(a_1\otimes \mI_2)=a_1\otimes \mI_2\ .
\end{eqnarray}
It is the simplest example of a $\sigma$-preserving conditional expectation \cite{entanglemententropy_2020}.

The conditional expectations in (\ref{Condexpsigma}) are labelled by density matrices $\sigma$ on $\mA_2$. In fact, these are the only conditional expectations from $\mA_1\otimes \mA_2$ to $\mA_1\otimes \mI_2$. To see this, we use the bimodule property:
\begin{eqnarray}
 \mE(a_1\otimes a_2)=(a_1\otimes \mI)\mE(\mI\otimes a_2)=\mE((\mI\otimes a_2)(a_1\otimes \mI))=\mE(\mI\otimes a_2)(a_1\otimes \mI)\ .
\end{eqnarray}
Therefore, $\mE(\mI\otimes a_2)$ commutes with all $a_1\otimes \mI$ and has to take the form
\begin{eqnarray}
 \mE_\epsilon(a_1\otimes a_2)=(a_1\otimes \mI)\: \epsilon(a_2),
\end{eqnarray}
where $\epsilon(a_2)$ is a unital CP map from $\mA_2\to \mathbb{C}$ which is in one-to-one correspondence with density matrices on $\mA_2$:\footnote{$\epsilon(a_2)$ is a continuous linear functional on $\mA_2$ which by Riesz representation theorem can be associated with a unique vector $\ket{\epsilon}\in\mK_2$ such that $\epsilon(a_2) = \braket{\epsilon|a_2} $.}
\begin{eqnarray}
 \mE_\sigma(a_1\otimes a_2)=(a_1\otimes \mI)\: \tr(\sigma a_2)\ .
\end{eqnarray}

The conditional expectation $\mE_\sigma$ preserves all states of the form $\rho\otimes \sigma$. Moreover, given a product state $\rho\otimes \sigma$ the conditional expectation $\mE_{\sigma}$ that preserves it is unique. However, for a generic $\sigma_{12}$ there does not exist a conditional expectation that preserves it.

To gain more intuition about conditional expectations $\mE:\mA\to \mA^I$ in finite dimensional matrix algebras consider their Kraus representation $\mE(a)=\sum_r V_r^\dagger a V_r$.
The Hilbert space $\mK$ decomposes as $\mK=\oplus_q \mK_1^q\otimes \mK_2^q$ such that
\begin{eqnarray}
 &&c=\oplus_q c^q\otimes \mI_2^q\qquad \forall c\in \mA^I\nn\\
 &&V_r=\oplus_q \mI_1^q\otimes V_r^q\qquad \forall r\ .
\end{eqnarray}
A conditional expectation $\mE$ projects every operator in $\mA$ to its invariant subalgebra $\mA^I$. Denote the projection to the subspace $\mK_1^q\otimes \mK_2^q$ by $P^q$.
Since $P^q\in \mA^C$ from the bi-module property (\ref{bimodcond}) we have \cite{magan2020quantum}
\begin{eqnarray}
    &&\mE(a)=\mE\lb \sum_{q'q} P^{q'}a P^{q}\rb=\sum_q P^{q}\mE(a)P^{q}=\sum_q \mE^{q}(a)\nn\\
    &&\mE^{q}(a)=\mE\lb P^{q}aP^{q}\rb,
\end{eqnarray}
where we have used $P^{q'}c P^{q}=\delta_{q'q}c^q$ for all $c\in \mA^I$. As a result, every conditional expectation $\mE:\mA\to\mA^I$ decomposes as a sum of conditional expectations $\mE^q:B(\mK_1^q\otimes \mK_2^q)\to  B(\mK_1^q)\otimes \mI_2^q$. However, we already showed that the conditional expectations $\mE^q$ are labelled by density matrices $\sigma_2^q$:
\begin{eqnarray}
    \mE^q_{\sigma}(a_1^q\otimes a_2^q)=\tr_2\lb (\mI_1^q\otimes \sigma_2^q)(a^q_1\otimes a_2^q)\rb\ .
\end{eqnarray}
As a result, the conditional expectations from $\mA\to \mA^I$ are in one-to-one correspondence with unnormalized states $\sigma=\oplus_q \mI_1^{q}\otimes \sigma_2^{q}$ on the commutant $(\mA^I)'$:
\begin{eqnarray}
    \mE_{\sigma}(a)=\tr_2(\sigma a)\otimes \mI_2=\oplus_q \tr_2\lb (\mI_1^q\otimes \sigma_2^q)P^{q}a P^{q}\rb\otimes \mI_2
    ^q\ .
\end{eqnarray}
This conditional expectation preserves every state of the form $\rho=\oplus_q p_q \rho_1^{q}\otimes \sigma_2^q$:
\begin{eqnarray}
    &&\tr(\rho \mE_\sigma(a))=\sum_q \tr(\rho \mE_\sigma^q(a))=\sum_{q}p_q \tr\lb (\rho_1^{q}\otimes \sigma_2^q)a\rb=\tr(\rho a)\ .
\end{eqnarray}
If a state does not have the form we postulated for $\rho$ there exists no conditional expectation that preserves it. 
The restriction of the state $\rho$ to the subalgebra $\mA^I$ is
\begin{eqnarray}
    \rho_0=\oplus_q p_q\rho_1^{q}\otimes \mI_2^q\ .
\end{eqnarray}
The discussion above was restricted to finite dimensional matrix algebras. In section \ref{app:errorcorrection}, we show that the necessary and sufficient condition for the existence of a $\rho$-preserving conditional expectation is
\begin{eqnarray}
    \rho^{1/2}c\rho^{-1/2}=\rho_0^{1/2}c \rho_0^{-1/2}\ .
\end{eqnarray}
This condition holds trivially for $\sigma$ and $\sigma_0$ in the example above.

\section{GNS Hilbert space and Petz map}\label{sec:GNS}

The class of completely positive (CP) linear maps are of particular importance for error correction because they describe the dynamics of an open quantum system. Here, we review the representation of CP maps as contractions in the GNS Hilbert space. In particular, we discuss the Petz dual map that plays a central role in operator algebra quantum error correction.

In appendix \ref{sec:CPmaps&duals}, we used the trace to define an inner product and represent the algebra of finite dimensional complex matrices as a Hilbert space.
In some infinite dimensional systems the trace of the identity operator is infinite. As a result, the vector representative of the identity operator cannot be normalizable. 
Even worse, in some quantum systems such as the algebra of local observables in quantum field theory (QFT) there exists no trace.\footnote{Formally, trace is defined to be a CP map $\tr:\mA\to \mathbb{C}$ such that $\tr(a_1a_2)=\tr(a_2a_1)$ for all $a_1$ and $a_2$.} The Stinespring theorem gives us a hint as how to define a Hilbert space without using a trace. While this construction is fully general, here, we use the notation of matrix algebras that might be more accessible to physicists. We comment on a few subtleties in infinite dimensions.

Given a density matrix $\rho=\sum_i \lambda_i^2\ket{i}\bra{i}$ consider the CP map $\phi_\rho:\mA\to \mathbb{C}$ given by $\phi_\rho(a)=\tr(\rho a)$. If $\rho$ is full rank this map is faithful. The Hilbert space $\hat{\mH}$  we obtain in the Stinespring theorem is called the GNS Hilbert space and we denote it by $\mH_\rho$. 
It defines a map from $\mA\to \mH_\rho$ that replaces the unnormalized vector $\ket{e}$ in (\ref{evector}) with a normalized vector $\ket{\rho^{1/2}}$:
\begin{eqnarray}\label{rho1/2}
  &&a\to \ket{a}_\rho=(a\otimes \mathbb{I})\ket{\rho^{1/2}}= \sum_i \lambda_i(a\otimes \mathbb{I})\ket{i}\ket{i}\nn\\
&&\ket{\rho^{1/2}}=\sum_i \lambda_i\ket{i}\ket{i}\ .
\end{eqnarray}
The Hilbert space $\mH_\rho$ is simply the set of vectors $(a\otimes \mI )\ket{\rho^{1/2}}$ endowed with the inner product
\begin{eqnarray}\label{normGNS}
    \braket{a_1|a_2}_\rho\equiv \text{tr}(\rho a_1^\dagger a_2)=\braket{\rho^{1/2}|(a_1^\dagger a_2\otimes \mathbb{I})|\rho^{1/2}}\ .
\end{eqnarray}
As we saw in the Stinespring dilation, if $\rho$ is not full rank we first need to quotient by null vectors.
When $\rho$ is full rank the GNS Hilbert space is isomorphic to $\mK_A\otimes \mK'_A$. 
The vector $\ket{\rho^{1/2}}$ is a purification of the density matrix $\rho$ in $\mK_A \otimes \mK_A'$:
\begin{eqnarray}
    &&\braket{\rho^{1/2}|(a\otimes \mI)|\rho^{1/2}}=\tr(\rho a)
\end{eqnarray}

In the GNS Hilbert space of matrix algebras every operator $a\in \mA$ has a mirror operator $a_m\in \mA'$:
\begin{eqnarray}
    &&(a\otimes\mI)\ket{\rho^{1/2}}=(\mI\otimes a_m)\ket{\rho^{1/2}}\nn\\
    &&a_m=\rho^{1/2}a^T\rho^{-1/2}
\end{eqnarray}
where $a^T$ is the transpose of $a$ in the eigenbasis of $\rho$. 
If $V'\in \mA'$ is an isometry its mirror operator $(V')^m\in \mA$ acting on $\ket{\rho^{1/2}}$ gives another purification of $\rho$ in $\mH_\rho$. 
It is desirable to find a subset of vectors that is in one-to-one correspondence with density matrices. We define the anti-linear {\it modular conjugation} operator in the GNS Hilbert space as
\begin{eqnarray}\label{modularconj}
    J_\rho (a\otimes \mI)\ket{\rho^{1/2}}
    =(\mI\otimes (a^\dagger)^T)\ket{\rho^{1/2}}
\end{eqnarray}
where the transpose is in the eigenbasis of $\rho$. There is a unique purification of $\rho$ that is invariant under $J_\rho$.\footnote{ The set of all vectors that are invariant under $J_\rho$ is called the {\it natural cone} and are of the form $aJ_\rho a\ket{\rho^{1/2}}$. Given a vector in the natural cone there exists no isometry $V'\in \mA'$ that leaves the state invariant. See \ref{app:natural} for a discussion of the natural cone.} The modular conjugation $J_\rho$ acts as an anti-linear swap in the eigenbasis of $\rho$:
\begin{eqnarray}\label{modularconjug}
&&J_\rho\ket{\rho^{1/2}}=\ket{\rho^{1/2}}\nn\\
    &&J_\rho c_i\ket{i}\ket{j}=c_i^*\ket{j}\ket{i}
\end{eqnarray}
where $c_i$ is a complex number.

In the GNS Hilbert space of matrix algebras $M_n$, we have a one-to-one correspondence between vectors $\ket{a}_\rho$ and operators $a\in\mA$.\footnote{The reason is that if $\ket{\Psi}=a_1\ket{\rho^{1/2}}=a_2\ket{\rho^{1/2}}$ then $(a_1-a_2)\ket{\rho^{1/2}}=0$ which is impossible for an invertible $\rho$.} In infinite dimensions, to every operator corresponds a vector in the GNS Hilbert space but not every vector corresponds to an operator. This has to do with the fact that the GNS Hilbert space $\mH_\rho$ is not the set $a\ket{\rho^{1/2}}$ but its closure.

\subsection{Superoperators versus operators}\label{sec:supervsoperator}

In matrix algebras, there is also a one-to-one correspondence between the linear operators in $\mH_\rho$ and linear maps from $\mA$ to  $\mA$.\footnote{This follows from a straightforward generalization of the argument in section \ref{sec:CPmaps&duals}.} In a general von Neumann algebra, including the local algebra of QFT, every normal superoperator has a corresponding  operator in the GNS Hilbert space, however the converse does not hold; see section \ref{sec:QFT}. To prove statements about superoperator it is often easier to use their corresponding operators in the GNS Hilbert space.

Consider a general superoperator $\mT$. If it is unital its corresponding $T_\rho$ leaves $\ket{\rho^{1/2}}$ invariant: $T_\rho\ket{\rho^{1/2}}=\ket{\rho^{1/2}}$. If it is $\rho$-preserving the conjugate $T^\dagger_\rho$ leaves $\ket{\rho^{1/2}}$ invariant:
\begin{eqnarray}
\tr(\rho \mT(a))=\braket{\rho^{1/2}|\mT(a)\rho^{1/2}}=\braket{\rho^{1/2}|T_\rho a\rho^{1/2}}=\braket{T^\dagger_\rho \rho^{1/2}|a\rho^{1/2}}\ .
\end{eqnarray}
We show in (\ref{contraction}) that a unital CP map that preserves $\rho$ corresponds to a contraction in $\mH_\rho$; see appendix \ref{sec:CPmaps&duals} for the definitions. A $\rho$-preserving superoperator $\mE_\rho:\mA\to \mA^C$ that leaves every operator in $c\in \mA^C$ invariant corresponds to an operator $E_\rho:\mH_\rho\to \mH_C$ that satisfies $E_\rho^2=E_\rho$. Here, $\mH_C$ is the subspace spanned by $c\ket{\rho^{1/2}}$. If this superoperator is CP it is a conditional expectation. 
Then, from the bimodule property of conditional expectations in (\ref{bimod}) we have
\begin{eqnarray}
 &&\braket{a_1|E_\rho a_2}_\rho=\braket{\rho^{1/2}|a_1^\dagger \mE_\rho(a_2)\rho^{1/2}}=\braket{\rho^{1/2}|\mE_\rho(a_1^\dagger \mE_\rho(a_2))\rho^{1/2}}\nn\\
 &&=\braket{\rho^{1/2}|\mE_\rho(a_1^\dagger)\mE_\rho(a_2)\rho^{1/2}}=\braket{a_1|E_\rho^\dagger E_\rho a_2}_\rho
\end{eqnarray}
where we have used the notation from (\ref{normGNS}) for the inner product in the GNS Hilbert space. This implies $E_\rho=E_\rho^\dagger E_\rho$ which combined with $E_\rho=E_\rho^2$ implies that $E_\rho$ is an orthogonal projection. As a result, the GNS operators corresponding to a conditional expectation $\mE_\rho$ is simply the projection from $E_\rho:\mH_\rho\to \mH_C$. It follows that the $\rho$-preserving conditional expectation is unique, because we have
\begin{eqnarray}
    \lb E^{(1)}_\rho-E_\rho^{(2)}\rb \ket{c}_\rho=0
\end{eqnarray}
which implies that $E^{(1)}_\rho=E_\rho^{(2)}$ and $\mE^{(1)}_\rho=\mE_\rho^{(2)}$.\footnote{In infinite dimensions, since the map between operators and superoperators is not one-to-one one needs a more careful analysis, however the result remains the same \cite{takesaki1972}.}

An important anti-linear superoperator to consider in the algebra is the Tomita map $\mathcal{S}(a)=a^\dagger$ (see section \ref{sec:cp_maps}) \cite{witten2018aps}. Its corresponding operator in the Hilbert space  is the {\it Tomita} operator $S_\rho:\mH_\rho\to \mH_\rho$ that acts as
\begin{eqnarray}
    S_\rho(a\otimes \mI)\ket{\rho^{1/2}}=(a^\dagger\otimes \mI)\ket{\rho^{1/2}}\ .
\end{eqnarray}
We can also introduce an anti-linear superoperator $\mathcal{J}_\rho:\mA\to \mA'$ which establishes a one-to-one correspondence between operators in $\mA$ and $\mA'$:
\begin{eqnarray}
\mathcal{J}_\rho(\ket{i}\bra{j})=\ket{i}\bra{j}\in \mA'\ .
\end{eqnarray}
For a general $a\in \mA$ we have $\mathcal{J}_\rho(a)=(a^\dagger)^T\in \mA'$.
Its corresponding operator in $\mH_\rho$ is the modular conjugation operator we defined in (\ref{modularconj}).

\begin{table}[t]
\centering
\begin{tabular}{|c|c|c|}
     \hline
     \multicolumn{2}{|c|}{\bf Superoperator} & \bf GNS Operator  \\
     \hline
     \multicolumn{2}{|c|}{(anti-)linear $\mT$} & (anti-)linear T \\
     \hline
     \multirow{8}{*}{linear CP} & unital $\Phi$ & $F:\, (F-1)\ket{\rho^{1/2}}=0$ \\
     \cline{2-3}
     & $\rho$-preserving $\Phi$ & $F:\, (F^\dagger-1)\ket{\rho^{1/2}}=0$ \\
     \cline{2-3}
     & unital $\rho$-preserving $\Phi$ & contraction $\|F\|\leq 1$ \\
     \cline{2-3}
     & conditional expectation $\mE$ & projection $E^2=E$ \\
     \cline{2-3}
     & isometric embedding $\iota$ & isometry $W$ \\
     
     & (faithful representation) & $W^\dagger W=1$ \\
      \cline{2-3}
    & $\rho$-dual  $\Phi'_\rho$ & co-isometry $F^\dagger$ \\
    \cline{2-3}
     & Petz dual $\Phi^P_\rho$ & $J_B F^\dagger J_A$ \\
    \hline
     linear non-CP & relative modular operator $\mathcal{D}_{\sigma|\rho}$ & $\Delta_{\sigma|\rho} = \sigma\otimes\rho^{-1}$ \\

     \hline
     anti-linear & Tomita map $\mS$ & Tomita operator $S_\rho$ \\
     \cline{2-3}
     non-CP & modular conjugation $\mathcal{J}_\rho$ & modular conjugation $J_\rho$ \\
     \hline
\end{tabular}
\caption{Linear maps of the operator algebra (superoperators) correspond to operators in the GNS Hilbert space. Above is a list of some important superoperators and their corresponding operators. In matrix algebras, this correspondence is one to one.}
\label{tab:supop-op}
\end{table}

Another important superoperator is the {\it relative modular} operator defined as $\mathcal{D}_{\sigma|\rho}(a)=\sigma a \rho^{-1}$ for two invertible density matrices $\sigma$ and $\rho$. Its corresponding operator in the Hilbert space is $\Delta_{\sigma|\rho}=\sigma\otimes \rho^{-1}$:
\begin{eqnarray}
    (\mathcal{D}_{\sigma|\rho}(a)\otimes \mathbb{I})\ket{\rho^{1/2}}=\Delta_{\sigma|\rho}(a\otimes \mathbb{I})\ket{\rho^{1/2}}\ .
\end{eqnarray}
If both density matrices are the same this operator is called {\it modular operator} $\Delta_\rho\equiv\rho\otimes \rho^{-1}$ and corresponds to a symmetry of $\ket{\rho^{1/2}}$:
\begin{eqnarray}\label{modularop}
    \Delta_\rho^\alpha\ket{\rho^{1/2}}=\ket{\rho^{1/2}}
\end{eqnarray}
where $\alpha$ is any complex number. The modular map $\mathcal{D}_\rho(a)\equiv \rho a\rho^{-1}$ is multiplicative but does not respect the Hermitian conjugation: $\mathcal{D}_\rho(a^\dagger)=(\mathcal{D}_\rho^{-1}(a))^\dagger$.
The {\it modular flow} of an operator is a unital isometric CP map from the algebra to itself:
\begin{eqnarray}
    a_\rho(t)\equiv \Delta_\rho^{it}(a\otimes \mathbb{I})\Delta_\rho^{-it}=(\rho^{it}a\rho^{-it}\otimes \mathbb{I})
\end{eqnarray}
It is straightforward to check that $S_\rho=J_\rho \Delta_\rho^{1/2}$ and $J_\rho=\Delta_\rho^{1/2}S_\rho$:
\begin{eqnarray}
    &&J_\rho\Delta_\rho^{1/2}\ket{a}_\rho=J_\rho(\mI\otimes a^T)\ket{\rho^{1/2}}=(a^\dagger\otimes \mI)\ket{\rho^{1/2}}=S_\rho\ket{a}_\rho\\
    &&\Delta_\rho^{1/2}S_\rho\ket{a}_\rho=\Delta^{1/2}_\rho(a^\dagger\otimes \mI)\ket{\rho^{1/2}}=(\mI\otimes (a^\dagger)^T)\ket{\rho^{1/2}}=J_\rho\ket{a}_\rho\ . \nn
\end{eqnarray}
\subsection{Natural cone}\label{app:natural}

In this subsection, we characterize the vectors in $\mH_\rho$ that are invariant under $J_\rho$. The cone of such vectors is called the natural cone. Vectors in the natural cone are in one-to-one correspondence with the reduced states on $\mA$.

In matrix algebras, the modular conjugation operator $J_\rho$ is the anti-linear swap operator in the eigenbasis of $\rho$:
\begin{eqnarray}
   &&J_\rho\ket{k,k'}=\ket{k',k}\nn\\
   &&\ket{\rho^{1/2}}=\sum_k \sqrt{p_k}\ket{k,k}\ .
\end{eqnarray}
It corresponds to a superoperator that isometrically sends operators in $\mA$ to $\mA'$:
\begin{eqnarray}\label{Jaction}
   J_\rho(a\otimes \mI)J_\rho=(\mI\otimes (a^\dagger)^T)\ .
\end{eqnarray}
The vector $\ket{e_k}=(\rho^{-1/2}\otimes\mI)\ket{\rho^{1/2}}$ is the vector representative of the maximally mixed state in the natural cone. From the action in (\ref{Jaction}) it follows that every vector $(\psi^{1/2}\otimes \mI)\ket{e_k}$ is also in the natural cone because
\begin{eqnarray}
   J_\rho (\psi^{1/2}\otimes \mI)\ket{e_k}=(\mI\otimes (\psi^{1/2})^T)\ket{e_k}=(\psi^{1/2}\otimes \mI)\ket{e_k}
\end{eqnarray}
where we have used the fact that $a$ and $a^T$ are mirror operators of each other in $\mH_{e_k}$:
\begin{eqnarray}
   (a\otimes \mI)\ket{e_k}=(\mI\otimes a^T)\ket{e_k}\ .
\end{eqnarray}
A given state $\psi$ can be purified in any vector of the form
\begin{eqnarray}
   \ket{\Psi_u}=(\psi^{1/2}u\rho^{-1/2}\otimes \mI)\ket{\rho^{1/2}}=(\mathcal{D}_{\psi|\rho}^{1/2}(u)\otimes\mI)\ket{\rho^{1/2}}
\end{eqnarray}
where $u$ is a co-isometry, i.e. $uu^\dagger=\mI$, and $\mathcal{D}_{\psi|\rho}(a)=\psi a \rho^{-1}$ is the relative modular superoperator. 
As we saw above, the case $u=\mI$ is special in that the resulting vector is invariant under $J_\rho$:
\begin{eqnarray}
&&\ket{\psi^{1/2}}\equiv (\psi^{1/2}\rho^{-1/2}\otimes \mI)\ket{\rho^{1/2}}\ .
\end{eqnarray}
Written in the Hilbert space the unique representative of the state $\psi$ in the natural cone is 
\begin{eqnarray}
   \ket{\psi^{1/2}}=\Delta_{\psi|\rho}^{1/2}\ket{\rho^{1/2}}\ .
\end{eqnarray}
In infinite dimensions, all the purifications of $\psi$ correspond to $u'\ket{\rho^{1/2}}$ where $u'\in \mA'$ is an isometry and $\ket{\psi^{1/2}}$ is the natural cone representative of $\psi$:
\begin{eqnarray}
   J_\rho \ket{\psi^{1/2}}=\ket{\psi^{1/2}}=\Delta_{\psi|\rho}^{1/2}\ket{\rho^{1/2}}\ .
\end{eqnarray}

An alternative way to characterize the vectors in the natural cone is to note that the vector $aJ_\rho a\ket{\rho^{1/2}}$ is in the natural cone for any $a\in \mA$. We define $\tilde{a} = \mathcal{D}_\rho^{-1/4}(a)$ to write this vector as: 
\begin{eqnarray}
a J_\rho a \ket{\rho^{1/2}} = a S_\rho \Delta_\rho^{-1/4}\tilde{a}\ket{\rho^{1/2}} = \Delta_\rho^{1/4}\tilde{a}\tilde{a}^\dagger\ket{\rho^{1/2}}
\end{eqnarray}
where we have assumed $a$ is an analytic operator, meaning that $\mathcal{D}_\rho^\alpha(a)\in \mA$. 
The natural cone is identical the cone of vectors $\Delta_\rho^{1/4}aa^\dagger\ket{\rho^{1/2}}$.

In section \ref{sec:petz} we introduced the alternate inner product
\begin{eqnarray}
(a_1,a_2)_\rho=\braket{a_1\rho^{1/2}|\Delta_\rho^{1/2}a_2\rho^{1/2}}\ .
\end{eqnarray}
With the rewriting above, the alternate inner product is simply the norm of these natural cone vectors $(a_+|a_+)_\rho=\|\Delta_\rho^{1/4}a_+\ket{\rho^{1/2}}\|$. Note that every operator in matrix algebra is analytic. In infinite dimensions, the set of analytic operators is dense in $\mA$ \cite{bratteli2012operator}.
Table \ref{tab:supop-op} is a list of some important superoperators in this work, and their corresponding operators in the GNS Hilbert space.

\subsection{Fixed points}\label{sec:fixed}

Given a linear superoperator $\mathcal{T}:\mA\to \mA$ we can define its spectrum to be the values of $\lambda$ for which $\mathcal{T}-\lambda \,\text{id}$ is not invertible. It is convenient to think in terms of the spectrum of the GNS representation of $\mT$ in a Hilbert space $T:\mH_\rho\to \mH_\rho$, that is the set of all $\lambda$ for which $T-\lambda \mI$ is not invertible.
Since the unital CP maps correspond to contractions, their spectra satisfy $|\lambda|\leq 1$. 

In finite dimensions, the spectrum corresponds to the set of $\lambda$ for which there exists an operator $X$ such that $\mathcal{T}(X)=\lambda X$. The operator $T$ has a Jordan block representation
\begin{eqnarray}
T=X(\oplus_k J_k(\lambda_k))X^{-1},\qquad J_k(\lambda)=
\begin{pmatrix}
 \lambda &1& \dots&0\\
 0&\lambda& 1& &\\
 \vdots & & \ddots&1\\
 0&\dots &0 &\lambda
\end{pmatrix}\in M_{d_k}(\mathcal{C})\ .
\end{eqnarray}
We can split each Jordan block into a projection $P_k$ and a nilpotent part $N_k$\footnote{For a review of the spectral theory of quantum channels see \cite{wolf2012quantum}.}:
\begin{eqnarray}
 J_k(\lambda)=\lambda P_k+N_k,\qquad N_k^{d_k}=0,\qquad P_k P_l=\delta_{kl}P_k,\qquad \text{tr}(P_k)=d_k\ .
\end{eqnarray}
If a Jordan block $J_k$ is just a projection with no nilpotent part we call it trivial. For a unital CP map $\Phi$ the Jordan blocks corresponding to $|\lambda|=1$ are always trivial \cite{wolf2012quantum}. Therefore, we can define the projection $E$ to the set of invariant operators:
\begin{eqnarray}
 E=\sum_{\lambda_k=1}P_k\ .
\end{eqnarray}
As we will show in section \ref{app:errorcorrection}, this projection $E$ corresponds to a conditional expectation that one can canonically associate to a unital CP map $\Phi:\mA\to \mA$. It is given by $\mathcal{E}=\lim_{N\to \infty}\sum_{n=1}^N \Phi^n$ and projects to the subalgebra of operators invariant under $\Phi$. This subalgebra plays an important role in error correction.
The Jordan block form of a matrix has a generalization for compact operators in general Hilbert spaces.

\subsection{Petz dual map}\label{sec:petz}

Consider a linear superoperator $\mT:\mA\to \mB$ and the GNS Hilbert spaces $\mH_{\rho_A}$ and $\mH_{\rho_B}$. To simplify the notation, we denote the Hilbert spaces with $\mH_A$ and $\mH_B$, respectively. The operator corresponding to $\mathcal{T}$ is defined by
\begin{eqnarray}
\mT(a)\ket{\rho_B^{1/2}}=Ta\ket{\rho_A^{1/2}}\ .
\end{eqnarray}
It is tempting to define the dual map in the GNS Hilbert space by the equation
\begin{eqnarray}
    &&\braket{b \rho_B^{1/2} |\mT(a)\rho^{1/2}_B}=\braket{\mT_\rho^*(b)\rho_A^{1/2}|a\rho_A^{1/2}}
\end{eqnarray}
or equivalently
\begin{eqnarray}
    \braket{b|\mT(a)}_{\rho_B}=\braket{\mT^*_\rho(b)|a}_{\rho_A}
\end{eqnarray}
for operators $a\in\mA$ and $b\in\mB$. In the case of matrix algebras, we have
\begin{eqnarray}
      \tr(\rho_B b^\dagger \mT(a))=\tr(\rho_A \mT^*_\rho(b^\dagger) a) \ .
\end{eqnarray}
However, there is a problem with this definition that can be seen  by solving explicitly for $\mT^*_\rho$ in terms of $\mT^*$ defined in (\ref{transpose}):
\begin{eqnarray}
 \mT^*_\rho(b)=\rho_A^{-1}\mT^*(\rho_B b)\ .
\end{eqnarray}
Defined this way the dual of a CP map is not CP! 

If we think in terms of the GNS Hilbert space then the superoperator $\mT$ is represented by the operator $T$ whose conjugate is $T^\dagger$. 
The problem is that the superoperator $\mT^*_\rho$ we get by solving the equation 
\begin{eqnarray}
T^\dagger b\ket{\rho_B^{1/2}}=\mT^*_\rho(b)\ket{\rho_A^{1/2}}
\end{eqnarray}
is not CP.
However, since every vector in $\mH_B$ can also be written as $b'\ket{\rho_B^{1/2}}$ we could consider $T^\dagger$ as corresponding to a superoperator $\mT'_\rho$ from $\mB'\to \mA'$. If we consider the superoperator on the commutant that corresponds to $T^\dagger$ we get a dual map $\mT'_\rho:\mB'\to \mA'$
\begin{eqnarray}\label{rhodual}
   T^\dagger b' \ket{\rho_B^{1/2}}=\mT'_\rho(b') \ket{\rho_A^{1/2}}
\end{eqnarray}
that is positive if $\mT$ is positive.\footnote{This is the dual map of Accardi and Cecchini \cite{accardi1982conditional}; see theorem \ref{thm:rhodual} in section \ref{sec:QFT}. More formally, the existence of this map is guaranteed by the commutant Radon-Nikodym theorem (for instance see theorem 2.1 of \cite{albeverio1978frobenius}).} For a positive operator $b'_+\in\mB'$ we have
\begin{eqnarray}
   && \braket{a|\mT'_\rho(b_+')a}_{\rho_A}=\braket{\rho_A^{1/2}|a^\dagger \mT'_\rho(b'_+) a|\rho_A^{1/2}}=\braket{\rho_A^{1/2}| \mT'_\rho(b_+') a^\dagger a|\rho_A^{1/2}}
   =\braket{\rho^{1/2}_B| b'_+ \mT(a^\dagger a)\rho_B^{1/2}}\nn\\
   &&=\braket{\rho_B^{1/2}|(b'_+)^{1/2} \mT(a^\dagger a)(b')_+^{1/2}|\rho_B^{1/2}}=\braket{(b'_+)^{1/2}|\mT(a^\dagger a)(b'_+)^{1/2}}_{\rho_B}\geq 0
\end{eqnarray}
where we have used the fact that $[\mT_\rho'(b'),a]=0$. We call the map $\mT'_\rho$ the {\it $\rho$-dual} of $\mT$.

We saw that the modular conjugation map for cyclic and separating vectors is a unitary superoperator $\mathcal{J}:\mA\to \mA'$. We can use the modular conjugation and $\rho$-dual to associate to each linear CP map $\mT:\mA\to \mB$ a unique linear CP map $\mT^P_\rho:\mB\to \mA$ that we call  the {\it Petz dual map}:\footnote{Accardi and Cecchini call this map the bi-dual \cite{accardi1982conditional}.}
\begin{eqnarray}\label{Petzdualmap}
 \mT^P_\rho(b)=\mathcal{J}_A\circ \mT'_\rho\circ \mathcal{J}_B(b)=J_A\mT'_\rho(J_B b J_B)J_A\ .
\end{eqnarray}
Another way to understand the Petz dual map is to realize that it is the dual map defined with respect to an {\it alternate inner product}
\begin{eqnarray}\label{alteranteinner}
(a_1|a_2)_\rho\equiv \braket{\mathcal{J}_\rho(a_1^\dagger)\rho^{1/2}|a_2\rho^{1/2}}=\tr(\rho^{1/2} a_1^\dagger \rho^{1/2}a_2)\ .
\end{eqnarray}
Note that this is the Heisenberg picture of the Petz recovery map in equation 5 of \cite{junge2018universal}.
In the GNS Hilbert space, this inner product can be expressed using the modular operator\footnote{The alternate inner product can be understood as the GNS inner product in the natural cone where we choose the vector representative of a state to be invariant under modular conjugation $J_\rho$. See appendix \ref{app:natural}.}
\begin{eqnarray}
    (a_1|a_2)_\rho=\braket{\rho^{1/2}|a_1^\dagger\Delta_\rho^{1/2}a_2|\rho^{1/2}}\ .
\end{eqnarray}
The Petz dual is the dual of a CP map defined with the alternate inner product
\begin{eqnarray}\label{newinner}
   \tr(\rho_B^{1/2}b \rho_B^{1/2}\mT(a))=\tr(\rho_A^{1/2}\mT^P_\rho(b) \rho_A^{1/2}a)
\end{eqnarray}
which can be solved explicitly in terms of the standard trace-dual as
\begin{eqnarray}\label{rhodualnew}
    \mT_\rho^P(b)=\rho_A^{-1/2}\mT^*(\rho_B^{1/2}b\rho_B^{1/2})\rho_A^{-1/2}\ . 
\end{eqnarray}
This map is manifestly CP. Note that the Petz dual of a unital map is also unital. 

Consider the example of an isometric embedding $\iota_1(a_1)=a_1\otimes \mI_2$ with the GNS Hilbert space $\mH_{\rho_{12}}\simeq \mK_{12}\otimes \mK'_{12}$ where $\mK_{12}=\mK_1\otimes \mK_2$, the dual map $\iota_1^*$ is partial trace. If the state on $\mA_{12}$ is a full rank density matrix $\rho_{12}$, the reduced state on $\mA_1$ is $\rho_1$ and its corresponding GNS Hilbert space is $\mH_{\rho_1}\simeq \mK_1\otimes \mK'_1$. This embedding is isometric with respect to the GNS inner product
\begin{eqnarray}
    \braket{a_1|a_2}_{\rho_1}=\braket{(a_1\otimes \mathbb{I}_2)|(a_2\otimes \mathbb{I}_2)}_{\rho_{12}}\ .
\end{eqnarray}
Consider the isometry $V_\rho:\mH_{\rho_1}\to \mH_{\rho_{12}}$ defined by
\begin{eqnarray}\label{isomet}
    V_\rho a_1\ket{\rho_1^{1/2}}=(a_1\otimes \mI_2)\ket{\rho_{12}^{1/2}}
\end{eqnarray}
In this case, the Petz dual map is
\begin{eqnarray}
    \iota^P_\rho(a_1\otimes a_2)=\rho_1^{-1/2}\tr_2(\rho_{12}^{1/2}(a_1\otimes a_2)\rho_{12}^{1/2})\rho_1^{-1/2}\ .
\end{eqnarray}
The composite map $\mE^P_\rho=\iota_1\circ\iota^P_\rho$ in general does not leave $a_1\otimes \mathbb{I}_2$ invariant. It becomes a conditional expectation if and only if $\rho_{12}=\rho_1\otimes \rho_2$.

Consider an isometric embedding $\iota:\mA^C\to \mA$. We call the $\rho$-preserving CP map $\mE^P_\rho=\iota \circ \iota^P_\rho$ a {\it generalized conditional expectation} \cite{accardi1982conditional}. 
In the GNS Hilbert space the operator that corresponds to $\iota$ is an isometry $W$:
\begin{eqnarray}
    \iota(c)\ket{\rho^{1/2}}=Wc\ket{\rho_C^{1/2}}\ .
\end{eqnarray}
From (\ref{rhodual}) and (\ref{Petzdualmap}) we find that the Petz dual corresponds to $J_CW^\dagger J_A$:
\begin{eqnarray}
    \iota^P_\rho (a)\ket{\rho_C^{1/2}}=J_C W^\dagger J_A a\ket{\rho^{1/2}}\ .
\end{eqnarray}
As a result, in the GNS Hilbert space the composite maps $\iota^P_\rho\circ \iota:\mA^C\to \mA^C$ and $\iota\circ \iota^P_\rho:\mA\to \mA^C$ are represented by operators $J_CW^\dagger J_A W$ and $WJ_CW^\dagger J_A$, respectively. 
Consider the {\it Takesaki condition}
\begin{eqnarray}\label{TakesakiJ}
J_AW=WJ_C\ .
\end{eqnarray}
This is an operator constraint in the GNS Hilbert space, hence it is a constraint on the state $\rho$.
Assuming the Takesaki condition, we find that $\iota^P_\rho\circ \iota$ is the identity map and $\iota\circ\iota^P_\rho$ is a $\rho$-preserving conditional because it is represented by the projection $WW^\dagger$ on $\mH_\rho$. In section \ref{app:errorcorrection}, we will see that the Takesaki condition is necessary for the existence of a $\rho$-preserving conditional expectation.

To highlight the difference between the Petz dual map and the dual map defined with respect to the Hilbert-Schmidt inner product we work out an example from commuting algebras \cite{wilde_2013}. 
Consider a trace-preserving CP map $\mathcal{N}$ with Kraus operators $V_{\alpha k}:\mK_B\to \mK_A$ where $\{\ket{k}\}$ and $\{\ket{\alpha}\}$ are orthonormal bases of $\mK_A$ and $\mK_B$, respectively:
\begin{eqnarray}
&& \mathcal{N}(a)=\sum_{\alpha k}V_{\alpha k}^\dagger a V_{\alpha k }\nn\\
&&    V_{\alpha k }^\dagger=\sqrt{p(\alpha|k)}\ket{\alpha}\bra{k}
\end{eqnarray}
and $p(\alpha|k)$ is the conditional probability that the vector $\ket{k}$ evolves to $\ket{\alpha}$. Such map are called classical-to-classical channels because they preserve the orthogonality of the basis $\{\ket{k}\}$. This map evolves $\rho=\sum_k p_k\ket{k}\bra{k}$ to $\mathcal{N}(\rho)=\sum_\alpha p_\alpha \ket{\alpha}\bra{\alpha}$ with
\begin{eqnarray}
    p_\alpha=\sum_k p(\alpha|k)p_k\ .
\end{eqnarray}
The Kraus operators of the dual map are the complex conjugate 
\begin{eqnarray}
    &&\mathcal{N}^* (b)=\sum_{\alpha k}V_{\alpha k} b V_{\alpha k}^\dagger
\end{eqnarray}
whereas the Petz dual map is
\begin{eqnarray}
  &&  {\mathcal{N}}^P_\rho(b)=\sum_{\alpha k}\hat{V}_{\alpha k} b \hat{V}_{\alpha k}^\dagger\nn\\
    &&\hat{V}_{\alpha k}=\sqrt{p(k|\alpha)}\ket{k}\bra{\alpha}\ .
\end{eqnarray}
The Petz dual map undoes the evolution by sending vector $\ket{\alpha}$ to $\ket{k}$ with conditional probability $p(k|\alpha)$ which is obtained using the Bayes rule
\begin{eqnarray}
    p(k|\alpha)p_\alpha=p(\alpha|k)p_k\ .
\end{eqnarray}

\subsection{CP maps in infinite dimensions}

A CP map from the algebra to complex numbers $\rho:\mA\to \mathbb{C}$ is an un-normalized state. It is normalized if the map is unital. In infinite dimensions, it is convenient to restrict to the set of continuous states: $\rho(\lim_n a_n)=\lim_n \rho(a_n)$. Such states are called {\it normal}.
Given a normalized continuous state $\rho:\mA\to \mathbb{C}$ the GNS Hilbert space is formed by the vectors $\ket{a}_\rho=a\ket{\rho^{1/2}}$ with the inner product
\begin{eqnarray}
  \braket{a_1|a_2}_\rho=\rho(a_1^\dagger a_2)\ .  
\end{eqnarray}
If $\rho$ is not faithful one needs to quotient by the set of null vectors $\ket{a}_\rho$, i.e. $\rho(a^\dagger a)=0$ and then take the completion. In the example of matrix algebras, states are in one-to-one correspondence to density matrices $\rho(a)=\tr(\rho a)$.
A faithful state corresponds to a full rank density matrix. However, in QFT, not every vector in $\mH_\rho$ has a corresponding operator in $\mA$. Since the set $a\ket{\rho^{1/2}}$ is dense in the Hilbert space, some vectors correspond to the limit of operators in $\mA$. 
Similarly, not every operator in $\mA$ has a mirror in $\mA'$. To every vector $\ket{\Psi}\in \mH_\rho$ one can associate an operator $\Psi$ that satisfies:
\begin{eqnarray}\label{creation}
    \forall a'\in \mA':\qquad \Psi a'\ket{\rho^{1/2}}=a'\ket{\Psi}\ .
\end{eqnarray}
Clearly, for $\ket{a}_\rho$ this operator is $a\in \mA$. 
From the property above it follows that 
\begin{eqnarray}
    [\Psi,a'_1]a'_2\ket{\rho^{1/2}}=0\ .
\end{eqnarray}
Therefore, such a $\Psi$ commutes with all $a'_1$, therefore it is affiliated with $\mA$ (need not be bounded).

The Tomita map $\mathcal{S}(a)=a^\dagger$ is represented in $\mH_{\rho}$ with the Tomita operator:
\begin{eqnarray}
    S_\rho a\ket{\rho^{1/2}}=a^\dagger \ket{\rho^{1/2}}\ .
\end{eqnarray}
Since $\mA\ket{\rho^{1/2}}$ is dense in $\mH_\rho$ this defines the action of $S_\rho$ on a dense set of vectors. The closure of the modular operator has a polar decomposition $S_\rho=J_\rho\Delta_\rho^{1/2}$ where $J_\rho$ is the analog of the modular conjugation in equation (\ref{modularconjug}) and $\Delta_\rho=S^\dagger_\rho S_\rho$ is the analog of the modular operator in (\ref{modularop}).\footnote{All the equations in the appendices that we write in the GNS Hilbert space that do not involve the vector $\ket{e}$ continue to hold in QFT.}

The same superoperator $\mS(a)=a^\dagger$ can be represented as an operator from $\mH_\rho\to \mH_\psi$:
\begin{eqnarray}
    S_{\psi|\rho}a\ket{\rho^{1/2}}=a^\dagger\ket{\psi^{1/2}}\ .
\end{eqnarray}
If $\ket{\psi^{1/2}}$ belongs to $\mH_\rho$ then the equation above makes sense only if $\ket{\psi^{1/2}}$ is invariant under $J_\rho$ (belongs to the natural cone).\footnote{See appendix \ref{app:natural} for a review of the natural cone.}  Otherwise, as we will show below in equation (\ref{relTomitadepends}), the relative Tomita operator depends on the vector representative of $\psi$. 

We can generalize the definition above to a relative Tomita operator that depends on two arbitrary vectors
 \begin{eqnarray}
     S_{\Psi|\Omega}a\ket{\Omega}=a^\dagger\ket{\Psi}\ .
 \end{eqnarray}
If $u'\in \mA'$ is an isometry the vector $u'\ket{\psi^{1/2}}$ has the same reduced state $\psi$ on $\mA$. If we choose the vector $\ket{\Psi_u}=u'\ket{\psi^{1/2}}$ and $\ket{\Omega_v}=v'\ket{\omega^{1/2}}$ we find 
\begin{eqnarray}\label{relTomitadepends}
 S_{\Psi_u|\Omega_v}=u'S_{\psi|\omega}\:(v')^\dagger\ .
\end{eqnarray}
To avoid potential confusions about whether our vectors are in the natural cone or not, in the remainder of this subsection, we formulate our expressions in terms of general vectors. Hence, we rename $\ket{\omega^{1/2}}$, $S_\omega$, $J_\omega$ and $\Delta_\omega$ to the vector $\ket{\Omega}$, $S_\Omega$, $J_\Omega$ and $\Delta_\Omega$, correspondingly. Then the unbounded operator $\Psi=S^{\mA'}_{\Psi|\Omega}S_\Omega^{A'}$ satisfies the equation  (\ref{creation}). Since $S^{\mA'}_{\Psi|\Omega}=(S^{\mA}_{\Psi|\Omega})^\dagger$ and $J_{\Psi|\Omega}^\dagger=J_{\Omega|\Psi}$ we have
\begin{eqnarray}\label{operatorcreates}
    \Psi=\Delta^{1/2}_{\Psi|\Omega}J_{\Omega|\Psi}J_\Omega \Delta_\Omega^{-1/2}\ .
\end{eqnarray}
See section 2.2.2 of \cite{lashkari2018modular} for a review of these relations. In appendix \ref{app:natural}, we argue that the operator 
$\Delta_{\Psi|\Omega}^{1/2}\Delta_\Omega^{-1/2}$ acting on $\ket{\Omega}$ creates the vector representative of $\psi$ in the natural cone. Comparing this to the equation (\ref{operatorcreates}) we find that for vectors in the natural cone $J_{\Omega|\Psi}=J_\Omega=J_{\Psi|\Omega}$.  


In matrix algebras, an arbitrary vector is constructed by the action of the operator $\psi^{1/2}u^T\rho^{-1/2}$:
\begin{eqnarray}
    (\psi^{1/2} u^T \rho^{-1/2}\otimes\mI)\ket{\rho^{1/2}}=(\mI\otimes u)\ket{\psi^{1/2}}
\end{eqnarray}
where $\ket{\psi^{1/2}}$ is the vector representative of the $\psi$ in the natural cone \cite{Lashkari_2019}. Here, the transpose $u^T$ is defined in the eigenbasis of $\rho$. The operator $\psi^{1/2}u^T\rho^{-1/2}$ is the analog of (\ref{operatorcreates}) in matrix algebras.


In infinite dimensions, it is desirable to restrict to {\it normal} CP maps defined by their continuity properties. For an increasing bounded sequence of positive operators $a_n$, a CP map $\Phi$ is normal if $\Phi(\lim_n a_n)=\lim_n \Phi(a_n)$.\footnote{The limit is understood in the ultra-weak operator topology.}
In this work, we assume that all of our states and CP maps are normal. 
A unital CP map corresponds to a contraction $F$ that extends to a closed operator in the GNS Hilbert space
\begin{eqnarray}
    F\lim_n a_n\ket{\rho^{1/2}}=\Phi(\lim_n a_n)\ket{\rho^{1/2}}=\lim_n \Phi(a_n)\ket{\rho^{1/2}}=\lim_n Fa_n\ket{\rho^{1/2}}\ .
\end{eqnarray}
However, not every closed operator in the GNS Hilbert space corresponds to a normal superoperator. 

The simplest normal CP maps are normal states $\omega:\mA\to \mathbb{C}$. Consider the state
\begin{eqnarray}
    \omega_{a'}(a_1^\dagger a_2)=\braket{a' \rho^{1/2}|a_1^\dagger a_2|a' \rho^{1/2}}=\braket{a_1\rho^{1/2}|(a')^\dagger a'|a_2\rho^{1/2}}\ .
\end{eqnarray}
This is a generalization of the correspondence between un-normalized density matrices (states) and positive operators $(a')^\dagger a'\in \mA'$ mentioned in appendix \ref{sec:CPmaps&duals}. Since $b'\ket{\rho^{1/2}}$ is dense in the Hilbert space, any state $\psi$ corresponding to a vector $\ket{\Psi}\in \mH_\rho$ corresponds to a positive operator that is affiliated with $\mA'$ (commutes with all $a$):
\begin{eqnarray}\label{densitymatrixcomm}
    \psi(a_1^\dagger a_2)=\lim_n\lim_m\braket{a_1\rho^{1/2}|(a'_m)^\dagger a'_n|a_2\rho^{1/2}}=\braket{a_1\rho^{1/2}|(a'_\psi)^\dagger a'_\psi|a_2\rho^{1/2}}
\end{eqnarray}

\subsection{Kraus representation in infinite dimensions}\label{app:Krausinf}

To characterize the CP maps between infinite dimensional algebras it is convenient to start with the Stinespring dilation theorem. Consider a linear map $\Phi:\mA\to \mB$ with each algebra represented on GNS Hilbert spaces $\mH_{A}$ and $\mH_{B}$. 
Consider the space $\hat{\mH}=\mH_{A}\otimes \mH_{B}$ defined with the inner product 
\begin{eqnarray}
    \braket{a_1 \rho_A^{1/2}, b_1\rho_B^{1/2}|a_2\rho_A^{1/2},b_2\rho_B^{1/2}}_\Phi\equiv\braket{b_1\rho_B^{1/2}|\Phi(a_1^\dagger a_2)|b_2\rho_B^{1/2}}\ .
\end{eqnarray}
As before, if $\Phi$ is not faithful the vectors $\ket{a,\phi}$ with $\Phi(a^\dagger a)$ have zero norm and we quotient by them. After closure $\hat{\mH}$ becomes a Hilbert space.
Similar to the discussion of section \ref{sec:cp_maps} we define a representation $\pi(a)$ of $\mA$ in the Hilbert space $\hat{\mH}$ and the isometry $W:\mH_{B}\to \hat{\mH}$:
\begin{eqnarray}
    &&\pi(a_1)\ket{a_2\rho_A^{1/2},b\rho_B^{1/2}}=\ket{a_1a_2\rho_A^{1/2},b\rho_B^{1/2}}\nn\\
  &&  W\ket{b\rho_B^{1/2}}=\ket{\rho_A^{1/2},b\rho_B^{1/2}}\nn\\
  &&W^\dagger\ket{a\rho_A^{1/2},b\rho_B^{1/2}} = \Phi(a)\ket{b\rho_B^{1/2}}\ .
\end{eqnarray}
As a result, the CP map factors as
\begin{eqnarray}
    \Phi(a)=W^\dagger \pi(a)W\ . 
\end{eqnarray}
If $\Phi$ is faithful $\hat{\mH}=\mH_A\otimes \mH_B$ and $\pi(a)=a\otimes \mI_{BB'}$.
Since we are using a reducible representation on $\mH_B$ for $\mB$ the constraint that $\Phi(a)\in \mB \subset B(\mH_B)$ is non-trivial. We need to have $[\Phi(a),b']=0$ for all $b'\in \mB'$ which implies
\begin{eqnarray}\label{commutationcompl}
    [Wb'W^\dagger,P \pi(a) P]=0
\end{eqnarray}
where $P=WW^\dagger$ and  we have used $W^\dagger\pi(a)W=W^\dagger P\pi(a)P W$. When $\Phi$ is unital the projection $P$ leaves the states $\ket{\rho_A^{1/2}, b\rho_B^{1/2}}$ invariant as $P\ket{a\rho^{1/2}_A, b\rho_B^{1/2}} = \ket{\rho_A^{1/2}, \Phi(a)b\rho_B^{1/2}}$.

The algebra of all operators in the Hilbert space $\mH_B$ is an infinite dimensional matrix algebra,\footnote{It is a type I von Neumann factor. For a classification of von Neumann factors see \cite{witten2018aps}.} therefore we have a resolution of the identity operator in terms of orthogonal projections $\mI_{BB'}=\sum_r \ket{r}\bra{r}$ where $\ket{r}$ are vectors in $\mH_B$. When the CP map $\Phi$ is faithful the representation has the form $\pi(a)=a\otimes \mI_{BB'}$ and we can define $V_r=(1\otimes \bra{r})W$ to obtain a {\it generalized Kraus representation} of the CP map 
\begin{eqnarray}
    \Phi(a)=\sum_r W^\dagger (a\otimes \ket{r}\bra{r})W=\sum_r V_r^\dagger a V_r
\end{eqnarray}
where $V_r:\mH_{B}\to \mH_{A}$. Note that this is different from the standard Kraus representation where the Kraus operators are maps $V_r:\mK_B\to \mK_A$. If the algebra $\mB$ is type I we can take $\hat{\mH}=\mH_{A}\otimes \mK_B$ and we obtain the above representation with $V_r:\mK_B\to \mH_{A}$. It is only when both algebras are type I that we can take $\hat{\mH}=\mK_A\otimes\mK_B$ to obtain the standard Kraus representation.

The Stinespring dilation of unital CP maps discussed above involves the compression of an isometric embedding $\pi$ of $\mA$ in the Hilbert space $\mH_A\otimes \mH_B$. This Hilbert space seems too large. The algebras $\mA$ and $\mA'$ act on $\mH_A$ and $\mB$ and $\mB'$ act on $\mH_B$. In analogy with finite quantum systems, it is desirable to have a Hilbert space where only $\mA$ and $\mB$ act with $\mB$ playing the role of the commutant. 
In fact, given a unital CP map $\Phi$ one can construct a {\it bi-module Hilbert space} $\mH_\Phi$ where the algebra $\mA$ acts on the left and $\mB$ acts on the right. In finite dimensions, the left action $l(a)=a\otimes \mI_B$ and the right action is $r(b)=\mI_A\otimes b$. We have a cyclic and separating vector $\ket{\Omega_\Phi}$ for $\mA\vee \mB$ and 
\begin{eqnarray}
    \braket{\Omega_\Phi|l(a)|\Omega_\Phi}=\rho_B\circ \Phi(a), \qquad \braket{\Omega_\Phi|r(b)|\Omega_\Phi}=\rho_B(b)\ .
\end{eqnarray}
Here, $\rho_B$ is the state before the quantum channel and $\rho_B\circ \Phi$ is the state after the channel \cite{connes2014noncommutative}. Modular theory can be generalized to the bi-module Hilbert space $\mH_\Phi$ \cite{yamagami1994modular}. The Stinespring dilatation of a unital CP map $\Phi$ says that there exists a normal $*$-homomorphism $\rho:\mA\to \mB$ and an isometry $V\in \mB$ such that 
\begin{eqnarray}
    \Phi(a)=V^\dagger \rho(a) V\ .
\end{eqnarray}

\section{Operator algebra error correction}\label{app:errorcorrection}

Suppose we want to simulate a quantum system $\mB$ using the algebra of physical operators $\mA$. 
We encode $\mB$ as a subalgebra of $\mA$ using the isometric embedding map $\iota:\mB\to \mA$. We also have a decoding map $\alpha:\mA\to \mB$ such that $\alpha\circ \iota:\mB\to \mB$ is the identity map. 
The composite map $\iota\circ \alpha:\mA\to \iota(\mA)$ is a CP map that preserves every operator in $\iota(\mB)$. 
The set of states $\rho_A$ that are invariant under this map are the physical states that are decodable. 
Assume that $\rho_A$ is a decodable faithful state (full-rank density matrix) and $\rho_B=\rho_A\circ \iota$ is its restriction to $\mB$.
They can be represented as cyclic and separating vectors $\ket{\Omega_B}$ and $\ket{\Omega_A}$ in the GNS Hilbert spaces $\mH_{\rho_B}$ and $\mH_{\rho_A}$. We denote these Hilbert spaces by $\mH_B$ and $\mH_A$, respectively. 
The encoding map as a superoperator is represented by an isometry $W:\mH_B\to \mH_A$:
\begin{eqnarray}
    \iota(b)\ket{\Omega_A}=W b \ket{\Omega_B}\ .
\end{eqnarray}
Since we assumed that $\rho_A$ is decodable, this state is preserved under the conditional expectation $\iota\circ \alpha:\mA\to \iota(\mB)$. We will see in theorem \ref{thmTakesaki1} that this is equivalent to the Takesaki condition: $J_AW=WJ_B$, where $J_A$ and $J_B$ are the modular conjugation operators corresponding to $\rho_A$ and $\rho_B$. This implies that our decoding map corresponds to the GNS operator 
\begin{eqnarray}
\alpha(a)\ket{\rho_B^{1/2}}=J_B W^\dagger J_A a\ket{\rho_A^{1/2}}=W^\dagger a \ket{\rho_A^{1/2}}\ .
\end{eqnarray}
In other words, a state is decodable if it satisfies the Takesaki condition, in which case $\iota$ is the Petz dual of $\alpha$.

During the simulation, errors $V_r$ can occur that corrupt the physical states: 
\begin{eqnarray}
    a\ket{\rho_A^{1/2}}\to V_r a\ket{\rho_A^{1/2}}\ .
\end{eqnarray}
In particular, this corrupts our encoded states $\iota(b)\ket{\rho_A^{1/2}}$.
In the Heisenberg picture, the states do not change but the errors corrupt the physical operators. If there is only one error $V$ that occurs deterministically it has to be an isometry to preserve the norm of states and the error map in the Heisenberg picture is $a\to V^\dagger a V$. If there is a collection of errors $V_r$ the error map $a\to \Phi(a)=\sum_r V_r^\dagger a V_r$ is a unital CP map. As in (\ref{Krausrep}), we absorb the probability $p_r$ of error $V_r$ occurring in the definition of the Kraus operators.
The goal of the theory of quantum error correction is to find an encoding (faithful representation) of the algebra $\iota$ such that we can detect the errors $V_r$ and correct for them using correction operators $R_r$:
\begin{eqnarray}
    W^\dagger R_r V_r W b\ket{\rho_B^{1/2}}\propto \ket{\rho^{1/2}_B}\ .
\end{eqnarray}
It is convenient to absorb the encoding isometry in the definition of the error operator and the decoding co-isometry in the definition of the correction operators so that we have
\begin{eqnarray}
    &&R_rV_r b \ket{\Omega_B}\propto b \ket{\Omega_B},\nn\\
    &&R_r:\mH_A\to \mH_B, \qquad V_r:\mH_B\to \mH_A\ .
\end{eqnarray}
The new error map and recovery maps are
\begin{eqnarray}
    \Phi(a)=\sum_r V_r^\dagger a V_r\in \mB,\qquad \mathcal{R}(b)=\sum_r R_r^\dagger b R_r\in \mA\ .
\end{eqnarray}

\begin{figure}[t]
    \centering
    \includegraphics[width=0.25\linewidth]{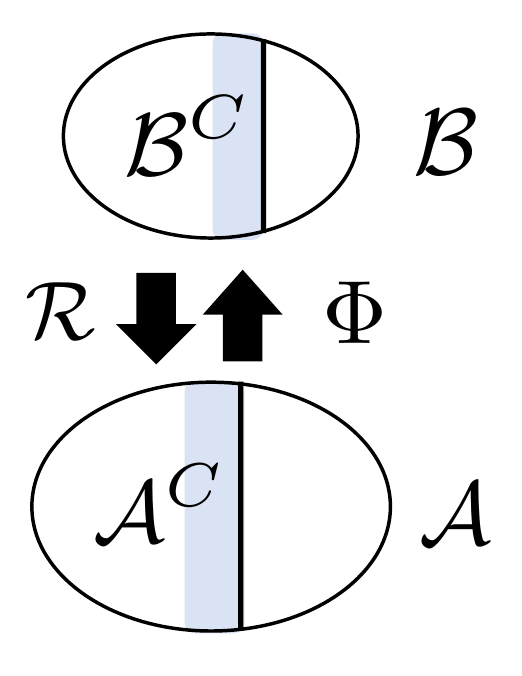}
    \caption{General setup for error correction: the error map is $\Phi:\mA\to \mB$ and the recovery map $\mR:\mB^C\to \mA^C$ recovers the operators in the correctable subalgebra $\mB^C$.}
    \label{fig13u}
\end{figure}

In the Heisenberg picture, instead of correcting states, we correct for operators. 
An operator $c\in \mB$ is called correctable if there exists a recovery map $\mR$ that satisfies $\Phi(\mR(c))=c$.\footnote{Note that in the Heisenberg picture, the order of actions is reversed.}
In addition to $c$, this recovery map corrects the whole algebra of operators invariant under $\Phi\circ \mR$. We say a subalgebra $\mB^C$ is correctable if there exists a recovery map $\mR$ such that $\Phi\circ \mR(c)=c$ for all $c\in \mB^C$; see figure \ref{fig13u}. We will show below that a subalgebra is correctable if and only if for all $c\in \mB^C$ and all errors $V_r$ and $V_s$ we have $[c,V_r^\dagger V_s]=0$. This is the same condition in (\ref{multisubalgeb}) for an operator that belongs to the multiplicative domain of $\Phi^*$.

With a recovery map in hand, we apply $\mR(c_1)\in\mA$ to the corrupted state that has the effect of $c_1$ in the presence of error:
\begin{eqnarray}\label{HilbertSpcorrection}
   \mR(c_1)V_r \ket{c_2}_{\rho_B}=V_r c_1\ket{c_2}_{\rho_B}\ .
\end{eqnarray}
Since the equation above should hold for all $c_2$ it implies an operator equation that we call the {\it recovery equation}:
\begin{eqnarray}\label{recoveryeq}
  \forall r,\qquad \mR(c)V_r=V_rc\ .
\end{eqnarray}

The recovery map $\mR:\mB^C\to \mA$ is a CP map which is unital if the kernel of the error map is empty.
Any operator $X$ that satisfies $XV_r=0$ for all errors $V_r$ can be added to $\mR$. If the span of the range of all $V_r$ is not the whole Hilbert space the error map $\Phi$ is not faithful. The information content of the operators in the kernel of $\Phi$ is forever lost and we cannot hope to recover them. It is convenient to truncate the physical algebra so that the error map $\Phi$ becomes faithful.
Define $P$ to be the projection to the span of $V_r\mK$ and replace $\mA$ by $P\mA P$. The projection $P$ projects down to the code subspace.
With this truncation the recovery equation uniquely fixes the recovery map. 
If $\mR_1$ and $\mR_2$ are two recovery maps we have $(\mR_1(c)-\mR_2(c))V_r=0$ and since the span of the range of all $V_r$ is the whole Hilbert space we find $\mR_1(c)=\mR_2(c)$ for all $c\in \mB^C$.
This unique recovery map $\mathcal{R}:\mB^C\to P\mA P$ is a representation because it satisfies
\begin{eqnarray}\label{multiplyV}
    \mR(c_1)\mR(c_2)V_r=\mR(c_1c_2)V_r
\end{eqnarray}
It is a faithful representation because none of the errors can kill code states.
A faithful representation establishes a $*$-isomorphism between the algebras $\mB^C$ and the subalgebra $\mA^C\equiv\mR(\mA^C)\subset \mA$. 
The superoperator $\mathcal{R}$ corresponds to an isometry $W:\mH_B\to \mH_A$ in the Hilbert space, and hence it is an isometric embedding of $\mB^C$ in $\mA$; see figure \ref{fig15u}.

The errors acting on $\mB^C$ are correctable if there exists a solution to the recovery equation (\ref{recoveryeq}). Consider a self-adjoint operator $c\in \mB^C$. If the equation holds (\ref{recoveryeq}) we find that
\begin{eqnarray}
      V_r^\dagger \mR(c) V_s=V_r^\dagger V_s c=c V_r^\dagger V_s\ .
 \end{eqnarray}
The self-adjoint operators in the code algebra satisfy the commutation relation $[c,V_r^\dagger V_s]=0$ for all $r,s$. 
We will see below the converse also holds and the correctable subalgebra can be defined as the commutant of the set of operators $V_r^\dagger V_s$ for all $r,s$. Note that the correctable algebra always includes the identity operator. If we pick our operators to be inside the correctable algebra we are guaranteed that there exist recovery maps that correct the errors.

In the case of a single error, it is straightforward to see that the dual map $\Phi^*$ is a recovery map because $V$ is an isometry. When there are several errors the dual map is $\Phi^*(c)=\sum_r V_r c V_r^\dagger$ satisfies
\begin{eqnarray}
   \Phi^*(c)V_r=\sum_sV_s c V_s^\dagger V_r=(\sum_s V_s V_s^\dagger)V_r c=\Phi^*(\mI)V_r c\ .
\end{eqnarray}
If $\Phi$ is faithful the dual map $\Phi^*$ is invertible and $\mR(c)=\Phi^*(\mI)^{-1}\Phi^*(c)$ solves the recovery equation in (\ref{recoveryeq}).\footnote{$\Phi$ is faithful, therefore there exists no projection $p\in \mA$ such that $\Phi(p)=0$. Since $\Phi(p)$ is a positive operator we have $\tr(\Phi(p))\neq 0$. This implies that $\tr(p \Phi^*(\mI))\neq 0$ for all projections $p$. In other words, $\Phi^*(\mI)$ is full rank.} Otherwise, we define $\Phi^*(\mI)$ on the orthogonal complement of the kernel of $\Phi$.
While not manifest from its form, this map is CP. It follows from the recovery equation that
\begin{eqnarray}
   \mR(c)\Phi^*(\mI)=\Phi^*(c)\ .
\end{eqnarray}
Therefore, $\mR(c)=\Phi^*(\mI)^{-1}\Phi^*(c)=\Phi^*(c)(\Phi^*(\mI))^{-1}$, and as a result  $[\Phi^*(c),(\Phi^*(\mI))^{-1}]=0$. To make the recovery map manifestly positive we write it in the form \cite{beny2007generalization}\footnote{If $A$ and $B$ are commuting positive matrices then $A$ and $B^{1/2}$ commute.}
\begin{eqnarray}\label{uniquerecovery}
   \mR(c)=(\Phi^*(\mI))^{-1/2}\Phi^*(c)(\Phi^*(\mI))^{-1/2}\ .
\end{eqnarray}
The map above is the unique recovery map $\mR:\mB^C\to P\mA P$.


\subsection{Passive error correction}\label{passiveQEC}

\begin{figure}[t]
    \centering
    \includegraphics[width=0.25\linewidth]{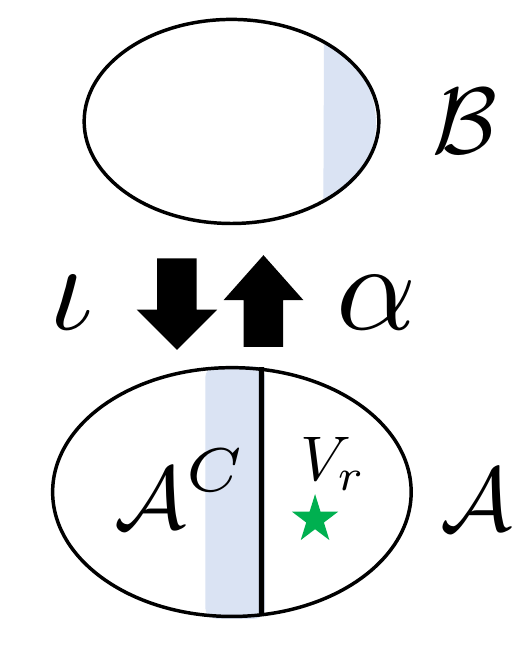}
    \caption{Passive error correction: The physical algebra $\mA$. The subalgebra $\mA^I\subset \mA$ of physical operators is used to simulate system $\mB$ because it is left invariant by the error map $\Phi$.}
    \label{fig14u}
\end{figure}

Perhaps the easiest way to protect against errors is to find an encoding of the algebra $\mB$ in the physical Hilbert space that is immune to errors so that we do not need to correct at all. 
We achieve this if we choose our code operators from the subalgebra $\mA^I$ that is invariant under the action of the error $\iota\circ \Phi:\mA\to \mA$; see figure \ref{fig14u}. For simplicity, in this case, we can absorb $\iota$ in $\Phi$ so that we get rid of $\mB$ all together. We have $\mA^C\equiv \iota(\mB)$ a subalgebra of $\mA$ and an error map that with a slight abuse of notation we write as $\Phi(a)=\sum_r V_r^\dagger cV_r$. 
As we showed in section \ref{sec:CPmaps&duals} an operator $c\in \mA^I$ if and only if $[c,V_r]=[c,V_r^\dagger]=0$ for all $r$; see figure \ref{fig14u}. 
The commutant algebra $(\mA^I)'$ is sometimes called the {\it interaction algebra}.\footnote{The interaction algebra is the double commutant of the set of errors $\{V_r,V_r^\dagger\}$.}

Encoding operators in the invariant subalgebra $\mA^I$ to protect them from an error map $\Phi$ has the advantage that we are simultaneously protected against any other error map whose Kraus operators are in the interaction algebra.
For instance, in matrix algebras, we are also protected against any error map
\begin{eqnarray}\label{manyerrors}
     &&\Phi_{\rho}(a)=\Phi(\rho)^{-1/2}\Phi(\rho^{1/2} a \rho^{1/2})\Phi(\rho)^{-1/2}
    \end{eqnarray}
where $\rho$ satisfies the condition:\footnote{In theorem \ref{thmTakesaki1} we show that this is equivalent to the Takesaki condition.}
\begin{eqnarray}\label{property}
 &&\rho^{1/2}c\rho^{-1/2}=\Phi(\rho)^{1/2}c\Phi(\rho)^{-1/2}\in \mA^I\ .
\end{eqnarray}
Note that the constraint above implies
\begin{eqnarray}
   \Phi(\rho^{1/2}c \rho^{1/2})=\sum_r V_r^\dagger \rho^{1/2}c\rho^{1/2}V_r=\Phi(\rho)\rho^{-1/2}c \rho^{1/2}\ .
\end{eqnarray}
Plugging this in the new error map gives
\begin{eqnarray}
   \Phi_\rho(c)=\Phi(\rho)^{1/2}\rho^{-1/2}c \rho^{1/2}\Phi(\rho)^{-1/2}=c
\end{eqnarray}
where we have used (\ref{property}) again.


In passive error correction the invariant subalgebra is the code subalgebra. Next, using the intuition from section \ref{sec:fixed}, for any error map we construct an explicit conditional expectation that projects down to the invariant subalgebra.

Given a unital CP map $\Phi$ that preserves some faithful state $\rho$, the $\rho$-preserving map
\begin{eqnarray}\label{vNergodic}
    \mE_\rho(a)=\lim_{n\to \infty}\frac{1}{n}(a+\Phi(a)+\Phi^2(a)+\cdots \Phi^{n-1}(a))
\end{eqnarray}
is a conditional expectation that projects to the invariant subalgebra of $\Phi$. To see this, consider the Stinespring representation $\Phi(a)=W^\dagger \pi(a) W$. Since $\Phi$ is unital, $W$ is an isometry. Both the representation map $a\to \pi(a)$ and the compression $\pi(a)\to W^\dagger \pi(a)W$ are norm non-increasing, hence $\|\Phi(a)\|\leq \|a\|$, and
\begin{eqnarray}
   \| \Phi(\mE_\rho(a))-\mE_\rho(a)\|=\|\lim_{n\to \infty}\frac{1}{n}(\Phi^n(a)-a)\|&&\leq\lim_{n\to\infty}\frac{1}{n}\lb\|\Phi^n(a)\|+\|a\|\rb\nn\\
   &&\leq  \lim_{n\to\infty}\frac{2}{n}\|a\|=0\ .
\end{eqnarray}
We find that the range of $\mE_\rho$ is $\mA^I$. This map is evidently CP and leaves every operator in $\mA^I$ invariant; therefore it is a $\rho$-preserving conditional expectation.

\subsection{Active error correction}\label{sec:activeEC}

\begin{figure}[t]
    \centering
    \includegraphics[width=0.7\linewidth]{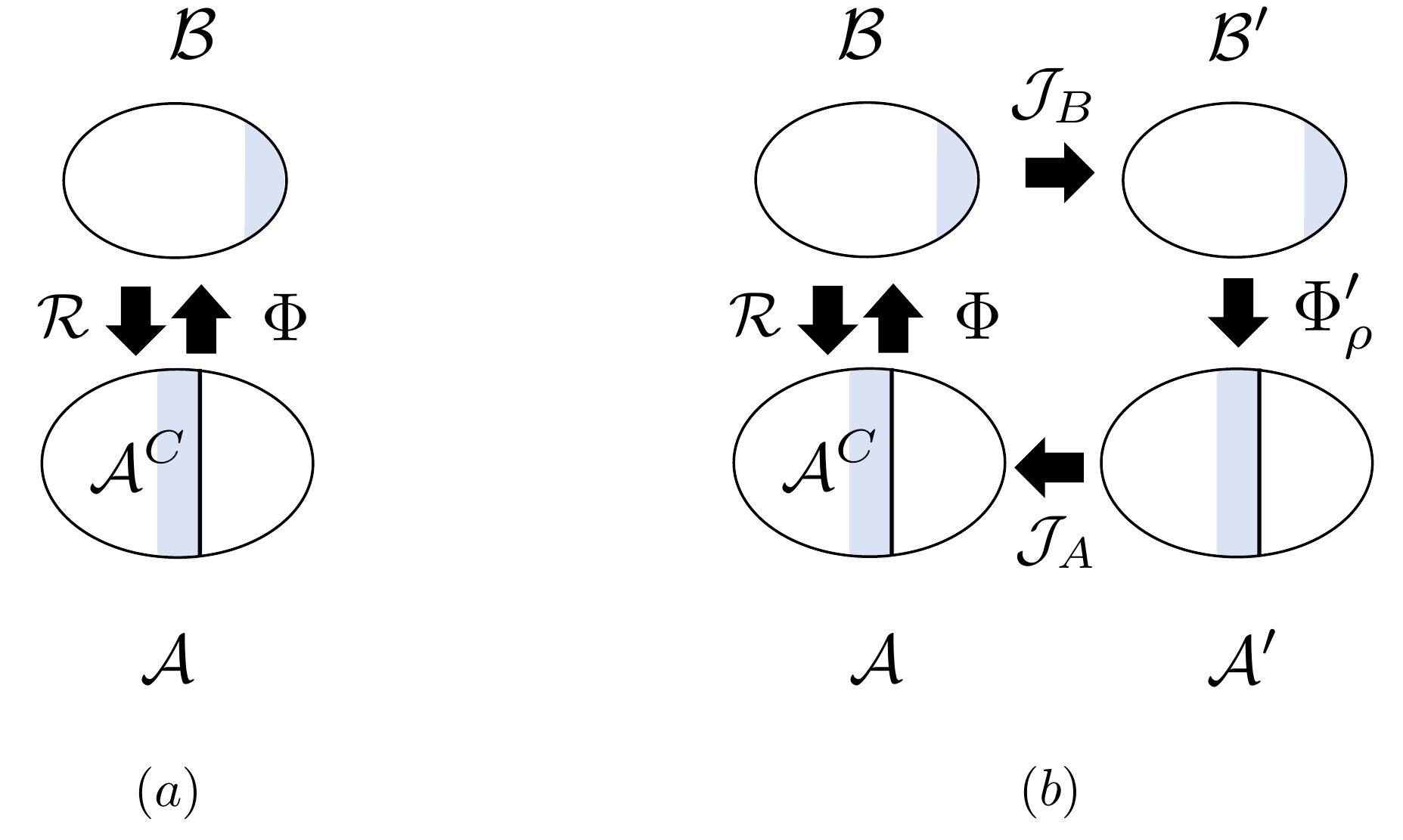}
    \caption{(a) Active error correction where the whole algebra $\mB$ is correctable: The action of the error map $\Phi:\mA\to \mB$ on the correctable subalgebra $\mA^C$ can be undone using the recovery map $\mR$. (b) The Petz dual map as the recovery map.}
    \label{fig15u}
\end{figure}

Passive error correction is convenient when there are a few types of errors. If we have a large set of errors we might not have the luxury of finding a large  invariant subalgebra to encode all our operators. Then, we have to apply recovery map to correct errors. 
The recovery equation in (\ref{recoveryeq}) only fixes the action of $\mR$ on $\mB^C$. For simplicity, in the remainder of this section, we focus on the case where the whole algebra $\mB$ is correctable. We will also assume that the kernel of the error map is trivial so that the recovery map is unital.
In this case, the composite map $\mathcal{R}\circ \Phi:\mA\to \mA^C$ is a conditional expectation. A state $\rho_A$ is correctable if it is invariant under this conditional expectation. The theorem \ref{thmTakesaki1} below characterizes all states that are preserved under a conditional expectation from $\mA$ to $\mA^C$. Given such a correctable state, the recovery map is the Petz dual of the error map which corrects the errors on a set of states; see figure \ref{fig15u}. Such states are called {\it sufficient} with respect to this error map. We will see that the relative entropy of any pair of sufficient states remains unchanged under the error map.

In matrix algebras,
there always exists a trace-preserving conditional expectation $\mE_e:\mA\to \mA^C$ if $\mA^C$ contains the identity operator.
To show this, we start with the orthogonal projection $P_e$ in the Hilbert space $\mH_e$
that projects down to $\mH_C$ that is the span of $\mA^C\ket{e}$. We show that the superoperator that is associated with it is a trace-preserving conditional expectation. Since $P_ec\ket{e}=c\ket{e}$ the superoperator $\mE_e$ satisfies $\mE_e(c)=c$ for all $c\in\mA^C$. Furthermore, we have
\begin{eqnarray}
 \braket{e|\mE_e(a)|e}=\braket{e|P_e a|e}=\braket{P_ee|a|e}=\braket{e|a|e},
\end{eqnarray}
therefore $\mE_e$ is trace-preserving. We only need to prove it is CP.

To show that $\mE_e(a_+)$ is positive we need to show the matrix element
\begin{eqnarray}
  \braket{a_2|\mE_e(a_+)|a_2}=\braket{a_2|P_ea_+ a_2}=\braket{P_ea_2|a_+a_2}
\end{eqnarray}
is positive.
It is clear that if $\ket{a}\in (P_e)_\perp$ this matrix element is zero, therefore we only need to consider $\braket{c|\mE(a_+)|c}$ for $c\in \mA^C$. The inner product in the Hilbert space $\mH_e$ has the special property that 
\begin{eqnarray}\label{propHe}
   \braket{a_1|a_2 a_1}=\tr(a_1^\dagger a_2 a_1)=\tr(a_1 a_1^\dagger a_2)=\braket{a_1^\dagger a_1|a_2}
\end{eqnarray}
where we have used the cyclicity of trace.
Therefore,
\begin{eqnarray}
   &&\braket{c|\mE_e(a_+)|c}=\braket{c^\dagger c|Pa_+}=\braket{P c^\dagger c|a_+}=\braket{c^\dagger c|a_+}=\braket{c|a_+|c}\geq 0\ .
\end{eqnarray}
Therefore, $\mE_e$ is a positive map. Similarly, the map $\mE_e\otimes \text{id}_n$ corresponds to $P_e\otimes \mI_n$ in the Hilbert space $\mH_e\otimes \mK_n$ which is also positive by the same argument, therefore $\mE_e$ is CP. The superoperator $\mE_e$ is the unique trace-preserving conditional expectation from $\mA\to \mA^C$.\footnote{The bi-module property follows from
\begin{eqnarray}\label{arg2}
\tr(c_1 \mE_e(c_2 a))=\braket{c_1^\dagger|P_e
    c_2 a}=\braket{P_e  c_1^\dagger|c_2 a}=\braket{c_2^\dagger P_ec_1^\dagger|a}=\braket{P_e c_2^\dagger c_1^\dagger| a}=\tr(c_1 c_2 \mE_e(a))\nn\ .
\end{eqnarray}}
If a density matrix $\rho$ satisfies the Takesaki condition the conditional expectation in (\ref{vNergodic}) that corresponds to $\Phi_\rho$ in (\ref{manyerrors}) preserves $\rho$. 
In fact, we can explicitly write down the $\rho$-preserving conditional expectation in terms of the trace-preserving one:
\begin{eqnarray}\label{condrho}
    \mE_\rho(a)=\rho_C^{-1/2}\mE_e(\rho^{1/2}a \rho^{1/2})\rho_C^{-1/2}
\end{eqnarray}
where $\rho_C$ is the restriction of $\rho$ to the subalgebra $\mA^C$.
These maps are the same as the $\rho$-preserving conditional expectations constructed in section \ref{sec:examplesoofCPmaps}.

We now prove that the Takesaki condition in (\ref{property}) is the necessary and sufficient condition for a state for the existence of a $\rho$-preserving conditional expectation. The argument trivially generalizes to infinite dimensions \cite{takesaki1972}.
\begin{theorem}[Takesaki's condition: matrix algebras]\label{thmTakesaki1}
The following statements are equivalent:
\begin{enumerate}
    \item There exists a $\rho$-preserving conditional expectation $\mE_\rho:\mA\to \mA^C$.
    \item For all $c\in \mA^C$ we have $\rho^{1/2}c\rho^{-1/2}\in \mA^C$.
    \item For all $c\in \mA^C$ we have $\rho^{1/2}c\rho^{-1/2}=\rho_C^{1/2}c\rho_C^{-1/2}$.
\end{enumerate}
Here, $\rho_C$ is the restriction of $\rho$ to $\mA^C$.
\end{theorem} 
{\bf Proof:}

{\bf (2 $\to$ 1):} Repeating the argument above for the projection $P_\rho$ in the GNS Hilbert space to the subspace $\mH_C$ spanned by $\mA^C\ket{\rho^{1/2}}$ reveals why there might not exist a $\rho$-preserving conditional expectation for an arbitrary $\rho$. By the same argument, the projection $P_\rho$ corresponds to a superoperator $\mE_\rho:\mA\to \mA^C$ that preserves $\rho$ and satisfies $\mE_\rho(c)=c$. However, in general, it will not be CP because there is no analog of the property (\ref{propHe}) in the GNS Hilbert space $\mH_\rho$. Instead, we have
\begin{eqnarray}
 \braket{a_1|a_2a_1}_\rho=\tr(a_1\rho a_1^\dagger a_2)=\tr(\rho (\rho^{-1} a_1\rho)a_1^\dagger a_2)=\braket{a_1\mathcal{D}_\rho(a_1^\dagger)|a_2}_\rho
\end{eqnarray}
where $\mathcal{D}_\rho(a)=\rho a\rho^{-1}$ is the modular superoperator we introduced in section (\ref{sec:supervsoperator}). If $\mathcal{D}_\rho(c)\in \mA^C$ we can repeat the argument above to show
\begin{eqnarray}\label{positiveGNSop}
 \braket{c|\mE_\rho(a_+)c}_\rho&=&\braket{c\mathcal{D}_\rho(c^\dagger)|P_\rho a_+}_\rho=\braket{P_\rho c\mathcal{D}_\rho(c^\dagger)|a_+}_\rho\nn\\
    &=&\braket{c\mathcal{D}_\rho(c^\dagger)|a_+}_\rho=\braket{c|a_+c}_\rho\geq 0\ .
\end{eqnarray}
Therefore, if $\mathcal{D}_\rho(c)\in \mA^C$ the superoperator $\mE_\rho(c)$ is CP and hence it is the unique $\rho$-preserving conditional expectation from $\mA$ to $\mA^C$.
If $\mathcal{D}^{1/2}_\rho(c)\in \mA^C$ so is $\mathcal{D}_\rho(c)\in \mA^C$, therefore the condition in (2) is sufficient for (1).

{\bf (1$\to$ 2):}
Assume that $\mE_\rho$ exists and $P_\rho$ is its corresponding projection operator in $\mH_\rho$. Consider the Tomita superoperator 
$\mS(a)=a^\dagger$.
Since $\mE_\rho$ is a positive map we have $\mE_\rho(a^\dagger)=\mE_\rho(a)^\dagger$ which implies $\mE_\rho(\mS(a))=\mS(\mE_\rho(a))$. In the GNS Hilbert space, this implies $[P_\rho, S_\rho]=0$. Since $P_\rho$ is self-adjoint when $\mE_\rho$ is $\rho$-preserving we also have $[P_\rho,S_\rho^\dagger]=0$. Therefore, we find  $[P_\rho,\Delta_\rho]=0$, where  $\Delta_\rho=S^\dagger_\rho S_\rho$ is the modular operator of $\rho$. Since both operators are positive we have $[P_\rho, \Delta_\rho^{1/2}]=0$, and 
using the superoperator representation we obtain $\mE(\mathcal{D}^{1/2}_\rho(a))=\mathcal{D}^{1/2}_\rho(\mE(a))$.
For any $c\in\mA^C$:
\begin{eqnarray}
\mE_\rho(\mathcal{D}^{1/2}_\rho(c)) = \mathcal{D}_\rho^{1/2}(\mE_\rho(c))=\mathcal{D}^{1/2}_\rho(c)\ . 
\end{eqnarray}
Therefore, $\mathcal{D}_\rho(c)=\rho^{1/2} c\rho^{-1/2}\in \mA^C$. 

{\bf (1$\to$ 3):} We saw that (1) implies the commutation relation $[P_\rho,\Delta_\rho]=0$. Define the state $\rho_C$ on the subalgebra $\mA^C$ as the restriction $\tr(\rho_C c)=\tr(\rho c)$.\footnote{Note that $\rho_C=\mE_e(\rho)$ because $\tr(c \rho_C)=\tr(c \rho)=\tr(\mE_e(c \rho))=\tr( c \mE_e(\rho))$.}
Consider its GNS Hilbert space $\mH_C$ spanned by $c\ket{\rho_C^{1/2}}$ and the linear map $W:\mH_C\to \mH_A$:
\begin{eqnarray}\label{isomembed}
    W c\ket{\rho_C^{1/2}}=c\ket{\rho^{1/2}}\ .
\end{eqnarray}
It follows from the definition of $\rho_C$ that this linear map is an isometry and $W\mA^CW^\dagger$ is an isometric embedding of $\mA^C$ in $\mA$. Acting with the modular operator we find
\begin{eqnarray}
    S_\rho Wc\ket{\rho^{1/2}_C}=c^\dagger \ket{\rho^{1/2}}=WS_C c\ket{\rho_C^{1/2}}\ .
\end{eqnarray}
In other words, $S_\rho W=WS_C$ and as a result we have $W^\dagger \Delta_\rho W=\Delta_C$ and $P_\rho \Delta_\rho P_\rho=W\Delta_CW^\dagger$. When $[\Delta_\rho,P_\rho]=0$ we can take the square root of this equation to find
\begin{eqnarray}
    P_\rho\Delta_\rho^{1/2}=W\Delta_C^{1/2}W^\dagger
\end{eqnarray}
or equivalently\footnote{We act with $W^\dagger$ on the left and take the Hermitian conjugate.}
\begin{eqnarray}
    \Delta_\rho^{1/2}W=W\Delta^{1/2}_C\ .
\end{eqnarray}
This together with $S_\rho W=WS_C$ gives the form of the Takesaki condition $J_\rho W=WJ_C$.
Then, the constraint that $\mathcal{D}_\rho(c)\in \mA^C$ becomes
\begin{eqnarray}
    \mathcal{D}_\rho^{1/2}(c)\ket{\rho^{1/2}}=P_\rho\Delta_\rho^{1/2}c\ket{\rho^{1/2}}=W\Delta_C^{1/2}c\ket{\rho_C^{1/2}}=W\mathcal{D}_C^{1/2}(c)\ket{\rho_C^{1/2}}=\mathcal{D}_C^{1/2}(c)\ket{\rho^{1/2}}\ .
\end{eqnarray}
As a result, we have
\begin{eqnarray}
    \rho^{1/2}c\rho^{-1/2}=\mathcal{D}_\rho^{1/2}(c)=\mathcal{D}_C^{1/2}(c)=\rho_C^{1/2}c\rho_C^{-1/2}
\end{eqnarray}
which is the condition in Takesaki's theorem.

{\bf (3$\to$ 1):}
Consider a subalgebra $\mA^C\subseteq \mA$ and the isometric embedding map $\iota:\mA^C\to \mA$. The Petz dual $\iota^P_\rho:\mA\to \mA^C$ is unital and CP. It follows from the definition of the alternate inner product in (\ref{alteranteinner}) that the Petz dual satisfies  
\begin{eqnarray}
    \braket{\iota^P_\rho(a)|\Delta^{1/2}_C c}_{\rho_C}=\braket{a| \Delta_\rho^{1/2} c}_\rho\ .
\end{eqnarray}
We now show that when (3) is satisfied this Petz dual map is a $\rho$-preserving conditional expectation. All we need to show is that $\iota^P_\rho(c)=c$:
\begin{eqnarray}
     \braket{\iota^P_\rho(c_1)|\Delta^{1/2}_C c_2}_{\rho_C}&=&\braket{c_1 |\Delta_\rho^{1/2}c_2}_\rho=\braket{c_1|\mathcal{D}_\rho^{1/2}(c_2)}_\rho\nn\\
     &=&\braket{c_1|\mathcal{D}_C^{1/2}(c_2)}_\rho=\braket{c_1|\Delta_C^{1/2}c_2}_{\rho_C}
\end{eqnarray}
where in the second line we have used (3) and $c_1^\dagger \mathcal{D}_C^{1/2}(c_2)\in \mA^C$. Since the isometric embedding is trivial in this case the composite map $\mE^P_\rho=\iota\circ \iota^P_\rho:\mA\to \mA^C$ is a $\rho$-preserving generalized conditional expectation that becomes a conditional expectation (3) is satisfied. $\Box$

All the steps of the arguments above can be repeated for an arbitrary von Neumann algebra with $\rho^{1/2}$ replaced with $\Delta_\rho^{1/2}$. The proof did not rely on the existence of a density matrix or a trace, and trivially generalizes to an arbitrary von Neumann algebra and its GNS Hilbert space representation:

\begin{theorem}[Takesaki's condition: von Neumann algebras]\label{thmTakesaki2}

Let $\mA^C\subset \mA$ be an inclusion of von Neumann algebras. Let $\rho_A$ be a faithful state of $\mA$ and $\rho_C$ be its restriction to $\mA^C$. Let $\ket{\rho_A^{1/2}}$ and $\ket{\rho_C^{1/2}}$ be the cyclic and separating vectors in $\mH_A$ and $\mH_C$. Define the isometry $W:\mH_C\to \mH_A$ as $Wc\ket{\rho_C^{1/2}}=c\ket{\rho_A^{1/2}}$ for all $c\in \mA^C$. The following statements are equivalent:
\begin{enumerate}
    \item There exists a $\rho_A$-preserving conditional expectation $\mE_\rho:\mA\to \mA^C$
    \item The modular conjugations $J_A$ and $J_C$ corresponding to $\ket{\rho_A^{1/2}}$ and $\ket{\rho_C^{1/2}}$ satisfy $J_AW=WJ_C$.
    \item $\Delta_A^{1/2}W=W\Delta_C^{1/2}$.
\end{enumerate} 
\end{theorem} 
Our next question is given a $\rho$-preserving conditional expectation what other states are also invariant under it?
To characterize all ``sufficient'' states of a $\rho$-preserving conditional expectation $\mE_\rho$ we show that it preserves another state $\omega$ if and only if the {\it sufficiency condition}
\begin{eqnarray}\label{suffcond}
\omega^{1/2}\omega_C^{-1/2}=\rho^{1/2}\rho_C^{-1/2}    
\end{eqnarray}
is satisfied \cite{petz1986sufficient,petz1988sufficiency}.
If we are given a $\rho$-preserving conditional expectation $\mE_\rho$ the map
\begin{eqnarray}\label{omegarho}
    \mE_\rho^\omega(a)=\omega_C^{-1/2}\rho_C^{1/2}\mE_\rho\lb \rho^{-1/2}\omega^{1/2}a\omega^{1/2}\rho^{-1/2}\rb\rho_C^{1/2}\omega_C^{-1/2}
\end{eqnarray}
is a $\omega$-preserving CP map from $\mA\to \mA^C$. If it preserves every operator in $c\in \mA^C$ it becomes an $\omega$-preserving conditional expectation. It is clear that if sufficiency condition in (\ref{omegarho}) holds it becomes an $\omega$-preserving conditional expectation $\mE_\omega=\mE_\rho$. Therefore, $\mE_\rho$ also preserves $\omega$.
We now prove the converse: the conditional expectation $\mE_\rho$ preserves $\omega$ only if the condition (\ref{suffcond}) holds. We basically repeat the proof of Takesaki's theorem for the relative Tomita operator $S_{\omega|\rho}a\ket{\rho^{1/2}}=a^\dagger \ket{\omega^{1/2}}$. The norm of this operator is the relative modular operator $\Delta_{\omega|\rho}:\mH_\rho\to \mH_\rho$. The superoperator corresponding to it is $\mathcal{D}_{\omega|\rho}(a)=\omega a \rho^{-1}$.
We repeat the argument for the Takesaki theorem with the relative modular map $\mathcal{D}_{\omega|\rho}(a)=\omega a \rho^{-1}$ to find $[P_\rho,\Delta_{\omega|\rho}^{1/2}]=0$. This implies 
\begin{eqnarray}
    \mE_\rho(\mathcal{D}_{\omega|\rho}^{1/2}(c))=\mathcal{D}_{\omega|\rho}^{1/2}(\mE_\rho(c))=D_{\omega|\rho}^{1/2}(c)\in \mA^C
\end{eqnarray}
We define the isometries 
\begin{eqnarray}
    &&W_\rho c\ket{\rho_C^{1/2}}=c\ket{\rho^{1/2}}\nn\\
    &&W_\omega c\ket{\omega_C^{1/2}}=c\ket{\omega^{1/2}}
\end{eqnarray}
so that
\begin{eqnarray}
    &&S_{\omega|\rho}W_\rho=W_\omega S_{\omega_C|\rho_C}\nn\\
     &&W^\dagger_\rho \Delta_{\omega|\rho}W_\rho=\Delta_{\omega_C|\rho_C}\ .
\end{eqnarray}
Since $[P_\rho,\Delta_{\omega|\rho}^{1/2}]=0$ we have
\begin{eqnarray}
  &&  P_\rho\Delta_{\omega|\rho}^{1/2}=W_\rho\Delta^{1/2}_{\omega_C|\rho_C}W_\rho^\dagger\ .
\end{eqnarray}
As a result, 
\begin{eqnarray}
    &&\mathcal{D}^{1/2}_{\omega|\rho}(c)\ket{\rho^{1/2}}=P_\rho  \Delta^{1/2}_{\omega|\rho}c\ket{\rho^{1/2}}=W_\rho\Delta^{1/2}_{\omega_C|\rho_C}c\ket{\rho_C^{1/2}}\nn\\
    &&=W_\rho\mcD^{1/2}_{\omega_C|\rho_C}(c)\ket{\rho_C^{1/2}}=\mcD^{1/2}_{\omega_C|\rho_C}(c)\ket{\rho^{1/2}}\ .
\end{eqnarray}
We obtain that 
\begin{eqnarray}
    \omega^{1/2}c\rho^{-1/2}=\mcD_{\omega|\rho}^{1/2}(c)=\mcD_{\omega_C|\rho_C}^{1/2}(c)=\omega_C^{1/2}c \rho^{-1/2}_C\ .
\end{eqnarray}
In other words,
\begin{eqnarray}
    \omega_C^{-1/2}\omega^{1/2}c \rho^{1/2}\rho_C^{-1/2}=c=\rho_C^{-1/2}\rho^{1/2}c\rho^{-1/2}\rho_C^{1/2}
\end{eqnarray}
which holds if and only if the sufficiency condition in (\ref{suffcond}) is satisfied.

The sufficiency condition can be expressed as
\begin{eqnarray}
    \Delta^{1/2}_{\omega|\rho}=W_\rho\Delta^{1/2}_{\omega_C|\rho_C}W_\rho^\dagger\ .
\end{eqnarray}
Using the integral representation of $X^\alpha$ for $\alpha \in (0,1)$
\begin{eqnarray}\label{power}
   X^{\alpha}=\frac{\sin(\pi\alpha)}{\pi}\int_0^\infty ds\: s^\alpha\lb\frac{1}{s}-\frac{1}{s+X} \rb
\end{eqnarray}
we find
\begin{eqnarray}
    \int_0^\infty ds\: s^{1/2}\lb \frac{1}{s+\Delta_{\omega|\rho}}-W_\rho\frac{1}{s+\Delta_{\omega_C|\rho_C}}W_\rho^\dagger\rb=0\ .
\end{eqnarray}
From the monotonicity of the relative modular operator \cite{witten2018aps,nielsen2004simple} we know that the operator in the integrand above is positive, therefore it has be zero:
\begin{eqnarray}
    \frac{1}{s+\Delta_{\omega|\rho}}=W_\rho\frac{1}{s+\Delta_{\omega_C|\rho_C}}W_\rho^\dagger
\end{eqnarray}
which implies 
\begin{eqnarray}
    \Delta^\alpha_{\omega|\rho}=W_\rho\Delta^\alpha_{\omega_C|\rho_C}W_\rho^\dagger\ .
\end{eqnarray}
Furthermore, for any continuous function $f$ we have
\begin{eqnarray}
   W_\rho f(\Delta_C)\ket{\rho^{1/2}_C}=f(\Delta)\ket{\rho^{1/2}}\ .
\end{eqnarray}
In particular, choosing $f(x)=x^{it}$ for $t\in\mathbb{R}$ we find that $\rho_C^{it}\omega_C^{-it}=\rho^{it}\omega^{-it}$.
This condition implies that the relative entropy for any pair of sufficient states $\rho$ and $\omega$:
\begin{eqnarray}
    S(\omega\|\rho)=S(\omega_C\|\rho_C)\ .
\end{eqnarray}

Intuitively, this says that a coarse-graining (conditional expectation) preserves a set of states $\{\rho_k\}$ (sufficient states) if and only if the distinguishability (relative entropy) of any pair of them remains the same.

\subsection{Condition for exact error correction}\label{app:QECintuition}

The purpose of this appendix is to provide an intuitive understanding of why the Takesaki condition $WJ_B=J_AW$ is the necessary and sufficient condition for exact quantum error correction.
To see how the subsystem error correction works in the language of algebras and operators consider a vector $\ket{\Omega_B}\in \mH_B$ that we encode in $\ket{\Omega_A}=W\ket{\Omega_B}$. The encoding is such that the actions of the errors $V_r$ in $A$ can be undone using the correction operators $R_r$: $R_rV_r\ket{\Omega_A}\propto\ket{\Omega_A}$.
We call the vector $\ket{\Omega_A}$ correctable. The goal is to protect the operators in $B'$ from the errors $V_r$ in $A$.
The encoding map sends all vectors $b'\ket{\Omega_B}\in \mH_B$ to $Wb'\ket{\Omega_B}$. The errors occur and we act with our correction operators $R_r$ to obtain the vector $R_rV_r Wb'\ket{\Omega_B}$. Since $Wb' W^\dagger$ 
is supported on $A'$ it commutes with the errors $V_r$ and $R_r$:
\begin{eqnarray}\label{recovercommut}
    R_r V_rWb'\ket{\Omega_B}=R_r V_r(Wb' W^\dagger)\ket{\Omega_A}=(Wb' W^\dagger) R_r V_r\ket{\Omega_A}\propto Wb' \ket{\Omega_B}\ .
\end{eqnarray}
Therefore, if $W\ket{\Omega_B}$ is correctable all vectors $Wb'\ket{\Omega_B}$ are correctable. 


The subsystem error correction above is rather trivial. 
To make it more interesting one would like to correct for errors $V'_r$ that occur inside $A'$. Assume for a moment that we have complementary recovery which means that for all operators in $B$ the encoding $W b W^\dagger$ are correctable and supported on $A$. Furthermore, assume that there exist swap operators $J_A$ and $J_B$ that swap $A\leftrightarrow A'$ and $B \leftrightarrow B'$.
Then, a simple idea is to first use the swap to encode $b'$ in $B$, and then use the map $W b W^\dagger$ that brings to $A$. Now, it commutes with all the errors $V'_r$. 

We use the swap $J_A$ to map an error $V'_r$ to $A$ so that we can correct the vector $\ket{\Omega_A}$ using our correction operator $R_r$ in $A$:
\begin{eqnarray}
    R_rJ_A V'_r\ket{\Omega_A}\propto\ket{\Omega_A}\ .
\end{eqnarray}
With the assumption of complementary recovery, it is easy to see that the correction $R_r$ can correct all vectors $WJ_B b'\ket{\Omega_B}$:
\begin{eqnarray}
&&R_rJ_A V'_r WJ_B b'\ket{\Omega_B}=R_r J_A V'_r(WJ_B b' J_BW^\dagger)\ket{\Omega_A}\nn\\
&&=R_rJ_A(WJ_B b' J_BW^\dagger)V'_r\ket{\Omega_A}=J_A(WJ_Bb'J_BW^\dagger)J_A R_rJ_AV'_r\ket{\Omega_A}\nn\\
&&\propto J_A W J_B b'\ket{\Omega_B}
\end{eqnarray}
if $J_AW=WJ_B$ holds as an operator statement. Here, we have assumed that $J_A$ and $J_B$ are symmetries of the vectors $\ket{\Omega_A}$ and $\ket{\Omega_B}$ and $J_A^2$ is identity, and $J_A=J_A^\dagger$.\footnote{Swapping twice is the identity operation.} 
The encoding map that played a central role in correcting the errors in $A'$ is called the Petz map:
\begin{eqnarray}
    \iota(b')=J_AWJ_B b' J_B W^\dagger J_A\ .
\end{eqnarray}
In the Heisenberg picture, in the absence of any errors this encoding map should satisfy
\begin{eqnarray}
    \alpha(\iota(b'))=b'\ .
\end{eqnarray}
This means that we should have
\begin{eqnarray}
    W^\dagger (J_A W J_B b' J_B  W^\dagger J_A) W=b'\ .
\end{eqnarray}
It is clear that since $J_A^2=\mI$ and $J_B^2=\mI$ the condition $J_A W=WJ_B$ is sufficient to satisfy the equation above. Theorem \ref{thmTakesaki2} shows that the Takesaki condition $J_AW=WJ_B$ is also necessary for a vector to be correctable.

\bibliographystyle{JHEP}
\bibliography{main}

\providecommand{\href}[2]{#2}\begingroup\raggedright\begin{thebibliography}{10}

\bibitem{kribs2005unified}
D.~Kribs, R.~Laflamme, and D.~Poulin, {\it Unified and generalized approach to
  quantum error correction},  {\em Physical review letters} {\bf 94} (2005),
  no.~18 180501.

\bibitem{beny2007quantum}
C.~B{\'e}ny, A.~Kempf, and D.~W. Kribs, {\it Quantum error correction of
  observables},  {\em Physical Review A} {\bf 76} (2007), no.~4 042303.

\bibitem{beny2007generalization}
C.~B{\'e}ny, A.~Kempf, and D.~W. Kribs, {\it {Generalization of quantum error
  correction via the Heisenberg picture}},  {\em Physical review letters} {\bf
  98} (2007), no.~10 100502.

\bibitem{almheiriharlowdong2015bulk}
A.~Almheiri, X.~Dong, and D.~Harlow, {\it {Bulk locality and quantum error
  correction in AdS/CFT}},  {\em Journal of High Energy Physics} {\bf 2015}
  (2015), no.~4 163.

\bibitem{Harlow2017}
D.~Harlow, {\it {The Ryu--Takayanagi formula from quantum error correction}},
  {\em Communications in Mathematical Physics} {\bf 354} (2017), no.~3
  865--912.

\bibitem{faulkner2020holographic}
T.~Faulkner, {\it The holographic map as a conditional expectation},  {\em
  arXiv preprint arXiv:2008.04810} (2020).

\bibitem{Furuya:2021lgx}
K.~Furuya, N.~Lashkari, and M.~Moosa, {\it {Renormalization group and
  approximate error correction}},  \href{http://arxiv.org/abs/2112.05099}{{\tt
  arXiv:2112.05099}}.

\bibitem{evenbly2011tensor}
G.~Evenbly and G.~Vidal, {\it Tensor network states and geometry},  {\em
  Journal of Statistical Physics} {\bf 145} (2011), no.~4 891--918.

\bibitem{swingle2012entanglement}
B.~Swingle, {\it Entanglement renormalization and holography},  {\em Physical
  Review D} {\bf 86} (2012), no.~6 065007.

\bibitem{kim2017entanglement}
I.~H. Kim and M.~J. Kastoryano, {\it Entanglement renormalization, quantum
  error correction, and bulk causality},  {\em Journal of High Energy Physics}
  {\bf 2017} (2017), no.~4 40.

\bibitem{hayden2019learning}
P.~Hayden and G.~Penington, {\it Learning the alpha-bits of black holes},  {\em
  Journal of High Energy Physics} {\bf 2019} (2019), no.~12 1--55.

\bibitem{akers2019large}
C.~Akers, A.~Levine, and S.~Leichenauer, {\it Large breakdowns of entanglement
  wedge reconstruction},  {\em Physical Review D} {\bf 100} (2019), no.~12
  126006.

\bibitem{longo1995nets}
R.~Longo and K.-H. Rehren, {\it Nets of subfactors},  {\em Reviews in
  Mathematical Physics} {\bf 7} (1995), no.~04 567--597.

\bibitem{casini2011wedge}
H.~Casini, {\it Wedge reflection positivity},  {\em Journal of Physics A:
  Mathematical and Theoretical} {\bf 44} (2011), no.~43 435202.

\bibitem{hartman2017averaged}
T.~Hartman, S.~Kundu, and A.~Tajdini, {\it Averaged null energy condition from
  causality},  {\em Journal of High Energy Physics} {\bf 2017} (2017), no.~7
  66.

\bibitem{knill-laflamme1997qec}
E.~Knill and R.~Laflamme, {\it Theory of quantum error-correcting codes},  {\em
  Phys. Rev. A} {\bf 55} (Feb, 1997) 900--911.

\bibitem{haegeman2013entanglement}
J.~Haegeman, T.~J. Osborne, H.~Verschelde, and F.~Verstraete, {\it Entanglement
  renormalization for quantum fields in real space},  {\em Physical review
  letters} {\bf 110} (2013), no.~10 100402.

\bibitem{zou2019magic}
Y.~Zou, M.~Ganahl, and G.~Vidal, {\it Magic entanglement renormalization for
  quantum fields},  {\em arXiv preprint arXiv:1906.04218} (2019).

\bibitem{vidal2007entanglement}
G.~Vidal, {\it Entanglement renormalization},  {\em Physical review letters}
  {\bf 99} (2007), no.~22 220405.

\bibitem{vidal2008class}
G.~Vidal, {\it Class of quantum many-body states that can be efficiently
  simulated},  {\em Physical review letters} {\bf 101} (2008), no.~11 110501.

\bibitem{nishioka2009holographic}
T.~Nishioka, S.~Ryu, and T.~Takayanagi, {\it Holographic entanglement entropy:
  an overview},  {\em Journal of Physics A: Mathematical and Theoretical} {\bf
  42} (2009), no.~50 504008.

\bibitem{czech2012gravity}
B.~Czech, J.~L. Karczmarek, F.~Nogueira, and M.~Van~Raamsdonk, {\it The gravity
  dual of a density matrix},  {\em Classical and Quantum Gravity} {\bf 29}
  (2012), no.~15 155009.

\bibitem{Pastawski-preskill2015toymodel}
F.~Pastawski, B.~Yoshida, D.~Harlow, and J.~Preskill, {\it Holographic quantum
  error-correcting codes: toy models for the bulk/boundary correspondence},
  {\em Journal of High Energy Physics} {\bf 2015} (2015), no.~6 149.

\bibitem{chen2020entanglement}
C.-F. Chen, G.~Penington, and G.~Salton, {\it {Entanglement wedge
  reconstruction using the Petz map}},  {\em Journal of High Energy Physics}
  {\bf 2020} (2020), no.~1 168.

\bibitem{penington2019replica}
G.~Penington, S.~H. Shenker, D.~Stanford, and Z.~Yang, {\it Replica wormholes
  and the black hole interior},  2019.

\bibitem{accardi1982conditional}
L.~Accardi and C.~Cecchini, {\it {Conditional expectations in von Neumann
  algebras and a theorem of Takesaki}},  {\em Journal of Functional Analysis}
  {\bf 45} (1982), no.~2 245 -- 273.

\bibitem{kosaki1986extension}
H.~Kosaki, {\it {Extension of Jones' theory on index to arbitrary factors}},
  {\em Journal of functional analysis} {\bf 66} (1986), no.~1 123--140.

\bibitem{connes1980spatial}
A.~Connes, {\it On the spatial theory of von neumann algebras},  {\em Journal
  of Functional Analysis} {\bf 35} (1980), no.~2 153--164.

\bibitem{entanglemententropy_2020}
K.~Furuya, N.~Lashkari, and S.~Ouseph, {\it Generalized entanglement entropy,
  charges, and intertwiners},  {\em Journal of High Energy Physics} {\bf 2020}
  (Aug, 2020).

\bibitem{petz2007quantum}
D.~Petz, {\em Quantum information theory and quantum statistics}.
\newblock Springer Science \& Business Media, 2007.

\bibitem{watrous_2018}
J.~Watrous, {\em The Theory of Quantum Information}.
\newblock Cambridge University Press, 2018.

\bibitem{kribs2003quantum}
D.~W. Kribs, {\it Quantum channels, wavelets, dilations and representations of
  $\mathcal{O}_{n}$},  {\em Proceedings of the Edinburgh Mathematical Society}
  {\bf 46} (2003), no.~2 421–433.

\bibitem{choi1974schwarz}
M.-D. Choi et~al., {\it {A Schwarz inequality for positive linear maps on
  $C^*$-algebras}},  {\em Illinois Journal of Mathematics} {\bf 18} (1974),
  no.~4 565--574.

\bibitem{magan2020quantum}
J.~M. Magan and D.~Pontello, {\it Quantum complementarity through entropic
  certainty principles},  {\em arXiv preprint arXiv:2005.01760} (2020).

\bibitem{takesaki1972}
M.~Takesaki, {\it {Conditional expectations in von Neumann algebras}},  {\em
  Journal of Functional Analysis} {\bf 9} (1972), no.~3 306 -- 321.

\bibitem{witten2018aps}
E.~Witten, {\it Aps medal for exceptional achievement in research: Invited
  article on entanglement properties of quantum field theory},  {\em Reviews of
  Modern Physics} {\bf 90} (2018), no.~4 045003.

\bibitem{bratteli2012operator}
O.~Bratteli and D.~W. Robinson, {\em Operator Algebras and Quantum Statistical
  Mechanics: Volume 1: C*-and W*-Algebras. Symmetry Groups. Decomposition of
  States}.
\newblock Springer Science \& Business Media, 2012.

\bibitem{wolf2012quantum}
M.~M. Wolf, {\it Quantum channels \& operations: Guided tour},  {\em Lecture
  notes available at http://www-m5. ma. tum. de/foswiki/pub M} {\bf 5} (2012).

\bibitem{albeverio1978frobenius}
S.~Albeverio and R.~H{\o}egh-Krohn, {\it Frobenius theory for positive maps of
  von neumann algebras},  {\em Communications in Mathematical Physics} {\bf 64}
  (1978), no.~1 83--94.

\bibitem{junge2018universal}
M.~Junge, R.~Renner, D.~Sutter, M.~M. Wilde, and A.~Winter, {\it {Universal
  Recovery Maps and Approximate Sufficiency of Quantum Relative Entropy}},
  {\em Annales Henri Poincare} {\bf 19} (2018), no.~10 2955--2978,
  [\href{http://arxiv.org/abs/1509.07127}{{\tt arXiv:1509.07127}}].

\bibitem{wilde_2013}
M.~M. Wilde, {\em Quantum Information Theory}.
\newblock Cambridge University Press, 2013.

\bibitem{lashkari2018modular}
N.~Lashkari, H.~Liu, and S.~Rajagopal, {\it Modular flow of excited states},
  {\em arXiv preprint arXiv:1811.05052} (2018).

\bibitem{Lashkari_2019}
N.~Lashkari, {\it Constraining quantum fields using modular theory},  {\em
  Journal of High Energy Physics} {\bf 2019} (Jan, 2019).

\bibitem{connes2014noncommutative}
A.~Connes, J.~Cuntz, M.~A. Rieffel, and G.~Yu, {\it Noncommutative geometry},
  {\em Oberwolfach Reports} {\bf 10} (2014), no.~3 2553--2629.

\bibitem{yamagami1994modular}
S.~Yamagami, {\it Modular theory for bimodules},  {\em Journal of Functional
  Analysis} {\bf 125} (1994), no.~2 327--357.

\bibitem{petz1986sufficient}
D.~Petz, {\it {Sufficient subalgebras and the relative entropy of states of a
  von Neumann algebra}},  {\em Communications in mathematical physics} {\bf
  105} (1986), no.~1 123--131.

\bibitem{petz1988sufficiency}
D.~Petz, {\it {Sufficiency of channels over von Neumann algebras}},  {\em The
  Quarterly Journal of Mathematics} {\bf 39} (1988), no.~1 97--108.

\bibitem{nielsen2004simple}
M.~A. Nielsen and D.~Petz, {\it A simple proof of the strong subadditivity
  inequality},  {\em arXiv preprint quant-ph/0408130} (2004).

\end{thebibliography}\endgroup
\end{document}